\documentclass[12pt,fleqn]{report}
\usepackage[T1]{fontenc}
\usepackage[latin1]{inputenc}
\usepackage{geometry}
\usepackage{amsmath}
\usepackage{graphics}
\usepackage{setspace}
\usepackage{amssymb}
\usepackage{subfigure}

\makeatletter

\providecommand{\LyX}{L\kern-.1667em\lower.25em\hbox{Y}\kern-.125emX\@}

 \usepackage{verbatim}

\usepackage{here}
\ifx\nopictures Y \else \input epsf.tex \fi

\usepackage{here}
\usepackage{citesort}

\def\xh{\widehat x}
\def\zh{\widehat z}
\def\qTh{\widehat q_T}

\def\alpi{\frac{\alpha_S}{\pi}}                         
\def\sFs{\frac{\sigma_0 F_l}{S_{eA}}}

\def\D0{D\0~}
\def\ov{\overline}
\def\ra{\rightarrow}
\def\cms{c.m. }
\def\eps{\epsilon}

\def\nlb{\nolinebreak}
\def\beq{\begin{equation}}
\def\eeq{\end{equation}}
\def\bea{\begin{eqnarray}}
\def\eea{\end{eqnarray}}

\makeatother

\begin{document}

\newcommand{\wh}[1]{\widehat{#1 }}

\newcommand{\prescr}[1]{{\, }^{#1 }}

\newcommand{\wt}[1]{\widetilde{#1 }}

\newcommand{\Oas}{{\cal O}(\alpha _{S})}

\newcommand{\Vh}{\wh{V}}

\newcommand{\sigh}{\wh{\sigma }}

\newcommand{\MSbar}{\overline{MS}}

\newcommand{\whwtW}{\wh{\widetilde{W}}}

\newcommand{\ASud}{{\cal A}}

\newcommand{\BSud}{{\cal B}}

{\par\centering \pagenumbering{roman} \thispagestyle{empty}
\begin{flushright}
CTEQ-108\\
MSUHEP-10810
\end{flushright}
\vspace{0.5in}
{\large \bf 
MULTIPLE PARTON RADIATION IN HADROPRODUCTION\\
AT LEPTON-HADRON COLLIDERS}\\
~\\
By\\
Pavel M. Nadolsky \\
\vspace{1.35in}A THESIS\\
~\\
Submitted to \\
Michigan State University\\
in partial fulfillment of the requirements\\
for the degree of \\
~\\
DOCTOR OF PHILOSOPHY\\
~\\
Department of Physics \& Astronomy\\
~\\
2001\par}
\newpage
\thispagestyle{empty}

{\par\centering \thispagestyle{empty}\par}

{\par\centering ABSTRACT\par}

{\par\centering 
{\large \bf MULTIPLE PARTON RADIATION IN HADROPRODUCTION\\
AT LEPTON-HADRON COLLIDERS}\\
~\\
By \\
~\\
Pavel M. Nadolsky\\
~\\
\par}
Factorization of long- and short-distance hadronic dynamics in perturbative
Quantum Chromodynamics (QCD) is often obstructed by the coherent partonic
radiation, which leads to the appearance of large logarithmic terms
in coefficients of the perturbative QCD series. In particular, large
remainders from the cancellation of infrared singularities distort theoretical
predictions for angular distributions of observed products of hadronic
reactions. In several important cases, the predictive power of
QCD can be restored through summation of large logarithmic terms to all orders
of the perturbative expansion. Here I discuss the impact of the 
the coherent parton radiation on semi-inclusive production of hadrons 
in deep inelastic scattering at lepton-proton colliders. Such
radiation can be consistently described in the $b$-space resummation
formalism, which was originally 
developed  to improve theoretical description of
production of hadrons at $e^+e^-$ colliders and electroweak vector bosons at
hadron-hadron colliders. I present the detailed derivation of the resummed 
cross section and the energy flow at the next-to-leading order of
perturbative QCD. The theoretical results are compared to the experimental
data measured at the $ep$ collider HERA. A good agreement is found
between the theory and experiment in the region of validity of the
\newpage \thispagestyle{empty}\noindent
resummation formalism. I argue that semi-inclusive deep
inelastic scattering (SIDIS) at lepton-hadron colliders offers exceptional
opportunities to study coherent parton radiation, which
are not available yet at colliders of other types. Specifically,
SIDIS can be used to test the factorization of hard
scattering and collinear contributions at small values of $x$ 
and to search for potential crossing symmetry relationships between the
properties of the coherent radiation in SIDIS, $e^+e^-$ hadroproduction and
Drell-Yan processes. 
%
%
\newpage
{\par\centering \pagestyle{plain}\quad\vspace{3in}\par}

{\par\centering To Sunny, my true love and inspiration\par}
\newpage

{\par\centering \textbf{\large ACKNOWLEDGMENTS} \textbf{\large }\\
\large \par}
My appreciation goes to many people who helped me grow as a physicist.
Foremost, I am deeply indebted to my advisors C.-P. Yuan and Wu-Ki Tung,
who made my years at MSU a truly enriching and enjoyable experience.  
I feel very fortunate that C.-P.  and Wu-Ki agreed to work with me when I first
asked them to. Traditionally {\nolinebreak C.-P.} warns each new graduate student about the
challenges that accompany the career in high energy physics.
In my own experience, I found that benefits and satisfaction from the
work in the team with C.-P. and Wu-Ki outweigh all possible drawbacks. 

Both of my advisors have spent a significant effort and time to 
teach me new valuable knowledge and skills. They also strongly 
supported my interest in the subjects of my research, 
both morally and through material means. 
I am wholeheartedly grateful for their guidance and support.     
The remarkable scope of vision, vigor, and patience of C.-P.,
together with  profound knowledge, careful judgment and precision of Wu-Ki,
inspire me as model personal qualities required for a scientist.  

I owe my deep appreciation to my coauthor Dan Stump,
who spent many focused hours working with me on the topics in this thesis.
Dan's proofreading was the main driving force behind the improvements in my
English. 
A significant fraction of my accomplishments is due to the possibility of
open and direct communication between the graduate students, 
MSU professors, and members of the CTEQ Collaboration.
My understanding of resummation formalism was significantly enhanced through
discussions with John Collins, Jianwei Qiu, Davison Soper, 
and George Sterman. At numerous occasions, I had useful exchanges of ideas 
with Edmond Berger, Raymond Brock, Joey Huston, 
Jim Linnemann, Fred Olness, Jon Pumplin, Wayne Repko, and Carl Schmidt.
Many tasks were made easy by the interest and help from  
my fellow graduate students and research associates, most of all from
Jim Amundson, Csaba Balazs, Qinghong Cao, Dylan Casey, Chris Glosser, 
Hong-Jian  He, Shinya Kanemura, Stefan Kretzer, Frank Kuehnel, Liang Lai, 
Simona Murgia, Tom Rockwell, Tim Tait, and Alex Zhukov. 

Throughout this work I was using the CTEQ Fortran libraries, which were
mainly developed by W.-K. Tung, H.-L. Lai and J. Collins.  
The numerical package for resummation in SIDIS was developed on the basis of 
the programs Legacy and 
ResBos written by C.~Balazs, G.~Ladinsky and C.-P.~Yuan. 
S. Kretzer has provided me with the Fortran
parametrizations of the 
fragmentation functions. Some preliminary calculations 
included in this thesis were done by Kelly McGlynn. 

I am grateful to Michael Klasen and Michael Kramer for physics discussions and
invitation to give a talk at DIS2000 Workshop.
I learned important information about the BFKL resummation formalism
from Carl Schmidt, J. Bartels and N.P. Zotov.  I thank 
Gunter Grindhammer and Heidi Schellman for explaining the details of 
SIDIS experiments at HERA and TEVATRON colliders. 
I also thank D. Graudenz for
the correspondence about the inclusive rate of SIDIS hadroproduction, and 
M. Kuhlen for the communication about the HZTOOL data package.
I enjoyed conversations about semi-inclusive hadroproduction with 
Brian Harris, Daniel Boer, Sourendu Gupta, Anna Kulesza, 
Tilman Plehn, Zack Sullivan, Werner Vogelsang, and other members of 
the HEP groups at Argonne and Brookhaven National
Laboratories, University of Wisconsin and Southern Methodist University.
I am grateful to Brage Golding, Joey Huston, Vladimir Zelevinsky
for useful advices and careful reading of my manuscript. 

I am sincerely grateful to Harry Weerts, who encouraged me to apply to the MSU
graduate program and later spent a significant effort to get me in.  
During my years at Michigan State University, I was surrounded by 
the friendly and productive atmosphere created for HEP graduate students 
by a persistent effort of many people, notably Jeanette Dubendorf,  
Phil Duxbury, Stephanie Holland, Julius Kovacs, 
Lorie Neuman, George Perkins,  Debbie Simmons, Brenda Wenzlick,
Laura Winterbottom and Margaret Wilke. My teaching experience was 
more pleasant due to the interactions with Darlene Salman and Mark Olson.

Perhaps none of this work would be completed without the loving care 
and enthu\-si\-astic encouragement from my wife Sunny, who brings the 
meaning and joy to each day of my life.  
The completion of this thesis is our joint achievement, of which
Sunny's help in the preparation of the manuscript is only the smallest
part.   
My deep gratitude also goes to my parents, 
who always support and love me despite my
being overseas. The memory of my grandmother who passed away during the
past year will always keep my heart warm. 

\newpage

{\par\centering \tableofcontents{}\par}
\newpage

{\par\centering \listoffigures{}\par}
\newpage

\chapter{Introduction\label{ch:Intro}}

\pagenumbering{arabic}

Since its foundation in 1970's, perturbative Quantum Chromodynamics (PQCD) has
evolved into a precise theory of energetic hadronic interactions. The success
of the QCD theory in the quantitative description of hadronic experimental data
originates from the following fundamental ideas:

\begin{enumerate}
\item Hadrons are not elementary particles. As it was first shown by the quark model
of Gell-Mann and Zweig \cite{Gell-MannZweig64}, basic properties of the observed
low-energy hadronic states are explained if hadrons are composed of a few ``constituent
quarks'' with spin 1/2, fractional electric charges and new quantum numbers
of \emph{flavor} and \emph{color} \cite{Color}. If the hadron constituents
(\emph{partons}) are bound weakly at some energy, they can possibly be detected
in scattering experiments. The parton model of Feynman and Bjorken
suggested that the pointlike hadronic constituents may reveal themselves in
the wide-angle scattering of leptons off hadronic targets~\cite{PartonModel}.
The first direct experimental proof of the hadronic substructure came from the
observation of the Bjorken scaling \cite{Bjorken69} in the electron-proton
deep-inelastic scattering \cite{ScalingExp}; subsequently the quantum numbers
of partons were tested in a variety of experiments \cite{ExpReview}.
\item The elementary partons of QCD are ``current'' quarks, which interact with
one another through mediation of non-Abelian gauge fields (\emph{gluons}) \cite{localSU3color}.
These gauge fields are introduced to preserve the local $ SU(3) $ symmetry
of the quark color charges, in accordance with the pioneering 
work on non-Abelian
gauge symmetries by C. N. Yang and R. L. Mills \cite{YangMills54}. Remarkably,
the QCD interactions weaken at small distances because of the anti-screening
of color charges by self-interacting gluons \cite{AsymFreedom}. Due to this
feature \emph{(asymptotic freedom}) of QCD, probabilities for parton interactions
at distance scales smaller than $ 1\mbox {\, GeV}^{-1} $ can be calculated
as a series in the small QCD running coupling $ \alpha _{S} $. In the opposite
limit of large distances, $ \alpha _{S} $ grows rapidly, so that the QCD
interactions become nonperturbative at the scale of about $ 0.2\mbox {\, GeV}^{-1} $.
Such scale dependence of the QCD coupling explains why the partons behave as
quasi-free particles when probed in the energetic collision, but eventually
are confined in colorless hadrons at the later stages of the scattering. 
\item Because of the parton confinement, quantitative calculations within QCD require
systematic separation of dynamics associated with short and long distance scales.
The possibility for such separation is proven by factorization theorems 
\cite{Amati78,LibbySterman78,Mueller78,Ellis78,Ellis79,EfremovRadyushkin80}.
With time, the factorization was proven for observables of increasing complexity.
In 1977, G.~Sterman and S.~Weinberg introduced a notion of infrared-safe observables,
which are not sensitive to the details of long-distance
dynamics \cite{StermanWeinberg}. A typical example of an infrared-safe
observable is the cross-section for the production of  well-separated hadronic
jets at an $ e^{+}e^{-} $ collider. It was shown that infrared-safe observables can be
systematically described by means of PQCD. As a next step, factorization was
proven for inclusive observables depending on \emph{one} large momentum scale
$ Q^{2} $. 
In the limit $ Q^{2}\rightarrow \infty  $, such observables can be factorized
into a perturbatively calculable hard part, describing energetic short-range
interactions of hadronic constituents, and several process-independent
nonperturbative functions, relevant to the complicated strong dynamics at large
distances.
\end{enumerate}

The proof of factorization is more involved for hadronic observables that depend
on several momentum scales (\emph{e.g.}, differential cross sections). The complications
stem not so much from the complex dependence of the cross sections on kinematical
variables, but from the presence of logarithms $ \ln r $, where $ r $
is some dimensionless function of the kinematical parameters of the system.
For instance, $ r $ may be a ratio of two momentum scales $ P_{1} $
and $ P_{2} $ of the system, $ r=P_{1}/P_{2} $. Near the boundaries of
the phase space, the ratio $ r $ can be very large or very small, in which
case the convergence of the series in the QCD coupling $ \alpha _{S} $ can
be spoiled by the largeness of terms proportional to powers of $ \ln r $.
Hence the factorization cannot be proven as straightforwardly as in the case
of the inclusive observables, because its most obvious requirement -- sufficiently
rapid convergence of the perturbative series \nolinebreak -- is violated. 

To restore the convergence of PQCD, one may have to sum the
large logarithmic terms through all orders of 
$\alpha_S$. This procedure is commonly
called \emph{resummation}. Logarithmic terms of one rather general class appear
due to the QCD radiation along the directions of the observed initial- or final-state
hadrons (\emph{collinear radiation}) or the emission of low-energy gluons
(\emph{soft radiation}). 
Such logarithms commonly affect observables sensitive to the angular
distribution of the hadrons. In several important processes,  
the soft and collinear logarithms can be consistently 
resummed through the use of the formalism developed by J. Collins, D. Soper
and G. Sterman (CSS).

\begin{figure}[H]
{\par\centering \subfigure[ ]
{\resizebox*{0.7\textwidth}{!}{\includegraphics{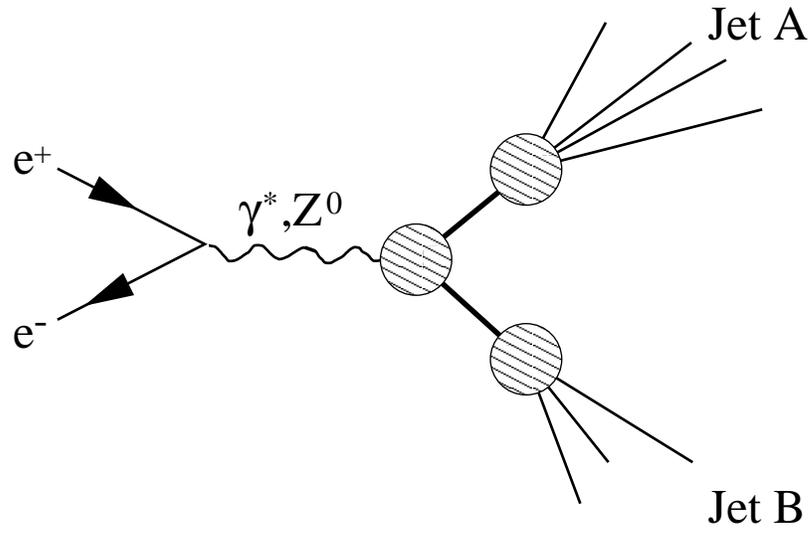}}} \par}
{\par\centering \subfigure[ ]
{\resizebox*{0.7\textwidth}{!}{\includegraphics{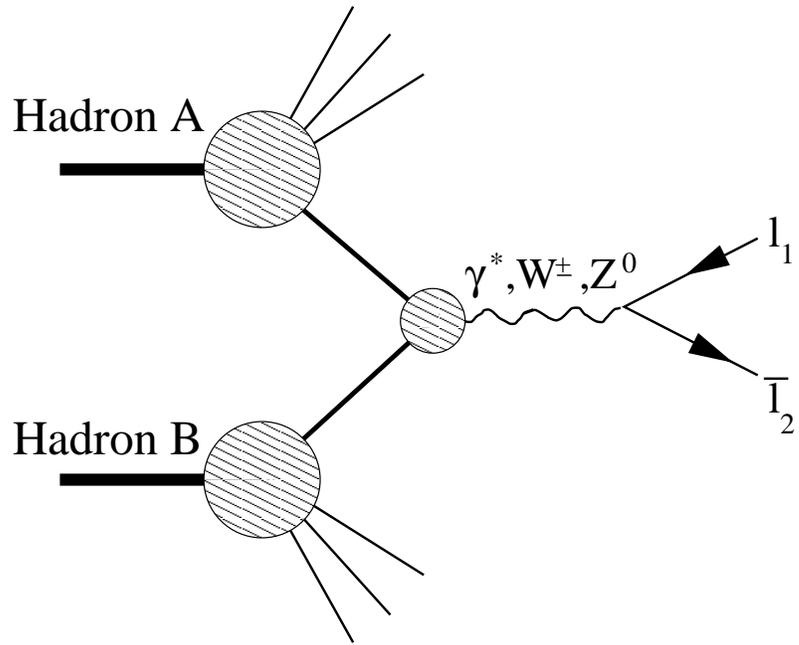}}} \par}
\caption{\label{fig:eeDY}(a) Production of hadronic jets at \protect$ e^{+}e^{-}\protect $colliders;
(b) Production of lepton pairs at hadron-hadron colliders}
\end{figure}

The original resummation technique was  
proposed  in  Ref.\,\cite{CS81} to
describe angular distributions of back-to-back jets produced 
at $e^+ e^-$ colliders (Fig.\,\ref{fig:eeDY}a). 
Subsequent developments of this technique and its
comparison to the data on the $e^+e^-$ hadroproduction 
were presented in Refs.\,\cite{CaoSoper83,CS85,CS87}. 
In Ref.\,\cite{CSS85} the resummation  formalism \, was
\newpage
\noindent 
extended to describe
transverse momentum distributions of lepton pairs produced 
at hadron-hadron colliders (Fig.\,\ref{fig:eeDY}b).   
In the subsequent publications 
\cite{DWS,ArnoldKauffman,LY,BY,BLLY,Landry2000}, 
this technique was developed to a high degree of numerical accuracy.
Currently the resummation analysis of this type
is employed in the measurements of the mass 
\cite{Wmass} and the width \cite{Wwidth} of the $ W $-bosons
produced at the $ p\bar{p} $ collider TEVATRON. 
With some modifications, this resummation formalism is also
used to improve PQCD predictions for the production
of Higgs bosons \cite{Higgs} and photon pairs \cite{BalazsDiphoton98} at the
Large Hadron Collider (LHC).

\begin{figure}[tb]
{\par\centering \resizebox*{!}{10cm}{\includegraphics{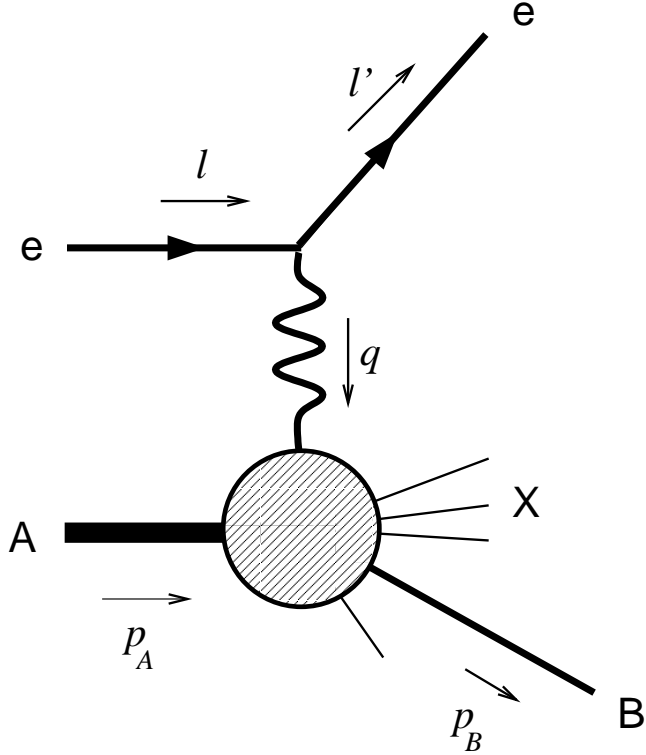}} \par}

\caption{\label{fig:SIDIS} Semi-inclusive deep inelastic scattering}
\end{figure}

The hadroproduction at $ e^{+}e^{-} $colliders and the lepton pair production
at hadron-hadron colliders (Drell-Yan process) are the simplest reactions that
require resummation of the soft and collinear logarithms. In both reactions,
the interaction between the leptons and two initial- or final-state hadronic
systems is mediated by an electroweak boson $ V $ with a timelike momentum.
The CSS resummation formalism can also be formulated for reactions with the
exchange of a spacelike electroweak vector boson \cite{Collins93,Meng2}. 
In this work, I discuss resummation
in the semi-inclusive production of hadrons in electron-hadron deep-inelastic
scattering, which is the natural analog of  $ e^{+}e^{-} $ hadroproduction
and Drell-Yan process in the spacelike channel. The reaction of semi-inclusive
deep-inelastic scattering (SIDIS) $ e+A\rightarrow e+B+X $, where $ A $
and $ B $ are the initial- and final-state hadrons, respectively, is shown in
Figure \ref{fig:SIDIS}. 

As in the other two reactions, in SIDIS the multiple parton radiation affects
angular distributions of the observed hadrons. The study of the resummation
in SIDIS has several advantages in comparison to the reactions in the timelike
channels. First, SIDIS is characterized by an obvious asymmetry between the
initial and final hadronic states, so that the dependence of the multiple parton
radiation on the properties of the initial state can be distinguished clearly
from the dependence on the properties of the final state. In contrast, in 
$ e^{+}e^{-} $ hadroproduction or the Drell-Yan process some details of the
dynamics may be hidden due to the symmetry between two external hadronic systems.
Notably, I will discuss the dependence of the resummed observables on the longitudinal
variables $ x $ and $ z $, which can be tested in SIDIS more directly
than in  $ e^{+}e^{-} $ hadroproduction or Drell-Yan process. 

Second, SIDIS can be studied in the kinematical region covered
by the measurements of the hadronic structure functions $ F_{i}(x,Q^{2}) $ in
completely inclusive DIS. The ongoing DIS experiments at the $ ep $ collider
HERA probe $ F_{i}(x,Q^{2}) $ at $ x $ down to $ 10^{-5} $, which are
much smaller than lower values of $ x $ reached at the existing hadron-hadron
colliders. The region of low $ x $, which is currently studied at HERA, will
also be probed routinely in the production of $ W^{\pm },Z^{0} $ and Higgs
bosons at the LHC. At such low values of $ x $, other dynamical mechanisms
may compete with the contributions from the soft and collinear radiation described
by the CSS formalism. The study of the existing SIDIS data provides a unique
opportunity to learn about the applicability of the CSS formalism in the low-$ x $
region and estimate robustness of theoretical predictions for the electroweak
boson production at the LHC. 

Last, but not the least, is the issue of  potential symmetry relations
between the resummed observables in SIDIS, $ e^{+}e^{-} $ hadroproduction
and Drell-Yan process. In SIDIS, the dynamics associated with the initial-state
radiation may be similar to the initial-state dynamics in the Drell-Yan process,
while the final-state dynamics may be similar to the final-state dynamics in
$ e^{+}e^{-} $hadroproduction. It is interesting to find out if the data
support the existence of such crossing symmetry. 

The results presented here were published or accepted for publication
in Ref.\,\cite{nsy1999,nsy2000,nsyasym}.
The remainder of the thesis is organized as follows. In Chapter\,\ref{ch:QCDbground},
I discuss the basics of factorization of mass singularities in hadronic
cross sections. Then I review the 
general properties of the Collins-Soper-Sterman
resummation formalism and illustrate some of its features with the example of
hadroproduction at $ e^{+}e^{-} $ colliders. 

In Chapter\,\ref{ch:Formalism},
I apply the resummation formalism to semi-inclusive deep inelastic scattering.
Guided by the similarities between SIDIS, $ e^{+}e^{-} $ hadroproduction
and Drell-Yan process, I introduce a set of kinematical variables that are
particularly convenient for the identification and subsequent summation of the
soft and collinear logarithms. I also identify observables
that are directly sensitive to the multiple parton radiation. In particular,
I argue that such radiation affects the dependence of SIDIS cross sections
and hadronic energy flow on the polar angle in the photon-proton center-of-mass
frame. Next I derive the $ {\cal O}(\alpha _{S}) $ cross section
and obtain the $ {\cal O}(\alpha _{S}) $ coefficients for the resummed
cross sections and the hadronic energy flow. 

In Chapter\,\ref{ch:Phenomenology},
I compare the results of the CSS resummation formalism and 
$ {\cal O}(\alpha _{S}) $
fixed-order calculation with the data from the $ ep $ collider HERA. 
I show that the CSS resummation
improves theoretical description of various aspects of these
data. I also discuss the dependence of the resummed
observables on the longitudinal variables $ x $ and $ z $. I show that
the HERA data are consistent with the rapid increase of nonperturbative contributions
to the resummed cross section at $ x\lesssim 0.01 $. I discuss the potential
dynamical origin of such low-$ x $ behavior of the CSS formula. 

Finally,
in Chapter\,\ref{ch:AzimuthalAsymmetries} I discuss the impact of the
multiple parton radiation on azimuthal asymmetries of the SIDIS cross sections.
I show that the CSS resummation formalism can be used to distinguish reliably
between perturbative and nonperturbative contributions to the azimuthal asymmetries.
I also suggest to measure azimuthal asymmetries of the transverse energy flow,
which provide a clean test of PQCD.

\chapter{\label{ch:QCDbground}Overview of the QCD factorization}
Perturbative calculations in  Quantum Chromodynamics rely on 
a systematic procedure for separation of
short- and long-distance dynamics in hadronic observables. 
The proof of feasibility of such procedure naturally leads to the
methods for improvement of the convergence of the perturbative
series when this convergence is degraded 
by infrared singularities of contributing subprocesses.  
Here I present the basics of the factorization procedure.
The omitted details can be found in standard textbooks
on the theory of strong interactions, {\it e.g.}, 
Refs.\,\cite{StermanBook,Yndurain,ChengLi,CollinsBook}.

\section{QCD Lagrangian and renormalization}

Low-energy hadronic states have internal substructure. They are composed of
more fundamental fermions (\emph{quarks}) that are bound together by non-Abelian
gauge forces. The quanta of the QCD gauge fields are called \emph{gluons}. Quantum
ChromoDynamics (QCD) is the theory that describes strong interactions between
the quarks. In the classical field theory, the QCD Lagrangian density in the
coordinate space is\begin{eqnarray}
{\cal L}_{QCD}(x) & = & \sum _{f}\bar{\psi }_{f}\left( i{\rlap {/}\partial }-gA_{a}\! \! \! \! \! \! /\, \, \, T_{a}-m_{f}\right) \psi _{f}\nonumber \\
 & - & \frac{1}{4}F^{\alpha \beta }_{a}F_{a\alpha \beta }-\frac{\lambda }{2}(\eta _{\alpha }A_{a}^{\alpha })^{2}+\bar{c}_{a}\left( \delta _{ad}\eta \cdot \partial -gC_{abd}\eta \cdot A_{b}\right) c_{d},\label{LQCD} 
\end{eqnarray}
where $ \psi _{fl}(x) $, $ A^{\alpha }_{a}(x) $ and $ c_{a}(x) $ are
the quark, gluon and ghost fields, respectively;
\begin{equation}
F_{a}^{\alpha \beta }(x)\equiv \partial ^{\alpha }A_{a}^{\beta }-\partial ^{\beta }A^{\alpha }_{a}-gC_{abc}A_{b}^{\alpha }A^{\beta }_{c}
\end{equation}
is the gauge field tensor; $ -\lambda (\eta _{\alpha }\cdot A_{a}^{\alpha })^{2}/2 $
is the term that fixes the gauge $ \eta \cdot A=0 $. The vector $ \eta ^{\alpha } $
is equal to the gradient vector $ \partial ^{\alpha } $ in covariant
gauges $ (\partial _{\alpha }A_{a}^{\alpha }=0) $ or an arbitrary vector
$ n^{\alpha } $ in axial gauges ($ n_{\alpha }A_{a}^{\alpha }=0 $).
The color indices $ l,m $ vary between 1 to $ N_{c} $ (where
$ N_{c}=3 $ is the number of colors), while the color indices $ a,b,c,d $
vary between 1 and $ N_{c}^{2}-1 $. The index $ f $ denotes the flavor
(\emph{i.e.}, the type) of the quarks, which is conserved in the strong
interactions. The remaining parameters in $ {\cal L}_{QCD}(x) $ are the QCD charge $ g $
and the masses of the quarks $ m_{f} $.

The QCD Lagrangian is invariant under the gauge transformations of the $ SU(N_c) $
color group: 
\begin{eqnarray}
\psi _{f}(x) & \rightarrow  & U(\theta (x))\psi _{f}(x);\\
T_{a}A_{a}^{\alpha }(x) & \rightarrow  & U(\theta (x))T_{a}A_{a}^{\alpha }(x)U^{-1}(\theta (x))+\frac{i}{g}\left( \partial ^{\alpha }U(\theta (x))\right) U^{-1}(\theta (x)),
\end{eqnarray}
where the $ x- $dependent unitary operator $ U(\theta (x)) $ is \begin{equation}
U(\theta (x))\equiv e^{-iT_{a}\theta _{a}(x)}.
\end{equation}
$ \left( T_{a}\right) _{lm} $ and $ C_{abd} $ are generator matrices and
structure constants of the color group. The commutators of
the matrices $ (T_{a})_{lm} $
are
\begin{equation}
\left[ T_{a},T_{b}\right] =iC_{abc}T_{c}.
\end{equation}
The quark fields $ \psi _{fl} $ and gauge fields $ A^{\alpha }_{a} $ are
vectors in the fundamental and adjoint representations of $ SU(N_c) $, respectively. 

In the quantum theory, $ \psi _{f},\, A^{\alpha },c_{a} $
are interpreted as ``bare'' (unrenormalized) operators of the corresponding fields; $ g $
and $ m_{f} $ are interpreted as the ``bare'' charge and masses. The
perturbative calculation introduces infinite ultraviolet corrections to these
quantities. In order to obtain finite theoretical predictions, $ {\cal L}_{QCD} $
has to be expressed in terms of the renormalized parameters, which are related
to the ``bare'' parameters through infinite multiplicative renormalizations. 

If the ultraviolet singularities 
are regularized by the continuation to $ n=4-2\eps$,  
$\eps\nolinebreak >\nolinebreak 0  $
dimensions \cite{HV}, the renormalized parameters (marked by the subscript
$ ``R'' $) are related to the bare parameters as\begin{eqnarray}
\psi _{fR}(\mu ) & = & Z_{\psi }^{-1}(\mu )\psi _{f},\label{BareRenorm1} \\
A^{\alpha }_{aR}(\mu ) & = & Z_{A}^{-1}(\mu )A^{\alpha }_{a},\\
c_{aR}(\mu ) & = & Z_{c}^{-1}(\mu )c_{a},\\
g_{R}(\mu ) & = & Z^{-1}_{g}(\mu )\mu ^{-\eps }g,\label{gR} \\
m_{fR}(\mu ) & = & Z^{-1}_{m}(\mu )m_{f},\label{mR} 
\end{eqnarray}
where $ Z_\psi, Z_A, Z_c, Z_g$, and $Z_m$ 
are perturbatively calculable renormalization constants.
In the dimensional regularization, the renormalized
parameters depend on an auxiliary momentum scale $ \mu _{n} $, which
is introduced to keep the charge $ g $ dimensionless in $ n\neq 4 $ dimensions. In Eqs.\,(\ref{BareRenorm1}-\ref{mR})
the renormalized parameters and the constants $ Z_{k} $ are expressed in
terms of another scale $ \mu  $, which is related to $ \mu _{n} $ as\begin{equation}
\label{mu2mun}
\mu ^{2}=4\pi e^{-\gamma _{E}}\mu _{n}^{2}.
\end{equation}
Here $ \gamma _{E}=0.577215... $ is the Euler constant.

\section{Asymptotic freedom }

The further improvement of the theory predictions for physical
observables is achieved by
enforcing their invariance under variations of the scale $ \mu  $, 
{\it i.e.}, by solving renormalization
group (RG) equations. Consider an observable $S$ that depends on
$N$ external momenta $p_i^{\mu},\quad i=1,\dots,N$. If the renormalized expression 
for $ S $ is\[
S\left( g_{R}(\mu ),\{m_{fR}(\mu )\},\{p_{i}\},\mu \right) \]
(where ``$\{\dots\}$''
denotes a set of parameters), 
then the RG-improved expression for $ S $ is\begin{equation}
\label{RGS}
S\left( \bar{g}(\mu ),\{\bar{m}_{f}(\mu )\},\{p_{i}\},\mu \right) ,
\end{equation}
where $ \bar{g}(\mu ) $ and $ \bar{m}_{f}(\mu ) $ are \emph{the running
QCD charge and quark masses}. By solving the equation for the independence of
$ S $ from $ \mu  $, \begin{equation}
\label{renorm1}
\mu \frac{d}{d\mu }S\left( \bar{g}(\mu ),\{\bar{m}_{f}(\mu )\},\{p_{i}\},\mu \right) =0,
\end{equation}
 we find the following differential equations for $ \bar{g}(\mu ) $ and $ \bar{m}_{f}(\mu ) $:
\newpage
\bea
\mu \frac{\partial \bar{g}(\mu )}{\partial \mu } &=&  \beta
(\bar{g}(\mu )),\label{runcoupling}\\ 
\mu \frac{\partial \bar{m}_{f}(\mu )}{\partial \mu } &=&  -\gamma _{mf}\left( \bar{g}(\mu )\right) \bar{m}_{f}(\mu ).\label{runmass} 
\eea
The \emph{approximate} expressions for the functions $ \beta (g) $ and $ \gamma _{m}(g) $
on the r.h.s. of Eqs.\,(\ref{runcoupling}) and (\ref{runmass}) are
found from the $ \mu - $dependence of the \emph{fixed-order} renormalized
charges and masses:\begin{eqnarray}
\beta (g_{R}(\mu )) & \equiv  & \mu \frac{\partial g_{R}(\mu )}{\partial \mu },\\
\gamma _{mf}(g_{R}(\mu )) & \equiv  & -\frac{1}{2m_{fR}^{2}(\mu )}\mu \frac{\partial m_{fR}^{2}(\mu )}{\partial \mu }.
\end{eqnarray}

The renormalization group analysis of the QCD Lagrangian suggests that the interactions
between the quarks weaken at high energies, \emph{i.e.}, that Quantum Chromodynamics
is \emph{asymptotically free} in this limit. Indeed, the perturbative series
for the function $ \beta (g) $ is \begin{equation}
\label{betafunc}
\beta (g)=-g\sum ^{\infty }_{k=1}\left( \frac{\alpha _{S}}{4\pi }\right) ^{k}\beta _{k},
\end{equation}
where $ \alpha _{S}\equiv g^{2}/4\pi  $ is the QCD coupling. In the modified
minimal subtraction ($ \MSbar  $) regularization scheme \cite{MSBar}, the
lowest-order coefficient $ \beta _{1} $ in Eq. (\ref{betafunc}) is given
by \begin{eqnarray}
\beta _{1} & = & \frac{11}{3}C_{A}-\frac{4}{3}T_{R}N_{f},
\end{eqnarray}
where $ N_{f} $ is the number of active quark flavors, $ C_{A}=N_{c}=3 $,
and $ T_{R}=1/2 $. By solving Eq.\,(\ref{runcoupling}), we find
that \begin{equation}
\label{AlphaS1}
\bar{\alpha }_{S}(\mu )=\frac{\bar{\alpha }_{S}(\mu _{0})}{1+\frac{\bar{\alpha }_{S}}{4\pi }\beta _{1}\ln \frac{\mu ^{2}}{\mu _{0}^{2}}}.
\end{equation}
This equation proves the asymptotic freedom of QCD interactions: 
for six known quark generations,
$ \beta _{1}>0 $ and \[
\lim _{\mu \rightarrow \infty }\bar{\alpha }_{S}(\mu )=0.\]
Higher-order corrections to the beta-function do not change this asymptotic
behavior. Eq.\,(\ref{AlphaS1}) also shows that $ \bar{\alpha }_{S}(\mu ) $
has a pole at some small value of $ \mu  $. The position of this pole can
be easily found from the alternative form of Eq.\,(\ref{AlphaS1}),
\begin{equation}
\label{AlphaS2}
\bar{\alpha }_{S}(\mu )=\frac{4\pi }{\beta _{1}\ln (\mu ^{2}/\Lambda _{QCD}^{2})}\left[ 1+\dots \right] .
\end{equation}
 In Eq.\,(\ref{AlphaS2}), $ \Lambda _{QCD} $ is a phenomenological 
parameter, which 
is found from the analysis of the experimental data. The most recent world average
value of $ \Lambda _{QCD} $ for  
$ N_{f}\nolinebreak =\nolinebreak 5 $ and ${\cal O}(\alpha_S^4)$ expression for the $\beta$-function is $ 208^{+25}_{-23} $
MeV \cite{PDG}. According to Eq.\,(\ref{AlphaS2}), $ \bar{\alpha }_{S}(\mu ) $
becomes infinite when $ \mu =\Lambda _{QCD} $. This feature of the 
QCD running coupling obstructs theoretical calculations for hadronic
interactions at low energies.

\section{Infrared safety\label{sec:InfraredSafety}}

Due to the asymptotic freedom, the calculation of QCD observables at large $ \mu  $
can be organized as a series in powers of the small parameter $ \bar{g}(\mu ) $.
To find out when the perturbative calculation may converge rapidly, consider
the formal expansion of the RG-improved expression
(\ref{RGS}) for the observable
$ S $ in the series of $ \bar{g}(\mu ) $:\begin{equation}
\label{Spert}
S=\Phi (\{p_{i}\},\{\bar{m}_f\},\mu )\sum ^{\infty }_{k=0}
S^{(k)}\left( \left\{\frac{p_{i}\cdot p_{j}}{\mu ^{2}}\right\},
\left\{\frac{\bar{m}_{f}(\mu )^{2}}{\mu ^{2}}\right\}\right) \bar{g}^{k}(\mu ).
\end{equation}
In this expression, the function $ \Phi (\{p_{i}\},\{ \bar
m_f \}, \mu ) $ includes all coefficients
that do not depend on the order of the perturbative calculation (for instance,
the phase space factors). The mass dimension of 
$ \Phi (\{p_{i}\},\{ \bar m_f\},\mu ) $
is equal to the mass dimension of $ S $. The sum over $k$ on the right-hand
side is dimensionless. The coefficients of the perturbative expansion $ S^{(k)} $
depend on dimensionless Lorentz-invariant combinations of the external momenta
$ p_{i}^{\mu } $, the mass parameter $ \mu  $, and the running quark masses
$ \bar{m}_{f}(\mu ) $. There are indications that the perturbative series
in Eq.\,(\ref{Spert}) are asymptotic \cite{AsymptoticSeries}, so that
it diverges at sufficiently large $ k $. However, the lowest few terms of
this series may approximate $ S $ sufficiently well if they do
not grow rapidly when $ k $ increases. 

The factors that control the convergence of Eq.\,(\ref{Spert}) 
can be understood
in a simpler case, when all Lorentz scalars $ p_{i}\cdot p_{j} $ in 
Eq.\,(\ref{Spert})
are of the same order $Q^{2}$. Then Eq.\,(\ref{Spert})
simplifies to \begin{equation}
\label{Spert2}
S=\Phi (\{p_{i}\},\{\bar{m}_f\},\mu )\sum ^{\infty }_{k=0}
S^{(k)}\left( \frac{Q^{2}}{\mu ^{2}},
\left\{\frac{\bar{m}_{f}(\mu )^{2}}{\mu ^{2}}\right\}\right) \bar{g}^{k}(\mu ).
\end{equation}
When $ Q^{2}\gg \Lambda ^{2}_{QCD} $, we can choose $ \mu \sim Q $ to
make $ \bar{g}(\mu ) $ small. This choice also eliminates potentially large
terms like $ \ln (Q^{2}/\mu ^{2}) $ from the coefficients $ S^{(k)} $.
In addition, let's assume that $ Q $ is much larger than any quark mass $ \bar{m}_{f}(\mu ) $
on which $ S $ depends. For instance, $ S $ may be dominated by contributions
from the $ u,\, d,\, s $ quarks, whose running masses are lighter than $ 200 $
MeV at $ \mu =2 $ GeV \cite{PDG}. At $ \mu '>2 $ GeV, the quarks become
even lighter due to the running of $ \bar{m}_{f} $: \begin{equation}
\bar{m}_{f}(\mu ')=\bar{m}_{f}(\mu )\exp \left\{ -\int ^{\mu '}_{\mu }\frac{d\bar{\mu }}{\bar{\mu }}\gamma _{mf}(\bar{\mu })\right\} <\bar{m}_{f}(\mu ),
\end{equation}
since in QCD\begin{equation}
\label{gammamf}
\gamma _{mf}(\mu )=\frac{3\bar{\alpha }_{S}(\mu )}{4\pi }C_{F}+{\cal O}(\bar{\alpha }_{S}^{2})>0.
\end{equation}
Here $ C_{F}\equiv (N_{c}^{2}-1)/(2N_{c})=4/3 $. 

When the quark masses vanish, many observables, which are finite if
$\bar m_{fi} \neq 0$, acquire infrared singularities.
These singularities are generated from the terms in the perturbative coefficients
that are proportional to the logarithms $ \ln (\bar{m}_{f}^{2}/\mu ^{2}) $.
The expansion in the 
perturbative series (\ref{Spert2}) makes sense only for those
observables
$S$ that remain finite when $ \bar{m}_{f}(\mu )^{2}/\mu ^{2}\rightarrow 0 $. 

There are two categories of observables for which the perturbative expansion
(\ref{Spert2}) is useful. In the first case, the coefficients $ S^{(k)} $
are finite and analytically calculable when $ \mu \rightarrow \infty : $\begin{equation}
S^{(k)}\left( \frac{Q^{2}}{\mu ^{2}},
\left\{\frac{\bar{m}_{f}(\mu )^{2}}{\mu ^{2}}\right\}\right) 
\rightarrow \wt{S}^{(k)}\left( \frac{Q^{2}}{\mu ^{2}}\right) 
+{\cal O}\left( \left\{\left( 
\frac{\bar{m}_{f}(\mu )^{2}}{\mu ^{2}}\right)^{a}\right\}\right) ,\quad a>0.
\end{equation}
Such observables are called \emph{infrared-safe} \cite{StermanWeinberg}\emph{.}
For instance, the total and jet production cross sections
in $ e^{+}e^{-} $ hadroproduction are infrared-safe. 
In this example, hadrons appear only
in the completely inclusive final state. According to the Kinoshita-Lee-Nauenberg (KLN)
theorem \cite{KLN}, such inclusive states are free of infrared
singularities,
so that  the finite expressions for the total and jet cross sections 
can be found from the massless perturbative calculation. 

\begin{figure}[H]
{\par\centering \resizebox*{!}{8cm}{\includegraphics{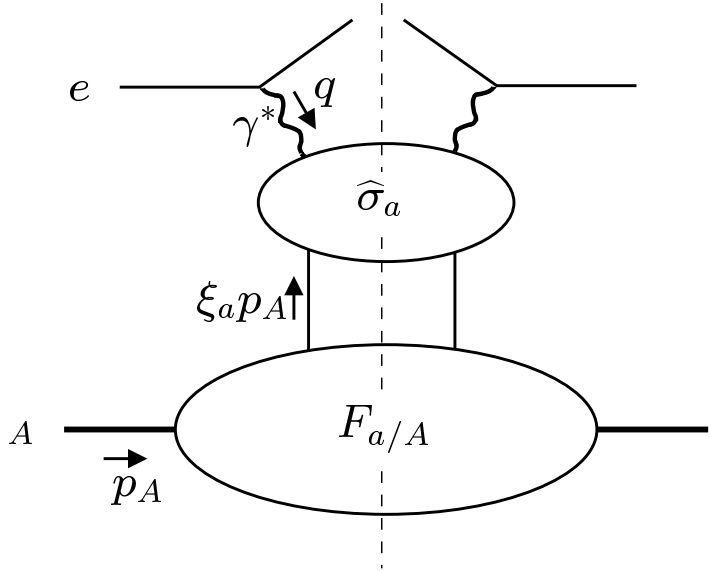}} \par}

\caption{\label{fig:DISfact}Factorization of collinear singularities in 
completely inclusive electron-hadron DIS}
\end{figure}

In the second case, $ S^{(k)} $ are not infrared-safe, but all mass singularities
of $ S^{(k)} $ can be absorbed (\emph{factorized}) into one or several process-independent
functions. These functions can be measured in one set of experiments and then
used to make predictions for other experiments.  

To understand which singularities should be factorized, notice
that there are two
classes of the infrared singularities in a massless gauge theory: soft singularities
and collinear singularities. The soft singularities occur in \emph{individual
Feynman diagrams} when the momentum $ k^{\mu } $ carried by some gluon line
vanishes ($ k^{\mu }\sim \lambda \kappa ^{\mu }, $ where $ \lambda \rightarrow 0 $
and $ \kappa ^{\mu } $ are fixed). The soft singularities cancel at each
order of $ \bar{\alpha }_{S}(\mu ) $ once all Feynman diagrams of this order
are summed over. 

In contrast, the collinear singularities occur when the momenta
$ p_{1}^{\mu } $ and $ p_{2}^{\mu } $ of two massless particles are collinear
to one another , \emph{i.e.}, when $ p_{1}\cdot p_{2}\rightarrow 0 $. Since
one or both collinear particles can be simultaneously soft, the class of the
collinear singularities partially overlaps with the class of the soft singularities.
The soft collinear singularities cancel in the complete fixed-order result
just as all soft singularities do. On the contrary, the singularities due to
the collinearity of the particles with non-vanishing momenta do not cancel and
should be absorbed in the long-distance phenomenological functions.

As an illustration of the factorization of the purely collinear singularities, consider
the factorized form for the cross section of inclusive deep inelastic
scattering $ e+A\stackrel{\gamma ^{*}}{\longrightarrow }e+X $ (where $ A $
is a hadron) in the limit $Q^2 \rightarrow \infty$: 
\begin{equation}
\label{DISfact}
\frac{d\sigma _{A}}{dxdQ^{2}}=\sum _{a}\int ^{1}_{x}d\xi
_{a}\frac{d\wh{\sigma }^{hard}_{a}}{dx\, dQ^{2}}\left( \bar{\alpha
}_{S}(Q),\, \frac{x}{\xi _{a}},\, \frac{Q}{\mu _{F}}\right)
F_{a/A}(\xi _{a},\mu _{F}) 
+{\cal O}\left(\frac{1}{Q^2}\right).
\end{equation}
This representation and notations for the particle momenta are illustrated in
Fig.\,\ref{fig:DISfact}.

In Eq.\,(\ref{DISfact}), $ Q^{2}\equiv -q^{2} $ is the large invariant
mass of the virtual photon $ \gamma ^{*}, $ $ x\equiv
Q^{2}/(2(p_{A}\cdot q)) $.
These variables are discussed in more detail in Subsection\,\ref{sub:LorentzScalars}.
$ d\wh{\sigma }^{hard}_{a}/\left( dx\, dQ^{2}\right)  $ is the
infrared-safe (``hard'')
part of the cross-section for the scattering $ e+a\rightarrow e+X $
of the electron on a parton $ a $. $ F_{a/A}(\xi _{a},\mu _{F}) $ is the
\emph{parton distribution function} (PDF), which absorbs the 
collinear singularities
subtracted from the full parton-level cross section to obtain 
$ d\wh{\sigma }^{hard}_{a}/(dx\, dQ^{2}) $.
In the inclusive DIS, all collinear singularities appear due to the radiation
of massless partons along the direction of the initial-state hadron $ A $.
The final state is completely inclusive; hence, 
by the KLN theorem, it is finite.

The collinear radiation in the initial state depends only on the types of $ a $
and $ A $ and does not depend on the type of the particle reaction. Therefore,
$ F_{a/A}(\xi _{a},\mu _{F}) $ is process-independent. It can be
interpreted as a probability of finding a massless parton $ a $ with the
momentum $ \xi _{a}p_{A}^{\mu } $ in the initial hadron with the momentum
$ p_{A}^{\mu } $. To obtain the complete hadron-level cross section, we sum
over all possible types of $ a $ ($ a=g,u,\bar{u},d,\bar{d},\dots ) $
and integrate over the allowed range of the momentum fractions $ \xi _{a} $
$ (0<x\leq \xi _{a}\leq 1) $. 

In Eq.\,(\ref{DISfact}), both the ``hard'' cross sections 
$ d\wh{\sigma }^{hard}_{a}/(dx\, dQ^{2}) $
and the parton distribution functions $ F_{a/A}(\xi _{a},\mu _{F}) $ depend
on an arbitrary factorization scale $ \mu _{F} $, which appears due to
some freedom
in the separation of the collinear contributions included in $ F_{a/A}(\xi _{a},\mu _{F}) $
from the ``hard'' contributions included in $ d\wh{\sigma }^{hard}_{a}/(dx\, dQ^{2}) $.
Of course, the complete hadron-level cross section on the l.h.s. of Eq.\,(\ref{DISfact})
should not depend on $ \mu _{F}. $ Hence the  $ \mu _{F} $-dependence of 
$ F_{a/A}(\xi _{a},\mu _{F}) $ should cancel the  
$ \mu _{F} $-dependence of the hard cross section. This requirement is 
used to find Dokshitser-Gribov-Lipatov-Altarelli-Parisi
(DGLAP) differential equations \cite{DGLAP}, which describe the dependence
of $ F_{a/A}(\xi _{a},\mu _{F}) $ on $ \mu _{F} $: \begin{equation}
\label{DGLAPEqsF}
\mu _{F}\frac{dF_{a/A}(\xi _{a},\mu _{F})}{d\mu _{F}}=\sum _{b}\left( {\cal P}^{S}_{ab}\otimes F_{b/A}\right) (\xi _{a},\mu _{F}).
\end{equation}
Here $ {\cal P}_{ab}^{S}(\xi ,\mu ) $ are ``spacelike'' splitting functions
that are currently known up to $ {\cal O}(\alpha _{S}^{2}) $~\cite{TwoLoopPab}.
They describe evolution of partons with spacelike momenta. 
The convolution in Eq.\,(\ref{DGLAPEqsF}) is defined as\begin{equation}
(f\otimes g)(x,\mu )\equiv \int _{x}^{1}f(x/\xi ,\mu )g(\xi ,\mu )\frac{d\xi }{\xi }.
\end{equation}

A similar approach can be used to derive factorized cross sections for 
reactions
with observed outgoing hadrons. Such cross sections depend on fragmentation
functions (FFs) $ D_{B/b}(\xi ,\mu _{D}) $, which absorb the 
singularities
due to the collinear radiation in the final state. The fragmentation function
can be interpreted as the 
probability of finding the hadron $ B $ among the products
of fragmentation of the parton $ b $. The variable $\xi$ is the fraction
of the momentum of $b$ that is carried by $B$.
In the presence of FFs,
the hadron-level cross section becomes dependent on yet another factorization
scale $ \mu _{D} $. Similarly to the PDFs, the dependence of the FFs on $ \mu _{D} $
is described by the DGLAP evolution equations: \begin{equation}
\label{DGLAPEqsD}
\mu _{D}\frac{dD_{B/b}(\xi _{b},\mu _{D})}{d\mu _{D}}=\sum _{a}\left( D_{B/a}\otimes {\cal P}^{T}_{ab}\right) (\xi _{b},\mu _{D}),
\end{equation}
where $ {\cal P}_{ab}^{T}(\xi ,\mu ) $ are the ``timelike'' splitting
functions.

As in the case of the renormalization scale $ \mu  $, it is natural to choose
the factorization scales $ \mu _{F} $ and $ \mu _{D} $ of order
$ Q $ to avoid the appearance of the potentially large logarithms $ \ln \left( Q/\mu _{F}\right)  $
and $ \ln \left( Q/\mu _{D}\right)  $ in the ``hard'' cross section. I should
emphasize that the factorized cross sections are derived under the assumption
that all Lorentz scalars $ p_{i}\cdot p_{j} $ are of order $ Q^{2} $,
so that $ x $ in Eqs.\,(\ref{DISfact}) is sufficiently
close to unity. When some scalar product $ p_{i}\cdot p_{j} $ is much larger
or smaller than $ Q^{2} $, the convergence of the perturbative series for
the hard cross section is worsened due to the large logarithms of the ratio
$ p_{i}\cdot p_{j}/Q^{2} $. This is a general observation that applies to
any PQCD calculation. In some cases, the predictive power of the theory can
be restored by the summation of the large logarithms through all orders of the
perturbative expansion. In particular, the resummation of the large logarithms
is required for the accurate description of the angular distributions of the
final-state particles, including angular distributions of the final-state hadrons
in SIDIS. In the next Section, 
I discuss general features of such resummation on the example
of angular distributions of the jets in  $ e^{+}e^{-} $ hadroproduction.

\newpage

\begin{figure}[H]
{\par\centering \resizebox*{0.87\textwidth}{!}{\includegraphics{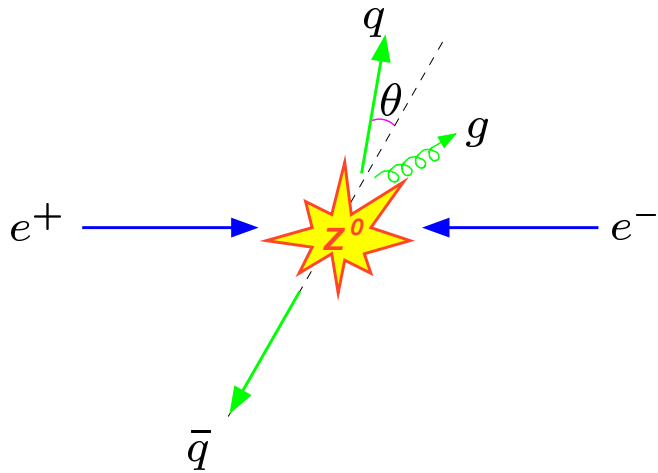}} \vspace{1cm}\par}

\caption{The space-time picture of hadroproduction at \protect$ e^{+}e^{-}\protect $
colliders\label{fig:eespace}}
\end{figure}
 
\newpage

\section{\label{sec:TwoScale}Two-scale problems}

\subsection{Resummation of soft and collinear logarithms \label{sub:ResumSoftColLogs}}

To understand the nature of the problem, consider
the process $ e^{+}e^{-}\rightarrow Z^{0}\rightarrow jets $
(Fig.\,\ref{fig:eeDY}a). The space-time picture of this process is
shown in Figure \ref{fig:eespace}. Let us assume that the $ Z^{0}
$-bosons
are produced at the resonance ($ E_{e^{+}}+E_{e^{-}}=M_{Z} $) at rest in
the laboratory frame. In $ e^{+}e^{-} $ hadroproduction, the hadronic decays
are initiated predominantly by the direct decay 
of the $ Z^{0} $-boson into a quark-antiquark pair. The QCD radiation off
the quarks produces hadronic jets, which are registered in the detector. 

If no additional hard QCD radiation is present (Fig.\,\ref{fig:eespace}a),
the decay of the $ Z^{0} $ boson produces two narrow jets escaping in the
opposite directions in the lab frame. The typical angular width of each jet
is of the order $ \Lambda _{QCD}/E_{A,\, B}\ll 1 $, where $ E_{A,B}\approx M_{Z}/2 $
are the energies of the jets. The quarks may also emit energetic
gluons, in which case the angle between the jets is not equal to $ \pi  $
(Fig.\,\ref{fig:eespace}b). If the angle $ \theta  $ in Fig.\,\ref{fig:eespace}b
is large, the additional QCD radiation is described well
by the rapidly converging series in the small perturbative parameter\footnote{%
From here on, I drop the ``bar'' in the notation of the running $ \alpha _{S} $. 
} $ \alpha _{S}(M_{Z})/\pi  $. But when $ \theta \rightarrow 0 $, the higher-order
radiation is no longer suppressed, because the smallness of $ \alpha _{S}(M_{Z})/\pi  $
is compensated by large terms $ \ln ^{p}(\theta ^{2}/4)/\theta ^{2},\, p\geq 0 $
in the hard part of the hadronic cross section. Therefore, the calculation at
\emph{any fixed order} does not describe reliably the shape of the hadronic
cross section when $ \theta \rightarrow 0 $. 

To illustrate this point, consider the 
\emph{hadronic energy-energy correlation} \cite{Basham:1979zq}, 
defined as \begin{equation}
\frac{d\Sigma }{d\cos \theta }\equiv \frac{1}{M_{Z}^{2}}\int
^{M_{Z}/2}_{0}dE_{A}\int _{0}^{M_{Z}/2}dE_{B}\, E_{A}E_{B}\frac{d\sigma
}{dE_{A}dE_{B}d\cos \theta }.
\end{equation}
\\
In the limit $ \theta \rightarrow 0 $, but $ \theta \neq 0 $ $, d\Sigma /d\cos \theta  $
behaves as\begin{equation}
\label{ee_asym}
\left. \frac{d\Sigma }{d\cos \theta }\right| _{\theta \rightarrow 0}\approx \frac{1}{\theta ^{2}}\sum ^{\infty }_{k=1}\left( \frac{\alpha _{S}(M_{Z})}{\pi }\right) ^{k}\, \sum _{m=0}^{2k-1}c_{km}\ln ^{m}\left( \frac{\theta ^{2}}{4}\right) ,
\end{equation}
where $ c_{km} $ are calculable dimensionless coefficients. Additionally
there are virtual corrections to the lowest order cross section, which contribute
at $ \theta =0 $. Suppose we truncate the perturbative series in Eq.\,(\ref{ee_asym})
at $ k=N $. If $ N $ increases by 1 (that is, if we go to one higher order
in the series of $ \alpha _{S} $), the highest possible power of the logarithms
$ \ln ^{m}(\theta ^{2}/4) $ on the r.h.s. of Eq.\,(\ref{ee_asym})
increases by 2. Therefore, the theoretical prediction does not become more accurate
if the order of the perturbative calculation increases. Equivalently, 
the energy-energy correlation receives sizeable contributions from
arbitrarily high orders of $\alpha_S$.

To expose the two-scale nature of this problem, let us
introduce a spacelike four-vector $ q_{t}^{\mu } $ and a momentum scale $ q_{T} $
as\begin{eqnarray}
q_{t}^{\mu } & \equiv  & q^{\mu }-p_{A}^{\mu }\frac{q\cdot p_{B}}{p_{A}\cdot p_{B}}-p_{B}^{\mu }\frac{q\cdot p_{A}}{p_{A}\cdot p_{B}},\label{qTmu2} \\
q_{T}^{2} & \equiv  & -q_{t}^{\mu }q_{t\mu }>0,\label{qT22} 
\end{eqnarray}
where $ q^{\mu },\, p^{\mu }_{A},\, p_{B}^{\mu } $ are the momenta of the
$ Z^{0} $-boson and two jets, respectively. The vector $ q_{t}^{\mu } $ is interpreted
as the component of the  four-momentum $ q^{\mu } $ of the $Z^0$-boson that is
transverse to the four-momenta of the jets; 
that is, 
\begin{equation}
q_{t}\cdot p_{A}=q_{t}\cdot p_{B}=0.
\end{equation}
 The orthogonality of $ q_{t}^{\mu } $ to both $ p_{A}^{\mu } $ and $ p_{B}^{\mu } $
follows immediately from its definition~(\ref{qTmu2}). 

In the laboratory frame,\begin{eqnarray}
q^{\mu } & = & (M_{Z},\, \vec{0});\\
p_{A}^{\mu } & = & E_{A}(1,\, \vec{n}_{A});\\
p_{B}^{\mu } & = & E_{B}(1,\, -\vec{n}_{B}),
\end{eqnarray}
where $ E_{A},\vec{n}_{A} $ and $ E_{B},-\vec{n}_{B} $ are the energies
of the jets and the unity vectors in the directions of the jets, respectively.
The large invariant mass $ q^{2}=M_{Z}^{2} $ of the $ Z^{0} $-boson can
be associated with the QCD renormalization scale $ Q^{2} $. Let the $ z $-axis
be directed along $ \vec{n}_{A} $. Then $ q_{T} $ coincides with the length
of the transverse component $ \vec{q}_{T} $ of $ q_{t}^{\mu } $:\[
q_{t}^{\mu }=\left( -M_{Z}\tan \frac{\theta }{2},q_{T},0,-M_{Z}\tan \frac{\theta }{2}\right) .\]
At the same time\begin{equation}
\frac{q_{T}^{2}}{Q^{2}}=\frac{1-\cos \theta }{1+\cos \theta },
\end{equation}
and
\begin{equation}
\label{thetaee}
\lim _{\theta \rightarrow 0}\frac{q_{T}^{2}}{Q^{2}}=\frac{\theta ^{2}}{4}\left( 1+\frac{\theta ^{2}}{6}+\dots \right) .
\end{equation}

We see that the problems at $ \theta \rightarrow 0 $ arise due to the large
logarithmic terms $ \ln ^{m}(q_{T}^{2}/Q^{2})/q_{T}^{2} $ when $ q_{T}^{2}/Q^{2}\ll 1 $:
\begin{eqnarray}
\left. \frac{d\Sigma }{d\cos \theta }\right| _{\theta \rightarrow 0} & \approx  & \frac{Q^{2}}{2}\left. \frac{d\Sigma }{dq_{T}^{2}}\right| _{q_{T}\rightarrow 0}=\nonumber \\
 & = & \frac{1}{q_{T}^{2}}\sum ^{\infty }_{k=1}\left( \frac{\alpha _{S}(Q)}{\pi }\right) ^{k}\, \sum _{m=0}^{2k-1}c^{\prime }_{km}\ln ^{m}\left( \frac{q_{T}^{2}}{Q^{2}}\right) ,\label{ee_asym2} 
\end{eqnarray}
where \begin{equation}
c_{km}^{'}=\frac{Q^{2}}{4}c_{km}.
\end{equation}

The origin of these logarithms can be traced back to the presence of infrared
singularities in the QCD theory. Before considering these singularities, notice
that the energy-energy correlation is sufficiently inclusive to be 
infrared-safe. Therefore, the complete expression for the energy-energy
correlation is finite 
at each order of $ \alpha _{S}(\mu ) $. On the other hand,
the infrared singularities do appear in \emph{individual Feynman diagrams}.
According to the discussion in Section\,\ref{sec:InfraredSafety},
these singularities are due to the emission of soft gluons.\footnote{%
The purely collinear singularities do not appear because of the overall infrared
safety of the energy-energy correlation.
} Although the soft singularities cancel in the sum of all Feynman diagrams 
at the given order of $ \alpha _{S} $, this cancellation leaves large
remainders $ \ln ^{m}(q_{T}^{2}/Q^{2})/q_{T}^{2} $ if $q_T$ is small.

\begin{figure}[tb]
{\par\centering \resizebox*{!}{8cm}{\includegraphics{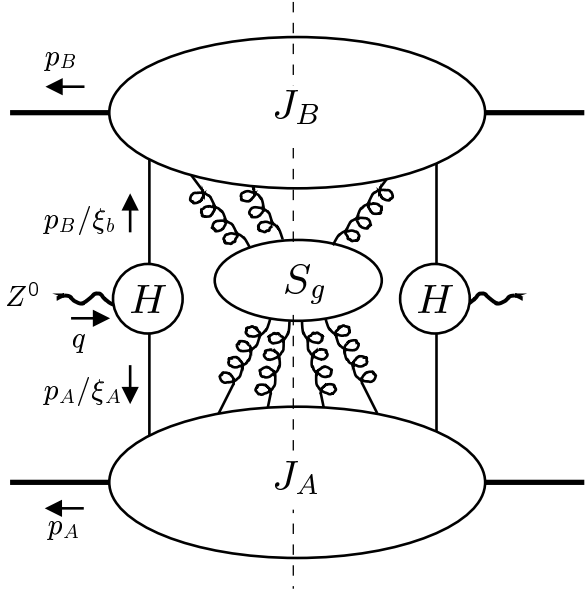}} \par}

\caption{\label{fig:ee_singularities}The structure of infrared singularities in a cut
diagram \protect$ D\protect $ for the energy-energy correlation in the axial
gauge}
\end{figure}

Fortunately, not all coefficients $ c^{\prime }_{km} $ in Eq.\,(\ref{ee_asym2})
are independent. Refs.\,\cite{DDT78,ParisiPetronzio79} suggested that 
the {\it leading} logarithmic subseries in Eq.\,(\ref{ee_asym2}) and
in analogous expressions in SIDIS and
Drell-Yan process can be summed
through all orders of $ \alpha _{S} $. The possibility to sum {\it
all} logarithmic subseries in Eq.\,(\ref{ee_asym2}) and restore the
convergence of the series in $\alpha_S$ was proven 
by J. Collins and D. Soper \cite{CS81}. 
Schematically, Eq.\,(\ref{ee_asym2}) can be written as
\cite{ArnoldKauffman}\begin{eqnarray}
\left. \frac{d\Sigma }{dq_{T}^{2}}\right| _{q_{T}\rightarrow 0} & \approx  & \frac{1}{q_{T}^{2}}\Bigl \{\nonumber \\
 &  & \alpha _{S}\left( L+1\right) \nonumber \\
 & + & \alpha ^{2}_{S}\left( L^{3}+L^{2}+L+1\right) \nonumber \\
 & + & \alpha _{S}^{3}\left( L^{5}+L^{4}+L^3 + L^2 + L +1\right) \nonumber \\
 & + & \dots \Bigr \},
\end{eqnarray}
where $ L\equiv \ln (q_{T}^{2}/Q^{2}) $, and the coefficients 
$ 2c'_{km}/(\pi ^{k} Q^2)$
are not shown. This series can be reorganized as\[
\left. \frac{d\Sigma }{dq_{T}^{2}}\right| _{q_{T}\rightarrow 0}\approx \frac{1}{q_{T}^{2}}\left\{ \alpha _{S}Z_{1}+\alpha ^{2}_{S}Z_{2}+\dots \right\} ,\]
where\begin{eqnarray}
\alpha _{S}Z_{1} & \sim  & \alpha _{S}(L+1)+\alpha ^{2}_{S}(L^{3}+L^{2})+\alpha ^{3}_{S}(L^{5}+L^{4})+\dots \hspace {30pt}\left| \, A_{1},B_{1},{\cal C}_{0}\right. ;\nonumber \\
\alpha ^{2}_{S}Z_{2} & \sim  & \hspace {77pt}\alpha ^{2}_{S}(L+1)+\alpha ^{3}_{S}(L^{3}+L^{2})+\dots \hspace {30pt}\left| \, A_{2},B_{2},{\cal C}_{1}\right. ;\nonumber \\
\alpha ^{3}_{S}Z_{3} & \sim  & \hspace {155pt}\alpha ^{3}_{S}(L+1)+\dots \hspace {30pt}\left| \, A_{3},B_{3},{\cal C}_{2}\right. ;\nonumber \\
 & \dots  & \label{aZ} 
\end{eqnarray}
In Eq.\,(\ref{aZ}), the right-hand side shows the new coefficients
$ A_{k},B_{k,}{\cal C}_{k-1} $ that are required to calculate each new subseries
$ \alpha ^{k}_{S}Z_{k}. $ The complete subseries $ \alpha _{S}^{k}Z_{k} $
can be reconstructed as soon as the coefficients $ A_{k},B_{k,}{\cal C}_{k-1} $
are known from the calculation of the term $ \alpha _{S}^{k}(L+1) $. Each
successive subseries $ \alpha _{S}^{k}Z_{k} $ in Eq.\,(\ref{aZ})
is smaller by $ \alpha _{S} $ than its predecessor, so that 
$ \alpha _{S} $ regains its role of
the small parameter of the perturbative expansion.

The rule that makes the resummation of the subseries $ \alpha _{S}^{k}Z_{k} $
possible follows from (a) the analysis of the structure of the infrared
singularities in the contributing Feynman diagrams \emph{at any order of $ \alpha _{S}(\mu ) $}
and (b) the requirement that the full energy-energy correlation is infrared-safe
and gauge- and renormalization-group invariant. 

The structure of the infrared singularities can be identified from the analysis
of analytic properties of the Feynman diagrams with the help of the Landau equations
\cite{LandauEqs,BjorkenDrell,AnalyticSMatrix} and the infrared power counting
\cite{Sterman78,Ellis79,StermanBook}. This structure for some contributing
cut diagram $ D $ is illustrated by Figure\,\ref{fig:ee_singularities}.
Throughout this discussion the axial gauge $ \zeta \cdot A=0 $ is used.\footnote{%
The discussion of the infrared singularities in covariant gauges can be found,
for instance, in Ref.\,\cite{StermanBook}.
} In $ D $ we can identify two jet parts $ J_{A},J_{B} $, the hard vertex
$ H $, and possibly the soft subdiagram $ S_{g} $. By their definition,
the jet parts $ J_{A} $ or $ J_{B} $ are the connected subdiagrams of
$ D $ that describe the propagation of nearly \emph{on-shell} massless particles
inside the observed jets. 
Each of the particles in the jet part $ J_{A} $
has a four-momentum $ p_{i}^{\mu } $ that is proportional to the momentum
of the jet $ A $:\begin{equation}
p_{i}^{\mu }=\beta _{i}p_{A}^{\mu },
\end{equation}
where \begin{eqnarray}
 &  & 0\leq \beta _{i}\leq 1,\mbox {\quad and\quad }\sum _{i}\beta _{i}=1.
\end{eqnarray}
Similar relations hold for the momenta of the particles in the jet part $ J_{B} $.

\begin{figure}[H]
{\par\centering \resizebox*{1\textwidth}{!}{\includegraphics{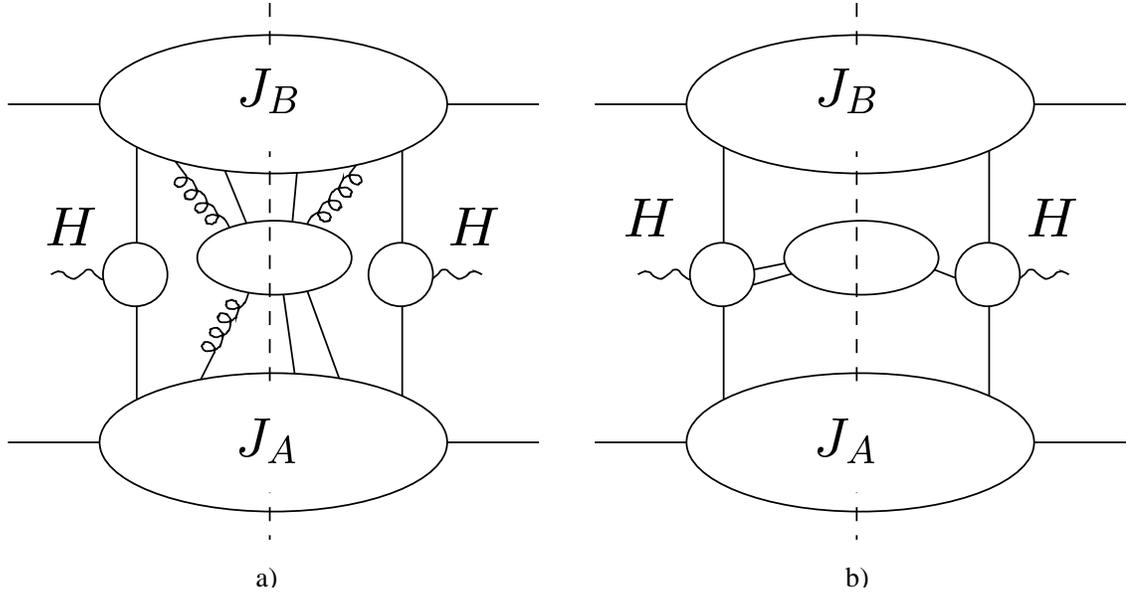}} \par}

\caption{\label{fig:ee_soft_finite}Examples of the finite soft subdiagrams: (a) the
subdiagrams that are connected to \protect$ J_{A},\, J_{B}\protect $ by one
or several quark lines; (b) the subdiagrams that are connected to \protect$ H\protect $}
\end{figure}

Both jets originate from the hard vertex $ H $ that contains contributions
from the highly off-shell particles. In the axial gauge the 
jet parts are connected
to $ H $ only through the single quark lines. Since the hard scattering happens
practically at one point, $ H $ depends only on $ Q^{2} $ and not on $ q_{T}^{2} $.

After the jets are created, they propagate in different directions with
the speed of light. Due to the Heisenberg uncertainty principle, these jets,
which are separated by large distances,
do not interact with one another except by the exchange of
low momentum (soft) particles.
The infrared singularities,
which are associated with the long-distance dynamics, can occur only in the
jet parts or the soft subdiagram. This
observation can be refined by the dimensional analysis of the Feynman integrals
in the infrared limit \cite{Sterman78,Ellis79,StermanBook}, which shows that
the infrared singularities are at most logarithmic. Also,
those soft subdiagrams that
are attached to the jet parts $ J_{A}$ and $J_{B} $ with one or more \emph{quark} propagators
(Fig.\,\ref{fig:ee_soft_finite}a) or  
 are connected to $ H $
(Fig.\,\ref{fig:ee_soft_finite}b) are finite. 

To summarize, the infrared singularities of any individual Feynman diagram reside
in the soft parts of the jets $ J_{A},J_{B} $ and in subdiagrams $ S_{g} $
that are connected to $ J_{A},\, J_{B} $ by soft gluon lines (cf. Fig.\,\ref{fig:ee_singularities}).
Both types of singularities contribute at $ q_{T}=0 $ (\emph{i.e.}, $ q^{\mu }=p_{A}^{\mu }+p_{B}^{\mu } $),
in agreement with the expectation that the small-$ q_{T} $ logarithms 
are remainders from the cancellation of such singularities. 
Therefore, 
at small $ q_{T} $ the distribution $ d\Sigma /dq_{T}^{2} $ 
naturally factorizes as
\begin{equation}
\left. \frac{d\Sigma }{dq_{T}^{2}}\right| _{q_{T}\rightarrow 0}=
H(Q^{2})\,{\cal C}_{A}^{out}\,{\cal C}_{B}^{out}\,{\cal S}(q_{T}^{2},Q^{2}),
\end{equation}
where $ H(Q^{2}) $ is the contribution from the pointlike hard part, $ {\cal S}(q_{T}^{2},Q^{2}) $
is the all-order sum of the large logarithms, and $ {\cal C}_{A,B}^{out} $
collect finite contributions from the jet parts. Clearly, ${\cal
C}_A^{out}={\cal C}_B^{out}$
due to the symmetry between the jets. 

The factorized
formula is proven by considering the Fourier-Bessel transform of $ d\Sigma /dq_{T}^{2} $
to the space of the impact parameter $ \vec{b} $, which is conjugate to $ \vec{q}_{T} $.
Explicitly,\begin{equation}
\label{ee_cor_resum}
\left. \frac{d\Sigma }{dq_{T}^{2}}\right| _{q_{T}\rightarrow 0}=\frac{\sigma _{0}}{S_{e^{+}e^{-}}}\int \frac{d^{2}\vec{b}}{(2\pi )^{2}}e^{i\vec{q}_{T}\cdot \vec{b}}\wt{W}_{\Sigma }(b,Q),
\end{equation}
where\begin{equation}
\label{ee_W}
\wt{W_{\Sigma }}(b,Q)=\sum _{j}{\wt e}_{j}^{2}{\cal C}^{out}_{A}(C_{1},C_{2}){\cal C}^{out}_{B}(C_{1},C_{2})e^{-S(b,Q,C_{1},C_{2})}.
\end{equation}
In Eqs.\,(\ref{ee_cor_resum},\ref{ee_W}),\begin{equation}
S_{e^{+}e^{-}}=Q^{2}
\end{equation}
is the square of the center-of-mass energy of the initial-state electron and
positron; $ \sum _{j} $ denotes the summation over the active quark flavors
(\emph{i.e.}, $ j=u,\bar{u},d,\bar{d},\dots  $); ${\wt e}_{j} $ 
are the couplings of the quarks to the 
$Z^0$-bosons\footnote{For the up quarks, 
$${\wt e}_j =\frac{e}{\sin
2\theta_W}\left(\frac{1}{2}-\frac{4}{3}\sin^2\theta_W\right),$$
where $e$ is the charge of the positron and $\theta_W$ is the weak
mixing angle. For the down quarks,
$${\wt e}_j =\frac{e}{\sin
2\theta_W}\left(-\frac{1}{2}+\frac{2}{3}\sin^2\theta_W\right).$$
};
$ (\sigma _{0}/S_{e^{+}e^{-}})\sum _{j}{\wt e}_{j}^{2} $
is the Born approximation for the hard part $ H(Q^{2}) $. The Fourier-Bessel
transform of the shape factor $ {\cal S}(q_{T}^{2},Q^{2}) $ is given
by
$ e^{-S(b,Q,C_{1},C_{2})} $, where $ S(b,Q,C_{1},C_{2}) $ is
called the Sudakov function. At $ b^{2}\ll \Lambda _{QCD}^{-2} $ (\emph{i.e.},
in the region of applicability of perturbative QCD), the Sudakov function is
given by the integral between two momentum scales of the order $ Q $
and $ 1/b, $ respectively:
\begin{eqnarray}
\hspace{-7pt}\lim _{b\rightarrow 0}S(b,Q,C_{1},C_{2}) & = & \int
_{C_{1}^{2}/b^{2}}^{C_{2}^{2}Q^{2}}\frac{d\ov \mu ^{2}}{\ov \mu ^{2}}\left(
\ASud (\alpha _{S}(\ov \mu ),C_{1})\ln \frac{C_{2}^{2}Q^{2}}{\ov \mu
^{2}}+\BSud (\alpha _{S}(\ov \mu ),C_{1},C_{2})\right) ,\nonumber \\ &&
\label{ee_SP} 
\end{eqnarray}
where $ {\cal A} $ and $ {\cal B} $ can be calculated in PQCD.
$ C_{1} $ and $ C_{2} $ are arbitrary constants of the order 1 that determine
the range of the integration in $S(b,Q)$.
The undetermined values of these constants reflect certain freedom 
in separation of the collinear-soft contributions 
included in $ S^{P}(b,Q) $ from
the purely collinear contributions 
included in $ {\cal C} $-functions. 
At each order of $ \alpha _{S} $, changes
in $ S^{P}(b,Q) $ due to the variation of $ C_{1},\, C_{2} $ are compensated
by the opposite changes in the $ {\cal C} $-functions. Hence the perturbative
expansion of $ \wt{W}_{\Sigma} $ does not depend on these constants. However,
the complete form-factor $ \wt{W}_{\Sigma} $ in Eq.\,(\ref{ee_W})
does have residual dependence on
$ C_{1},C_{2} $ because of the exponentiation of the terms depending on
$C_1$ and $C_2$ in $ \exp{(-S(b,Q,C_1,C_2))}$. 
The variation of $ C_{1},C_{2} $ allows
us to test the scale invariance of the separation of soft and
collinear 
contributions in the  $ \wt{W}_{\Sigma} $-term.

At $ b^{2}\gg \Lambda _{QCD}^{-2} $,
the behavior of $ S $ is determined by complicated nonperturbative dynamics,
which remains intractable at the current level of the development of the 
theory. At large $ b $ the Sudakov function $ S $ is parametrized by a phenomenological
function $ S^{NP}(b,Q) $, which has to be found from the comparison with
the experimental data. When $ Q\rightarrow \infty $, the sensitivity of the resummation
formula to the nonperturbative part of $ S(b,Q) $ is expected to decrease.

Now suppose that the experiment identifies a hadron 
$ H_{A} $ in the jet $ J_{A} $ and
a hadron $ B $ in the jet $ J_{B} $. Let $ z_{A,B} $ be the fractions
of the energies of the jets $ J_{A} $ and $ J_{B} $ carried by $ H_{A} $
and $ H_{B} $, respectively. The cross section 
of the process $ e^{+}e^{-}\stackrel{Z^{0}}{\longrightarrow }H_{A}H_{B}X $ 
is no longer infrared-safe because of the collinear singularities due to the
fragmentation into the hadrons $H_A$ and $H_B$. Nonetheless,
in the limit $q_T\rightarrow 0$ 
the  cross section $ d\sigma /(dz_{A}dz_{B}dq_{T}^{2}) $ factorizes
similarly to Eqs.\,(\ref{ee_cor_resum},\ref{ee_W}):
\begin{equation}
\label{ee_CSS}
\left. \frac{d\sigma _{H_{A}H_{B}}}{dz_{A}dz_{B}dq_{T}^{2}}\right| _{q_{T}\rightarrow 0}=\frac{\sigma _{0}}{S_{e^{+}e^{-}}}\int \frac{d^{2}\vec{b}}{(2\pi )^{2}}e^{i\vec{q}_{T}\cdot \vec{b}}\wt{W}_{H_{A}H_{B}}(b,z_{A},z_{B}),
\end{equation}
where at $ b\rightarrow 0 $
\begin{eqnarray}
&&\wt{W}_{H_{A}H_{B}}(b,z_{A},z_{B}) =  \sum _{j}e_{j}^{2}\times \nonumber \\
&&\left( \sum _{a}D_{H_{A}/a}\otimes {\cal C}^{out}_{aj}\right) (z_{A},b,\mu_D )
\left( \sum _{b}D_{H_{B}/b}\otimes {\cal C}^{out}_{bj}\right) (z_{B},b,\mu_D )
e^{-S};\nonumber\\
 && a,b  =g,\stackrel{{(-)}}{u},\stackrel{{(-)}}{d},\dots ;\nonumber\\
 && j =\stackrel{{(-)}}{u},\stackrel{{(-)}}{d},\dots .
\label{ee_cs_resum}
\end{eqnarray}
The only major difference between the form-factor $ \wt{W}_{H_{A}H_{B}} $
for the hadron pair production cross section $ d\sigma /(dz_{A}dz_{B}dq_{T}^{2}) $
and the form-factor $ \wt{W}_{\Sigma } $ for the energy-energy correlation
$ d\Sigma /dq_{T}^{2} $ is the presence of the fragmentation functions $ D_{H/a}(\xi ,\mu ) $,
which absorb the collinear singularities due to the final-state fragmentation into
the observed hadrons $ H_{A},H_{B} $. The FFs are convolved with the
coefficient functions \linebreak 
${\cal C}_{ab}^{out}(\xi,C_1,C_2,\mu_D,b)$, which 
absorb finite contributions due to the perturbative collinear
radiation. 

\begin{figure}[H]
{\par\centering \resizebox*{0.9\textwidth}{!}{\includegraphics{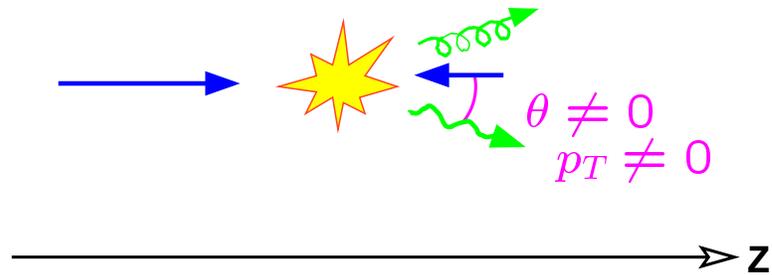}} \par}

\caption{The space-time picture of Drell-Yan process\label{fig:DYspace}}
\end{figure}

The same resummation technique can also be applied to the production of vector
bosons (\emph{e.g.}, virtual photons $\gamma^*$, which decay into
lepton-antilepton pairs) at hadron-hadron colliders (Fig.\,\ref{fig:eeDY}b).
In this process, the four-vector $ q_{t}^{\mu } $ is introduced using the
same definition (\ref{qTmu2}), where now $ p_{A} $ and $ p_{B} $ denote
the momenta of the initial hadrons $A$ and $B$. The scale $ q_{T} $ is just the magnitude
of the transverse momentum $ p_{T} $ of $\gamma^*$ in the center-of-mass
frame of the hadron beams (Fig.\,\ref{fig:DYspace}), since in this
frame\[
q_{t}^{\mu }=(0,p_{T},0,0).\]
Therefore, the $ b $-space resummation formalism \cite{CSS85}
applies to the 
production of vector bosons with small transverse momenta. The cross
section for the production of the virtual photon $\gamma^*$ 
at $ q_{T}\rightarrow 0 $ can be factorized as\begin{equation}
\left. \frac{d\sigma _{\gamma}}{dQ^{2}dydq_{T}^{2}}\right|
_{q_{T}\rightarrow 0}=\frac{\sigma '_{0}}{S_{AB}}\int \frac{d^{2}\vec{b}}{(2\pi )^{2}}e^{i\vec{q}_{T}\cdot \vec{b}}\wt{W}_{\gamma}(b,x_{A},x_{B}),
\end{equation}
where $ Q^{2} $ and $ y $ are the virtuality and rapidity of
$ \gamma^* $ in the lab frame, 
$ x_{A,B}\equiv \frac{Q^{2}}{S_{AB}}e^{\pm y}, $
and\begin{eqnarray}
\left. \wt{W}_{\gamma}(b,x_{A},x_{B})\right| _{b\rightarrow 0} & = & 
\sum _{a,b,j}e_j^2
\left( {\cal C}^{in}_{ja}\otimes F_{a/A}\right) (x_{A},b,\mu )
\left( {\cal C}^{in}_{jb}\otimes F_{b/B}\right) (x_{B},b,\mu )e^{-S}.\nonumber\\
&&\label{WV} 
\end{eqnarray}
In $ \wt{W}_{\gamma} $, $e_j$ are fractional electric charges of the
quarks 
($e_j=2/3$ for up
quarks and $-1/3$ for the down quarks).
$ \left( {\cal C}^{in}_{ja}\otimes F_{a/A}\right) (x_{A},b,\mu ) $
and $ \left( {\cal C}^{in}_{jb}\otimes F_{b/B}\right) (x_{B},b,\mu ) $ are
the jet parts corresponding to the \emph{incoming} hadrons $ A $ and $
B $.
They are constructed from the perturbatively calculable coefficient functions
$ {\cal C}_{ab}^{in}(\xi ,b,\mu ) $ convolved
with the PDFs for the relevant partons. The perturbative part of the Sudakov
function in Eq.\,(\ref{WV}) has the same functional dependence as in
Eq.\,(\ref{ee_SP}) for $e^+e^-$-hadroproduction. 
As in the case of $\wt{W}_\Sigma$ and $\wt{W}_{H_A H_B}$,
the large-$b$ behavior of $ \wt{W}_\gamma$ should be 
parametrized by a phenomenological
function.

To conclude, the $ b- $space resummation formalism was originally
derived to describe the production of hadrons at $ e^{+}e^{-} $ colliders
\cite{CS81} and production of electroweak vector bosons at hadron-hadron colliders
\cite{CSS85}. The possibility to apply the same formalism to SIDIS 
relies on close similarities between the three processes.
First, hadronic interactions in all three processes are described by the same
set of Feynman diagrams in different crossing channels. Second, multiple
parton radiation dominates each of the three processes when the
final-state particle
escapes closely to the direction predicted by the leading-order
kinematics. 
The formalism for the resummation of such radiation can be formulated in
Lorentz-invariant notations, so that it can be 
continued from one process to another.

\newpage

\newpage

\begin{figure}[H]
{\par\centering \resizebox*{!}{8cm}{\includegraphics{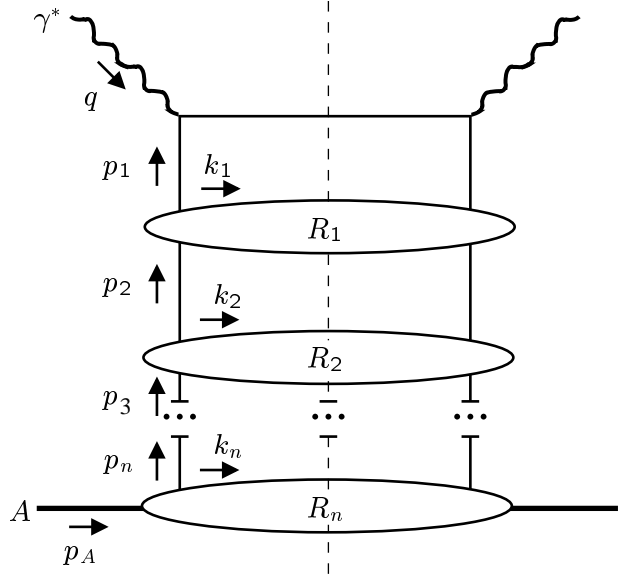}} \par}

\caption{\label{fig:Ladder}The ladder structure of the DIS cut diagrams}
\end{figure}

\subsection{QCD at small \protect$ x\protect $\label{sub:BFKL}}

According to the discussion in Section\,\ref{sec:InfraredSafety}, the
convergence of the series in $ \alpha _{S}(\mu ) $ depends on the absence
of very large or 
small dimensionless quantities in the perturbative coefficients. 
In particular, the dimensionless variable $ x $ 
in the inclusive DIS cross section should not be too close to zero: otherwise
the hard part of the DIS cross section contains large logarithms $ \ln ^{m}(1/x) $, 
which compensate for the smallness
of $ \alpha _{S}(Q) $. These logarithms are different from the logarithms
$ \ln ^{m}\mu ^{2} $ resummed by the DGLAP evolution equations. 
As a result, the factorization of the DIS cross section
in the hard cross section and PDFs (cf. Eq.\,(\ref{DISfact}))
may experience difficulties  at small~$x$. 

The large logarithms $ \ln^m (1/x) $ are resummed
in the formalism of Balitsky, Fadin, Kuraev and Lipatov (BFKL) \cite{BFKL}.
The BFKL and DGLAP pictures for the history of the parton probed in the hard
scattering are quite different. Both types of formalisms resum contributions from
the cut ladder diagrams shown in Figure\,\ref{fig:Ladder}.
In this Figure, each ``rung'' $ R_{k} $ is a two-particle irreducible subdiagram
that corresponds to the radiation off the probed parton line 
(see Refs.\,\cite{StermanBook,DKMT} for more details). 
The vertical propagators correspond to the quarks or the gluons that are parents
to the probed quark. The momenta $p_{1,2,\dots ,n}^\mu$ flow from 
the parent hadron to the probed quark. The momenta 
$k_{1,2,\dots ,n}^\mu$ flow through the 
rungs and are sums of the momenta of the radiated particles. 
The conservation of the momentum in each rung implies that 
\begin{equation}
p_{i}^\mu = p_{i+1}^\mu - k_{i}^\mu,\quad\quad i =1,\dots,n,  
\end{equation} 
where $p_{n+1}^\mu \equiv p_A^{\mu} $. In the reference frame where
the hadron $A$ moves at the speed of light along the $z$-axis, 
$x$ coincides with the ratio of the plus components of $p_0^\mu$ and $p_A^\mu$:
\begin{equation}
x = \frac{p_0^+}{p_A^+},
\end{equation}
where 
\begin{equation}
k^{\pm} \equiv k^0 \pm k^3. 
\end{equation}

The DGLAP equation arises from the resummation of the ladder diagrams corresponding
to the collinear radiation along the direction of the hadron $A$. The radiating parton
remains highly boosted at each rung of the ladder. At the same time, the transverse momentum
carried away by the radiation grows rapidly from the bottom to the top of the ladder.
The DGLAP equation corresponds to the strong ordering of the transverse momenta flowing through
the rungs $R_i$: that is,
\begin{equation}
Q^{2}\gg k^{2}_{T1}\gg k^{2}_{T2}\gg \dots \gg k^{2}_{Tn}\gg \Lambda _{QCD}^{2},
\end{equation}
while 
\begin{equation}
p_{1}^+\sim p_{2}^+\sim \dots \sim p_{n}^+\sim p_A^+ \gg 0
\end{equation}
and
\begin{equation}
p_{1}^-\sim p_{2}^-\sim \dots \sim p_{n}^-\sim p_A^- \sim 0.
\end{equation}

On the other hand, the BFKL formalism describes the situation in which
the QCD radiation carries 
away practically all energy of the probed parton. In this case, 
\begin{equation}
p_{1}^+\ll p_{2}^+\ll \dots \ll p_{n}^+\ll p_A^+, 
\end{equation}
and
\begin{equation}
p_{1}^-\gg p_{2}^-\gg \dots \gg p_{n}^-\gg p_A^-.
\end{equation}
In addition, the BFKL picture imposes no ordering on
the transverse components of~$k_i^\mu$:
\begin{equation}
k_{T1}^{2}\sim k_{T2}^{2}\sim \dots \sim k_{Tn}^{2}\gg \Lambda ^{2}_{QCD}.
\end{equation} 
As a result, the probed quark is likely to have a significant
transverse momentum throughout the whole process of evolution,
which is impossible in the DGLAP picture. Due to its large $k_T$, 
the radiating parton is off its mass shell at any moment of its
evolution history, so that the  BFKL radiation
cannot be factorized from the hard scattering. 
As another consequence of the $k_T$-unordered radiation, 
the BFKL picture implies broad angular distributions of the
final-state hadrons, while in the DGLAP picture the hadrons are more
likely to belong to the initial- and final-state jets.   

Since the BFKL approach applies to the 
limit $x\rightarrow 0$ and $k_{iT}^2 \gg \Lambda_{QCD}^2$,
it corresponds to asymptotically high energies of hadronic collisions.
So far, the experiments have produced no data that would definitely 
require the BFKL formalism to explain them.  
In particular, the behavior of the inclusive  
DIS structure functions in the low $x$ region at HERA agrees well with 
the ${\cal O}(\alpha_S^2)$ predictions of the traditional factorized
formalism  and disagrees with the steep power-law growth predicted by 
the leading-order solution of the BFKL equation \cite{BFKL}.  

The situation is not so clear for some less inclusive observables,
which deviate from the low-order predictions of PQCD. 
Specifically, SIDIS in the small-$x$ region is characterized by 
large higher-order corrections. Some of these corrections can be
potentially attributed to the enhanced
$k_T$-unordered radiation at $x \rightarrow 0$. If this is indeed the case, 
the effects of the $k_T$-unordered radiation may be identified by 
observing the changes in the angular distributions of the final-state 
hadrons or ``intrinsic $k_T$'' of the partons.  
In order to pinpoint these effects, good understanding of the angular
dynamics in the traditional DGLAP picture is needed. 
Such understanding can be achieved in the framework of the small-$q_T$ 
resummation formalism, which systematically describes
angular distributions 
of the hard, soft and collinear radiation. Hence it can be naturally used to 
organize our knowledge about the angular patterns of 
the DGLAP radiation and search for the effects from new low-$x$ QCD dynamics. 

\chapter{Resummation in semi-inclusive DIS: 
theoretical formalism \label{ch:Formalism}}

Deep-inelastic lepton-hadron scattering (DIS) is one of the cornerstone processes
to test PQCD. Traditionally, the experimental study of the fully inclusive DIS
process $ e+A\rightarrow e+X $, where $ A $ is usually a nucleon, and
$ X $ is any final state, is used to measure the parton distribution functions
(PDFs) for $ A $. These functions describe the long-range dynamics of hadron
interactions and are required by many PQCD calculations. During the 1990's,
significant attention has been also paid to various aspects of semi-inclusive
deep inelastic scattering (SIDIS), for instance, the semi-inclusive production
of hadrons and jets, $ e+A\rightarrow e+B+X $ and $ e+A\rightarrow e+jets+X $.
In particular, the H1 and ZEUS collaborations at HERA, European Muon Collaboration at CERN, and the E665 experiment at Fermi National Accelerator Laboratory 
performed extensive experimental
studies of the charged particle multiplicity \cite{H1chgdMisc,H1chgd97,ZEUSchgd96,ZEUSasym,EMC,E665}
and hadronic transverse energy flows \cite{H1z1,H1z2} at large momentum transfer
$ Q $. It was found that some aspects of the data, \emph{e.g.}, the Feynman
$ x $ distributions, can be successfully explained in the framework of PQCD
analysis \cite{GraudenzPLB,GraudenzThesis}. On the other hand, applicability
of PQCD to the description of other features of the process is limited. For
example, the perturbative calculation in lowest orders fails to describe the
pseudorapidity or transverse momentum distributions of the final hadrons. Under
certain kinematical conditions the whole perturbative expansion as a series
in the QCD coupling may fail due to the large logarithms discussed
in Section\,\ref{sec:TwoScale}.

To be more specific, consider semi-inclusive DIS production of hadrons of a
type~$ B $. At large energies, one can neglect the masses of the participating
particles. In semi-inclusive DIS at given energies of the beams, any event can
be characterized by two energy scales: the virtuality of the exchanged vector
boson $ Q $ and the scale $ q_{T} $ introduced
analogously to $ e^{+}e^{-} $ hadroproduction and Drell-Yan
process (cf. Section\,\ref{sec:TwoScale}). The scale $q_T$ is also related
to the transverse momentum of $B$.
The expansion in the series of $ \alpha _{S} $ is justified if at least one
of these scales is much larger than $ \Lambda _{QCD} $. However, the above
necessary condition does not guarantee fast convergence of perturbative series
in the presence of large logarithmic terms. If 
$ \Lambda ^{2}_{QCD}\ll Q^2,\, q^{2}_{T}\ll Q^{2} $, the
cross sections are dominated by the soft and collinear logarithms 
$ \log ^{m}\left( q^{2}_{T}/Q^{2}\right) , $
which can be resummed in the framework of the small-$ q_{T} $ resummation
formalism (Subsection\,\ref{sub:ResumSoftColLogs}). 
In the limit 
$ \Lambda ^{2}_{QCD}\ll q^2_T,\, Q^{2}\ll q^{2}_{T} $
(photoproduction region) PQCD may fail due to the large terms 
$ \log ^{m}\left( Q^{2}/q^{2}_{T}\right)  $,
which should be resummed into the parton distribution function of the virtual
photon \cite{Kramer}.
Finally, even in
the region $ \Lambda ^{2}_{QCD}\ll q^{2}_{T}\sim Q^{2} $ one may encounter
another type of large logarithms corresponding to events with large rapidity
separation between the partons and/or the hadrons. This type of large logarithms
can be resummed with the help of the Balitsky-Fadin-Kuraev-Lipatov (BFKL) formalism (Subsection\,\ref{sub:BFKL}). 

In this Chapter I discuss resummation of soft and collinear logarithms in SIDIS
hadroproduction $ e+A\rightarrow e+B+X $ in the limit 
$ \Lambda _{QCD}^2\ll Q^2,\, q_{T}^2\ll Q^2 $.
The calculations are based on the works by Meng, Olness, and Soper~\cite{Meng1,Meng2},
who analyzed the resummation technique for a particular energy distribution
function of the SIDIS process.\footnote{%
The general features of the resummation formalism in semi-inclusive DIS were
first discussed by J. Collins \cite{Collins93}. 
} This energy distribution function receives contributions from all possible
final-state hadrons and does not depend on the specifics of fragmentation.

Here the resummation is discussed in a more general context compared to \cite{Meng1,Meng2}:
namely, I also consider the final-state fragmentation of the partons. Using
this formalism, I discuss the impact of soft and collinear PQCD radiation
on a wide class of physical observables including particle multiplicities. The
calculations will be done in the next-to-leading order of PQCD. In the next
Chapter, I compare the resummation formalism with the H1 data on the pseudorapidity
distributions of the transverse energy flow~\cite{H1z1,H1z2} and 
ZEUS data on multiplicity of charged particles \cite{ZEUSchgd96} in the $ \gamma ^{*}p $
center-of-mass frame. Another goal of this study is to find in which regions
of kinematical parameters the CSS resummation formalism is sufficient to describe
the existing data, and in which regions significant contributions from other
hadroproduction mechanisms, such as the BFKL radiation \cite{BFKL}, higher-order
corrections including multijet production with \cite{Kramer} or without \cite{Catani,GraudenzMultijet}
resolved photon contributions, or photoproduction showering \cite{Jung}, cannot
be ignored.

\section{Kinematical Variables \label{sec:Formalism:KinemVar}}

I follow notations which are similar to the ones used in \cite{Meng1,Meng2}.
In this Section I summarize them.

I consider the process
\begin{equation}
\label{sdis2}
e+A\rightarrow e+B+X,
\end{equation}
 where $ e $ is an electron or positron, $ A $ is a proton (or other hadron
in the initial state), $ B $ is a hadron observed in the final state, and
$ X $ represents any other particles in the final state in the sense of inclusive
scattering (Fig.~\ref{fig:SIDIS}). I denote the momenta of $ A $ and $ B $
by $ p_{A}^{\mu } $ and $ p_{B}^{\mu } $, and the momenta of the electron
in the initial and final states by $ l^{\mu } $ and $ l^{\prime \mu } $.
Also, $ q^{\mu } $ is the momentum transfer to the hadron system, $ q^{\mu }=l^{\mu }-l^{\prime \mu } $.
Throughout all discussion, I neglect particle masses. 

I assume that the initial electron and hadron interact only through a single
photon exchange. Contributions due to the exchange of $ Z $-bosons or higher-order
electroweak radiative corrections will be neglected. Therefore, $ q^{\mu } $
also has the meaning of the 4-momentum of the exchanged virtual photon $ \gamma ^{*} $;
$ q^{\mu } $ is completely determined by the momenta of the initial- and
final-state electrons. In many respects, DIS behaves as scattering
of virtual photons on hadrons, so that the theoretical discussion of hadronic
interactions can often be simplified 
by considering only the photon-proton system.

\subsection{Lorentz scalars\label{sub:LorentzScalars}}

For further discussion, I define five Lorentz scalars relevant to the process
(\ref{sdis2}). The first is the center-of-mass energy of the initial hadron
and electron $ \sqrt{S_{eA}} $ where\begin{equation}
\label{SeA}
S_{eA}\equiv (p_{A}+l)^{2}=2p_{A}\cdot l.
\end{equation}
 I also use the conventional DIS variables $ x $ and $ Q^{2} $ which
are defined from the momentum transfer $ q^{\mu } $ by\begin{equation}
Q^{2}\equiv -q^{2}=2\ell \cdot \ell ^{\prime },
\end{equation}
\begin{equation}
\label{x}
x\equiv \frac{Q^{2}}{2p_{A}\cdot q}.
\end{equation}
 In principle, $ x $ and $ Q^{2} $ can be completely determined in an
experimental event by measuring the momentum of the outgoing electron.

Next I define a scalar $ z $ related to the momentum of the final hadron
state $ B $ by\begin{equation}
\label{z}
z=\frac{p_{B}\cdot p_{A}}{q\cdot p_{A}}=\frac{2xp_{B}\cdot p_{A}}{Q^{2}}.
\end{equation}
 The variable $ z $ plays an important role in the description of fragmentation
in the final state. In particular, in the quark-parton model (or in the leading
order perturbative calculation) it is equal to the fraction of the fragmenting
parton's momentum carried away by the observed hadron.

The next relativistic invariant $ q_{T}^{2} $ is the square of the component
of the virtual photon's 4-momentum $ q^{\mu } $ that is transverse to the
4-momenta of the initial and final hadrons:\begin{equation}
\label{qT2}
q_{T}^{2}=-q_{t}^{\mu }q_{t\mu },
\end{equation}
 where\begin{equation}
\label{qTmu}
q_{t}^{\mu }=q^{\mu }-p_{A}^{\mu }\frac{q\cdot p_{B}}{p_{A}\cdot p_{B}}-p_{B}^{\mu }\frac{q\cdot p_{A}}{p_{A}\cdot p_{B}}.
\end{equation}
 As discussed in Subsection\,\ref{sub:ResumSoftColLogs}, the momentum
$ q_{t}^{\mu } $ plays the crucial role in the resummation of the soft and
collinear logarithms. In particular, a fixed-order PQCD cross-section
is divergent when $ q_{T}\rightarrow 0 $,
so that all-order resummation is needed to make the theory predictions 
finite in this limit. According to Eqs.\,(\ref{z},\ref{qTmu}) $ q_{t}^{\mu }=0 $
if and only if $ p_{B}^{\mu }=z\left( xp_{A}^{\mu }+q^{\mu }\right) . $ Hence
the resummation is required when the final-state hadron
$ B $ approximately follows the direction of $ x\vec{p}_{A}+\vec{q}. $ 

In the analysis of kinematics, I will use three reference frames. The most obvious
frame is the laboratory frame, or the rest frame of the experimental detector.
The observables in this frame are measured directly, but the theoretical analysis is complicated due to the varying momentum of the
photon-proton system. Hence I will mostly use two other reference frames, the
center-of-mass frame of the initial hadron and the virtual photon (hadronic
c.m., or hCM frame), and a special type of Breit frame which I will call, depending
on whether the initial state is a hadron or a parton, the \textit{hadron} or
\textit{parton} frame. As was shown in Ref.\,\cite{Meng2}, the resummed
cross section can be derived naturally in the hadron frame. On the other hand,
many experimental results are presented for observables in the hCM~frame. These
observables are not measured directly; rather they are reconstructed from directly
measured observables in the laboratory frame. I will use subscripts $ h $,
$ cm $ and $ lab $ to denote kinematical variables in the hadron, hCM or
laboratory frame. Below I discuss kinematical variables in all three frames.

\subsection{Hadron frame}

\begin{figure}[H]
{\par\centering \resizebox*{0.8\textwidth}{!}{\includegraphics{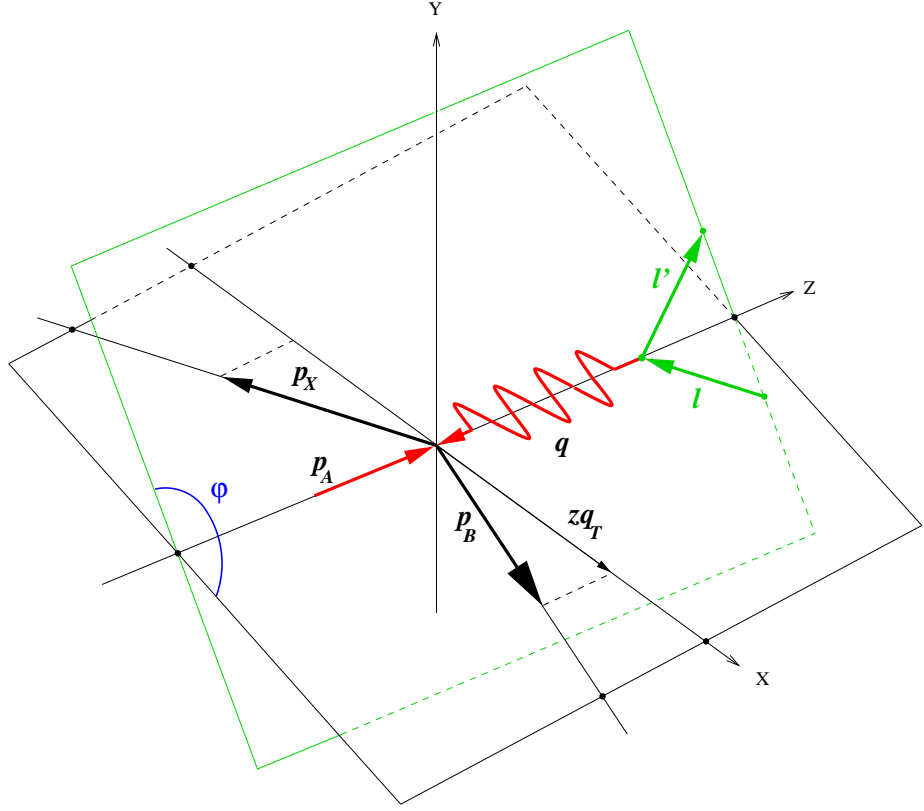}} \par}

\caption{Geometry of the particle momenta in the hadron frame\label{fig:HadronFrame}}
\end{figure}

Following Meng et al. \cite{Meng1,Meng2} the hadron frame is defined by two
conditions: (a) the energy component of the 4-momentum of the virtual
photon is zero, and
(b) the momentum of the outgoing hadron $ B $ lies in the $ xz $ plane.
The directions of particle momenta in this frame are shown in Fig.~\ref{fig:HadronFrame}.

In this frame the proton $ A $ moves in the $ +z $ direction, while the
momentum transfer $ \vec{q} $ is in the $ -z $ direction, and $ q^{0} $
is 0:\begin{eqnarray}
q^{\mu }_{h} & = & \left( 0,0,0,-Q\right) ,\label{qh} \\
p_{A,h}^{\mu } & = & \frac{Q}{2x}\left( 1,0,0,1\right) .\label{pAh} 
\end{eqnarray}

The momentum of the final-state hadron $ B $ is\begin{equation}
\label{pBh}
p_{B,h}^{\mu }=\frac{zQ}{2}\Bigl (1+\frac{q_{T}^{2}}{Q^{2}},\frac{2q_{T}}{Q},0,\frac{q_{T}^{2}}{Q^{2}}-1\Bigr ).
\end{equation}

The incoming and outgoing electron momenta in the hadron frame are defined in
terms of variables $ \psi  $ and $ \varphi  $ as follows \cite{phipsi}:
\begin{eqnarray}
\ell ^{\mu }_{h} & = & \frac{Q}{2}\left( \cosh \psi ,\sinh \psi \cos \varphi ,\sinh \psi \sin \varphi ,-1\right) ,\nonumber \\
\ell ^{\prime \mu }_{h} & = & \frac{Q}{2}\left( \cosh \psi ,\sinh \psi \cos \varphi ,\sinh \psi \sin \varphi ,+1\right) .\label{lh} 
\end{eqnarray}
 Note that $ \varphi  $ is the azimuthal angle of $ \vec{\ell }_{h} $
or $ \vec{\ell }^{\prime }_{h} $ around the $ Oz $-axis. $ \psi  $
is a parameter of a boost which relates the hadron frame to an electron Breit
frame in which $ \ell ^{\mu }=(Q/2,0,0,-Q/2) $. By (\ref{SeA}) and (\ref{lh})\begin{equation}
\cosh \psi =\frac{2xS_{eA}}{Q^{2}}-1=\frac{2}{y}-1,
\end{equation}
 where the conventional DIS variable $ y $ is defined as\begin{equation}
y\equiv\frac{Q^{2}}{xS_{eA}}.
\end{equation}
 The allowed range of the variable $ y $ in deep-inelastic scattering is
$ 0\leq y\leq 1 $ (see Subsection \ref{sub:LabFrame}); therefore $ \psi \geq 0 $.

The transverse part of the virtual photon momentum $ q_{t}^{\mu } $ has a
simple form in the hadron frame; it can be shown that\begin{equation}
q_{t,h}^{\mu }=(-\frac{q_{T}^{2}}{Q},-q_{T},0,-\frac{q_{T}^{2}}{Q}).
\end{equation}
 In other words, $ q_{T} $ is the magnitude of the transverse component of
$ \vec{q}_{t,h} $. The transverse momentum $ p_{T} $ of the final-state
hadron $ B $ in this frame is simply related to $ q_{T} $, by\begin{equation}
p_{T}=zq_{T}.
\end{equation}
 Also, the pseudorapidity of $ B $ in the hadron frame is\begin{equation}
\eta _{h}\equiv -\log \Bigl (\tan \frac{\theta _{B,h}}{2}\Bigr )=\log \frac{q_{T}}{Q}.
\end{equation}

The resummed cross-section will be derived using the hadron frame. To transform
the result to other frames, it is useful to express the basis vectors of the
hadron frame ($ T^{\mu },\, \, X^{\mu },\, \, Y^{\mu },Z^{\mu } $) in terms
of the particle momenta \cite{Meng1}. For an arbitrary coordinate frame,\begin{eqnarray}
T^{\mu } & = & \frac{q^{\mu }+2xp_{A}^{\mu }}{Q},\nonumber \\
X^{\mu } & = & \frac{1}{q_{T}}\Biggl (\frac{p^{\mu }_{B}}{z}-q^{\mu }-\Bigl [1+\frac{q_{T}^{2}}{Q^{2}}\Bigr ]xp_{A}^{\mu }\Biggr ),\nonumber \\
Y^{\mu } & = & \epsilon ^{\mu \nu \rho \sigma }Z_{\nu }T_{\rho }X_{\sigma },\nonumber \\
Z^{\mu } & = & -\frac{q^{\mu }}{Q}.
\end{eqnarray}
 If these relations are evaluated in the hadron frame, the basis vectors $ T^{\mu },\, \, X^{\mu },\, \, Y^{\mu },Z^{\mu } $
are $ (1,0,0,0),\, \, (0,1,0,0),(0,0,1,0),(0,0,0,1) $, respectively.

\begin{figure}[H]
{\par\centering \vspace{1\baselineskip}\resizebox*{\textwidth}{!}{\includegraphics{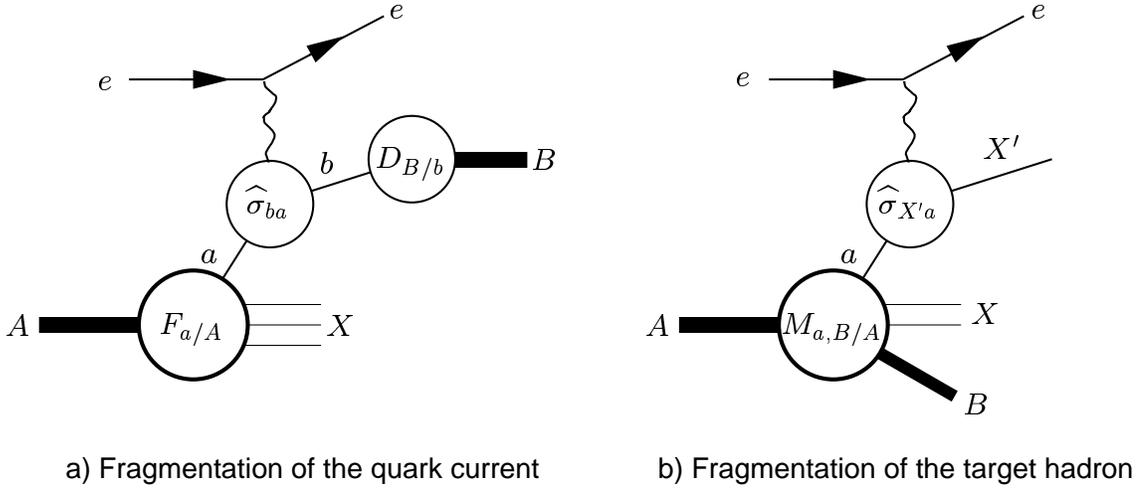}} \par}

\caption{\label{fig:FractureFns}(a) In the current fragmentation region, the hadron-level
cross section can be factorized into hard partonic cross sections \protect$ \widehat{\sigma }_{ba}\protect $,
parton distribution functions \protect$ F_{a/A}(\xi _{a},\mu _{F})\protect $,
and fragmentation functions \protect$ D_{B/b}(\xi _{b},\mu _{F})\protect $.
(b) In the target fragmentation region, the hadrons are produced through the
mechanism of diffractive scattering that depends on ``diffractive parton
distributions{}'' \protect$ M_{a,B/A}(\xi _{a},\zeta _{B},\mu _{F}).\protect $}
\end{figure}

The limit of small $ q_{T} $, which is the most relevant for our resummation
calculation, corresponds to the region of large \emph{negative} pseudorapidities
in the hadron frame. Hence the resummation affects the rate of the production of
the hadrons that follow closely the direction of the virtual photon. The region
of negative $ \eta _{h} $ is often called the \textit{current fragmentation
region,} since the final-state hadrons are produced due to the
interaction of the virtual photon with the quark current. In the current fragmentation
region, hadroproduction proceeds through independent scattering and subsequent
fragmentation of partons. Therefore, in this region the hadron-level 
cross section $ \sigma _{BA} $ can be factorized in
the cross sections $ \widehat{\sigma }_{ba} $ for the electron-parton scattering
$ e+a\rightarrow e+b+X $, the PDFs $ F_{a/A}(\xi _{a},\mu _{F}) $, 
and the FFs $ D_{B/b}(\xi _{b},\mu _{D}) $
(cf. Figure\,\ref{fig:FractureFns}a). 
The formal proof of the factorization in the current
region of SIDIS can be found in \cite{Collins93,LeveltMulders94}. 

In the opposite direction $ \eta _{h}\gg 0 $
($ q_{T}\rightarrow +\infty  $) contributions from the current fragmentation
vanish. Rather the produced hadron is likely to be a product of fragmentation
of the target proton, which moves in the $ +z $-direction
(cf. Eq.\,(\ref{pAh})).
According to Eq.\,(\ref{z}), such hadrons have $z\approx 0$.
The target fragmentation hadroproduction is described by a different approach,
which relies on factorization of the hadron-level cross section into cross sections
of parton subprocesses and
\textit{diffractive parton distributions}
$ M_{a,B/A}(\xi _{a},\zeta _{B},\mu _{F}) $
(cf. Figure\,\ref{fig:FractureFns}b). These distributions can be interpreted
as probabilities for the initial hadron $ A $ to fragment into 
the parton $ a $,
the hadron $ B $, and anything else. $ \xi _{a} $ and $ \zeta _{B} $
denote fractions of the momentum of $ A $ that are carried by the parton
$ a $ and the hadron $ B $, respectively. The distributions $ M_{a,B/A}(\xi _{a},\zeta _{B},\mu _{F}) $
(also called \emph{fracture functions}\textit{\emph{) were introduced in
Refs.\,\cite{TrentadueFractureFns,Berera94} and used in \cite{GraudenzThesis,GraudenzFractureFns,deFlorianFractureFns,GrazziniFractureFns}
to describe various aspects of SIDIS with unpolarized and polarized beams. The
factorization of cross sections in the target fragmentation region was formally
proven in the scalar field theory \cite{DiffractiveFactorizationPhi3} and in
full QCD \cite{Berera96,Collins98}. The recent experimental studies of the
diffractive scattering at HERA are reviewed in \cite{Newman2000}. The detailed
discussion of diffractive scattering and interesting models \cite{DiffractiveModels}
that are applied for its analysis is beyond the scope of this work.}}

\begin{figure}[H]
{\par\centering \resizebox*{1\textwidth}{!}{\includegraphics{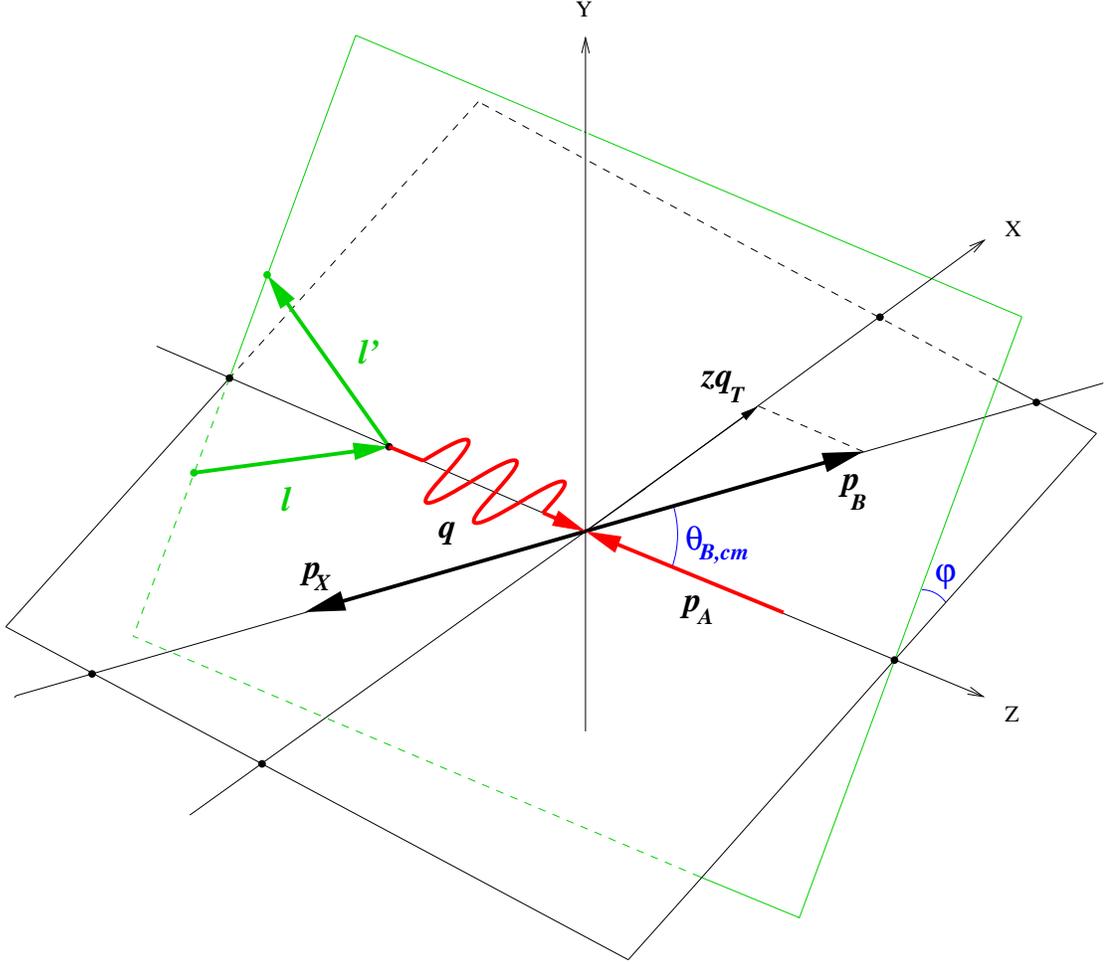}} \par}

\caption{Particle momenta in the hadronic center-of-mass (hCM) frame\label{fig:hCM}}
\end{figure}

\subsection{Photon-hadron center-of-mass frame}

The center-of-mass frame of the proton $ A $ and virtual photon $ \gamma ^{*} $
is defined by the condition $ \vec{p}_{A,cm}+\vec{q}_{cm}=0 $. The relationship
between particle momenta in this frame is illustrated in Fig.~\ref{fig:hCM}.
As in the hadron frame, the momenta $ \vec{q}_{cm} $ and $ \vec{p}_{A,cm} $
in the hCM frame are directed along the $ Oz $ axis. The coordinate transformation
from the hadron frame into the hCM frame consists of (a) a boost in the direction
of the virtual photon and (b) inversion of the direction of the $ Oz $ axis,
which is needed to make the definition of the hCM   frame consistent with the
one adopted in HERA experimental publications. In the hCM   frame the momentum
of $ \gamma ^{*} $ is\begin{equation}
q_{cm}^{\mu }=\Biggl (\frac{W^{2}-Q^{2}}{2W},0,0,\frac{W^{2}+Q^{2}}{2W}\Biggr ),
\end{equation}
 where $ W $ is the hCM  energy of the $ \gamma ^{*}p $ collisions,\begin{equation}
W^{2}\equiv (p_{A}+q)^{2}=Q^{2}\left( \frac{1}{x}-1\right) \geq 0.
\end{equation}
Since all energy of the $\gamma^*p$ system is transformed into the energy
of the final-state hadrons, $ W $ coincides with the invariant
mass of the $ B+X $ system. 

The momenta of the initial and final hadrons $ A $ and $ B $ are given
by\begin{equation}
\label{pAc}
p^{\mu }_{A,cm}=\Biggl (\frac{W^{2}+Q^{2}}{2W},0,0,-\frac{W^{2}+Q^{2}}{2W}\Biggr ),
\end{equation}
\begin{equation}
\label{pBc}
p^{\mu }_{B,cm}=\Biggl (E_{B},E_{B}\sin \theta_{B,cm},0,E_{B}\cos \theta_{B,cm}\Biggr ),
\end{equation}
where\begin{eqnarray}
E_{B} & = & z\frac{W^{2}+q_{T}^{2}}{2W},\\
\cos \theta_{B,cm} & = & \frac{W^{2}-q_{T}^{2}}{W^{2}+q_{T}^{2}}.
\end{eqnarray}

The hadron and hCM~frames are related by a boost along the $ z $-direction,
so that the expression for the transverse momentum of the final hadron $ B $ in the
hCM   frame is the same as the one in the hadron frame,\begin{equation}
\label{pTB}
p_{T}=zq_{T}.
\end{equation}
 Also, similar to the case of the hadron frame, the relationship between $ q_{T} $
and the pseudorapidity of $ B $ in the hCM   frame is simple,\begin{equation}
\label{qTetacm}
q_{T}=We^{-\eta _{cm}}.
\end{equation}

Since the directions of the $ z $-axis are opposite in the hadron frame and
the hCM frame, large \emph{negative} pseudorapidities in the hadron frame ($ q_{T}\rightarrow 0 $)
correspond to large \emph{positive} pseudorapidities in the hCM frame. Hence
multiple parton radiation effects should be looked for in SIDIS data at $ q_{T}/Q\lesssim 1 $,
or \begin{equation}
\eta _{cm}\gtrsim \ln \left( \sqrt{\frac{1-x}{x}}\right) >2.
\end{equation}

The boost from the hadron to the hCM frame also preserves the angle $ \varphi  $
between the planes of the hadronic and leptonic momenta, so that the momenta
$ l^{\mu },l^{\prime \mu } $ of the electrons in the hCM frame are
\begin{eqnarray}
&& l_{cm}^{\mu } =  \Biggl \{\frac{1}{4W}\biggl ((W^{2}+Q^{2})\cosh
\psi +W^{2}-Q^{2}\biggr ),\frac{Q}{2}\sinh \psi \cos \varphi ,
\nonumber \\
& &-\frac{Q}{2}\sinh \psi \sin \varphi ,
\frac{1}{4W}\biggl ((W^{2}+Q^{2})+(W^{2}-Q^{2})\cosh \psi \biggr )\Biggr \};\\
&&l_{cm}^{\prime \mu } =  \Biggl \{\frac{1}{4W}\biggl
((W^{2}+Q^{2})\cosh \psi -W^{2}+Q^{2}\biggr ),\frac{Q}{2}\sinh \psi
\cos \varphi ,\nonumber \\
& & -\frac{Q}{2}\sinh \psi \sin \varphi ,
\frac{1}{4W}\biggl (-W^{2}-Q^{2}+(W^{2}-Q^{2})\cosh \psi \biggr )\Biggr \}.
\end{eqnarray}

Finally I would like to mention two more variables, which are
commonly used in the experimental analysis.
The first variable is the flow of the transverse hadronic
energy
\begin{equation}
E_{T}\equiv E_{tot}\sin \theta_{cm},
\end{equation}
where $ E_{tot} $ is the total energy of the final-state hadrons registered
in the direction of the polar angle $ \theta_{cm} $. The measurement
of $E_T$ does not require identification of individual final-state hadrons;
hence $ E_{T} $ is less sensitive to the final-state fragmentation.
\\
\indent The second variable is Feynman $ x $, defined as\begin{equation}
\label{xF}
x_{F}\equiv \frac{2p_{B,cm}^{z}}{W}=z\left( 1-\frac{q_{T}^{2}}{W^{2}}\right) .
\end{equation}
 In (\ref{xF}) $ p_{B,cm}^{z} $ is the longitudinal component of the momentum
of the final-state hadron in some frame. For small values of $ q_{T} $,
\textit{i.e.,} in the region with the highest rate,\begin{equation}
x_{F}\approx z.
\end{equation}

\begin{figure}[H]
{\par\centering \resizebox*{1\textwidth}{!}{\includegraphics{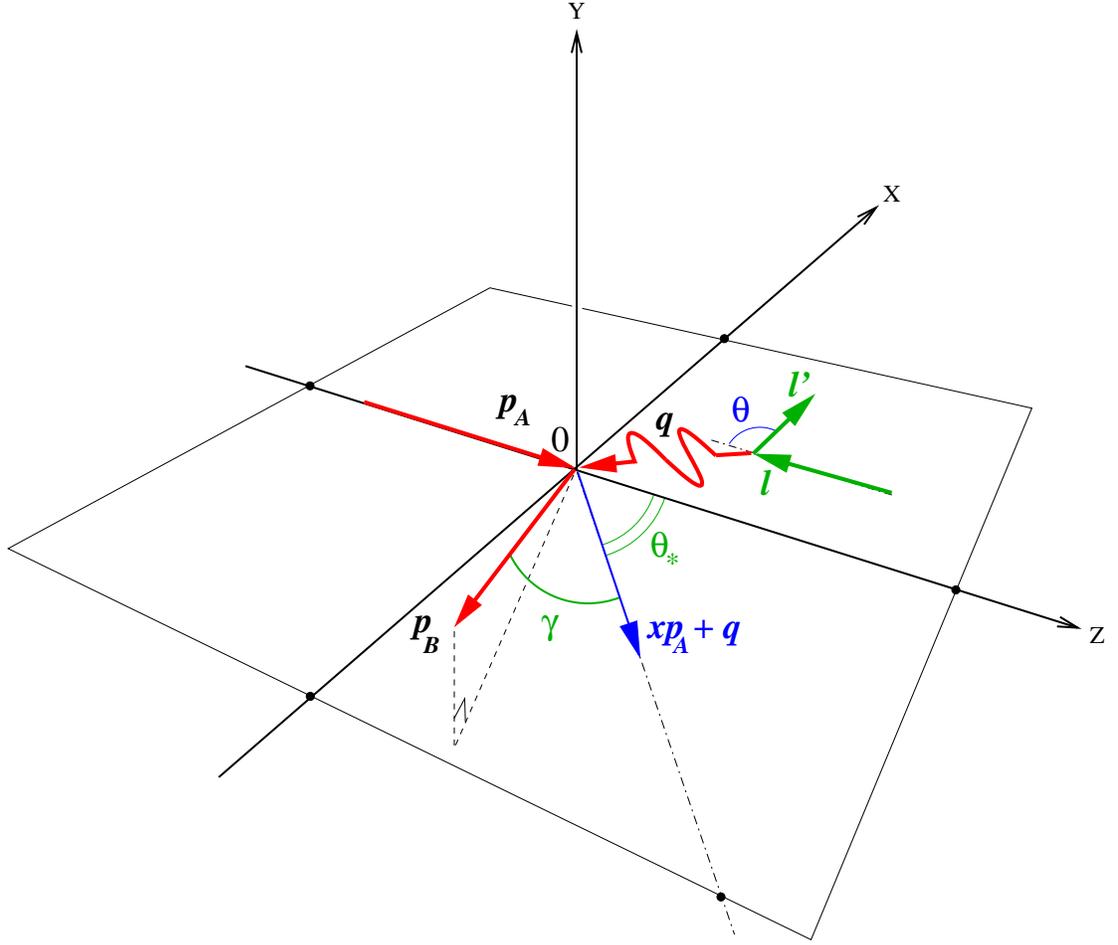}} \par}

\caption{\label{fig:LabFrame}Particle momenta in the laboratory frame}
\end{figure}

\subsection{Laboratory frame\label{sub:LabFrame}}

In the laboratory frame, the electron and proton beams are collinear to the
$ Oz $ axis. The definition of the HERA lab frame is that the proton ($ A $)
moves in the $ +z $ direction with energy $ E_{A} $, and the incoming
electron moves in the $ -z $ direction with energy $ E $. The momenta
of the incident particles are\begin{equation}
\label{pAlab}
p_{A,lab}^{\mu }=\left( E_{A},0,0,E_{A}\right) ,
\end{equation}
\begin{equation}
\label{llab}
l_{lab}^{\mu }=\left( E,0,0,-E\right) .
\end{equation}
We can use (\ref{SeA},\ref{pAlab}) and (\ref{llab}) to express
the Mandelstam variable $ S_{eA} $ in terms of the energies $ E_{A},E $
in the lab frame:\begin{equation}
S_{eA}=4E_{A}E.
\end{equation}

The outgoing electron has energy $ E^{\prime } $ and scattering angle $ \theta  $
relative to the $ -z $ direction. I define the $ Ox $-axis of the HERA
frame in such a way that the outgoing electron is in the $ Oxz $-plane; that is,\begin{equation}
l^{\prime \mu }_{lab}=\left( E^{\prime },-E^{\prime }\sin \theta ,0,-E^{\prime }\cos \theta \right) .
\end{equation}
 The four-momentum $ q^{\mu }=l^{\mu }-l^{\prime \mu } $ of the virtual photon
that probes the structure of the hadron is correspondingly\begin{equation}
q_{lab}^{\mu }=\left( E-E^{\prime },E^{\prime }\sin \theta ,0,-E+E^{\prime }\cos \theta \right) .
\end{equation}

The scalars $ x $ and $ Q^{2} $ are completely determined by measuring
the energy and the scattering angle of the outgoing electron:\begin{equation}
Q^{2}=2EE^{\prime }(1-\cos \theta ),
\end{equation}
\begin{equation}
x=\frac{EE^{\prime }(1-\cos \theta )}{E_{A}\left[ 2E-E^{\prime }(1+\cos \theta )\right] }.
\end{equation}
 Rather than working directly with $ E^{\prime } $ and $ \theta  $ (or $ Q^{2} $
and $ x $), it is convenient to introduce 
another pair of variables $ y $ and $ \beta  $:\begin{equation}
\label{y}
y\equiv \frac{Q^{2}}{xS_{eA}}=\frac{2E-E^{\prime }(1+\cos \theta )}{2E},
\end{equation}
and\begin{equation}
\beta \equiv \frac{2xE_{A}}{Q}=\frac{\sqrt{2EE^{\prime }(1-\cos \theta )}}{2E-E^{\prime }(1+\cos \theta )}.
\end{equation}
The variable $ y $ satisfies the constraints\begin{equation}
\label{yconstraints}
\frac{W^{2}}{S_{eA}}\leq y\leq 1,
\end{equation}
where $ W$ is defined in the previous subsection. 
The relationship (\ref{yconstraints})
can be derived easily by rewriting $ y $ as\begin{equation}
y=1-\frac{2(p_{A}\cdot l^{\prime })}{S_{eA}}=1+\frac{T_{eA}}{S_{eA}},
\end{equation}
where\begin{equation}
T_{eA}\equiv (p_{A}-l^{\prime })^{2}.
\end{equation}
Eq.\,(\ref{yconstraints}) follows from the geometrical constraints
on $ T_{eA} $ for the fixed invariant mass $ W^{2}$ of the final-state
hadrons: \begin{equation}
W^{2}-S_{eA}\leq T_{eA}\leq 0.
\end{equation}

The observed hadron ($ B $) has energy $ E_{B} $ and scattering angle
$ \theta _{B} $ with respect to the $ +z $ direction, and azimuthal angle
$ \varphi _{B} $; thus its momentum is\begin{equation}
\label{pTBl}
p_{B,lab}^{\mu }=(E_{B},E_{B}\sin \theta _{B}\cos \varphi _{B},E_{B}\sin \theta _{B}\sin \varphi _{B},E_{B}\cos \theta _{B}).
\end{equation}
The scalars $ z $ and $ q_{T}^{2} $ depend on the momentum of the outgoing
hadron:\begin{equation}
z=\frac{\beta E_{B}(1-\cos \theta _{B})}{Q},
\end{equation}
\begin{equation}
\label{qT2lab}
q_{T}^{2}=\frac{2E_{B}E_{0}}{z}\Biggl [1-\cos \gamma \Biggr ].
\end{equation}
 In Eq.\,(\ref{qT2lab}) $ \gamma  $ is the angle between $ \vec{p}_{B} $
and $ x\vec{p}_{A}+\vec{q} $ (cf. Fig.\,\ref{fig:LabFrame});
\begin{equation}
E_{0}\equiv \frac{Q\left( 1+(1-y)\beta ^{2}\right) }{2\beta }
\end{equation}
is the energy component of $ xp_{A}^{\mu }+q^{\mu } $. 
Define $ \theta _{*} $ to be the polar angle of $ xp_{A}^{\mu }+q^{\mu }: $\begin{equation}
xp^{\mu }_{A,lab}+q_{lab}^{\mu }\equiv E_{0}\left( 1,\sin \theta _{*},0,\cos \theta _{*}\right) ,
\end{equation}
where\begin{equation}
cot\frac{\theta _{*}}{2}= \beta \sqrt{1-y}.
\end{equation}
The angle $ \gamma  $ in Eq.\,(\ref{qT2lab}) can be easily expressed
in terms of the angles $ \theta _{*}, $ $ \theta _{B} $, and $ \varphi _{B} $,
as\begin{equation}
\label{cosgamma}
\cos \gamma =\cos \theta _{*}\cos \theta _{B}+\sin \theta _{*}\sin \theta _{B}\cos \varphi _{B}.
\end{equation}

\begin{figure}[H]
{\par\centering \resizebox*{0.9\textwidth}{!}{\includegraphics{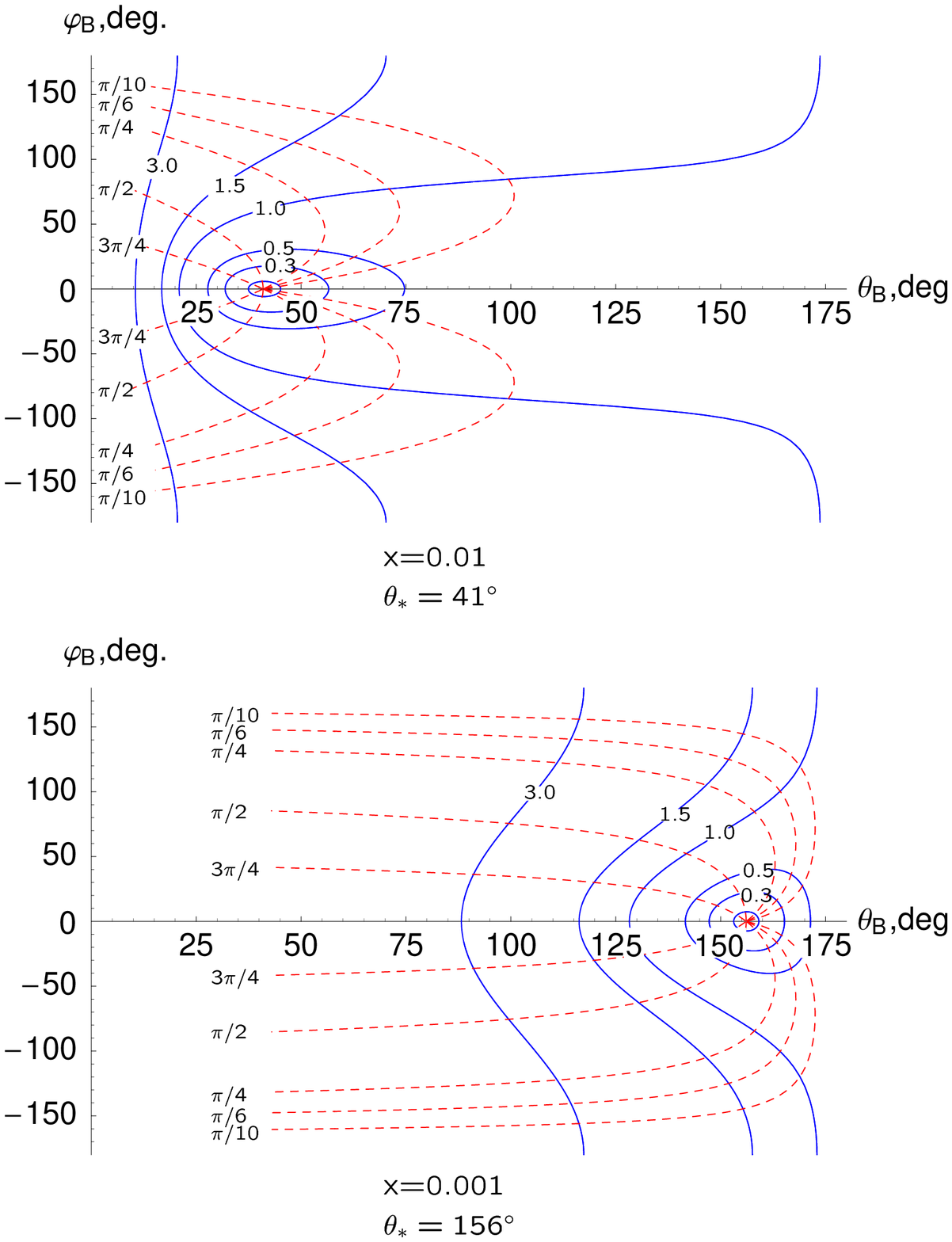}} \vspace{-0.5cm}\par}

\caption{\label{fig:thetaBphiB}The variables \protect$ q_{T}\protect $ and \protect$ \varphi \protect $
as functions of the angles \protect$ \theta _{B},\, \varphi _{B}\protect $.
Solid lines are contours of constant \protect$ q_{T}\protect $ for \protect$ q_{T}/Q\protect $
ranging from 0.1 (the innermost contour) to 3.0. Dashed lines are contours of
constant \protect$ \varphi \protect $ for \protect$ \varphi \protect $
ranging from \protect$ \pi /10\protect $ to \protect$ 3\pi /4\protect $.
The contour \protect$ \varphi =\pi \protect $ coincides with the \protect$ \theta _{B}\protect $-axis.
The plots correspond to \protect$ E_{A}=820\protect $ GeV, 
\protect$ E=27\protect $ GeV,
\protect$ Q=6\protect $ GeV, \protect$ x=0.01\protect $ (upper plot) and
\protect$ x=0.001\protect $ (lower plot).}
\end{figure}

Finally, the azimuthal angle $ \varphi  $ of the lepton plane in the hadron
frame (cf. Eqs.\,(\ref{lh})) is related to the lab frame variables
as\begin{equation}
\label{philab}
\cos \varphi =\frac{Q}{2q_{T}}\frac{1}{\sqrt{1-y}}\left[ 1-y+\frac{q_{T}^{2}}{Q^{2}}-\frac{1}{\beta ^{2}}\coth ^{2}\frac{\theta _{B}}{2}\right] .
\end{equation}
Figure \ref{fig:thetaBphiB} shows contours of constant $ q_{T} $ and $ \varphi  $
in the plane of the angles $ \theta _{B} $ and $ \varphi _{B} $. The point
$ q_{T}=0 $ corresponds to $ \theta _{B}=\theta _{*}, $ $ \varphi _{B}=0 $,
in agreement with Eqs.\,(\ref{qT2lab},\ref{cosgamma}). According to
these equations, $ q_{T} $ depends on $ \varphi _{B} $ through $ \cos \varphi _{B} $,
which is a sign-even function of $ \varphi _{B} $. Thus each pair of $ q_{T},\, \varphi  $
determines $ \varphi _{B} $ up to the sign, so that the contours in Figure
\ref{fig:thetaBphiB} are symmetric with respect to the replacement $ \varphi _{B}\rightarrow -\varphi _{B} $. 

\newpage

\newpage

\subsection{Parton kinematics}

The kinematical variables and momenta discussed so far are all hadron-level
variables. Next, I relate these to parton variables.

Let $ a $ denote the parton in $ A $ that participates in the hard scattering,
with momentum \begin{equation}
\label{pa}
p_{a}^{\mu }=\xi _{a}p_{A}^{\mu }.
\end{equation}
Let $ b $ denote the parton of which $ B $ is a fragment, with momentum
\begin{equation}
\label{pb}
p_{b}^{\mu }=p_{B}^{\mu }/\xi _{b}.
\end{equation}
The momentum fractions $ \xi _{a} $ and $ \xi _{b} $ range from 0 to 1.
At the parton level, I introduce the Lorentz scalars $ \widehat{x},\, \, \widehat{z},\, \, \widehat{q}_{T} $
analogous to the ones at the hadron level\begin{equation}
\widehat{x}=\frac{Q^{2}}{2p_{a}\cdot {q}}=\frac{x}{\xi _{a}},
\end{equation}
\begin{equation}
\widehat{z}=\frac{p_{b}\cdot p_{a}}{q\cdot p_{a}}=\frac{z}{\xi _{b}},
\end{equation}
\begin{equation}
\qTh ^{2}=-\widehat{q}_{t}^{\mu }\widehat{q}_{t\mu }.
\end{equation}
 Here $ \widehat{q}_{T}^{\mu } $ is the component of $ q^{\mu } $ which
is orthogonal to the parton 4-momenta $ p^{\mu }_{a} $ and $ p^{\mu }_{b} $,\[
\widehat{q}_{t}\cdot p_{a}=\widehat{q}_{t}\cdot p_{b}=0.\]
 Therefore,\begin{equation}
\label{qTmup}
\widehat{q}_{t}^{\mu }=q^{\mu }-p_{a}^{\mu }\frac{q\cdot p_{b}}{p_{a}\cdot p_{b}}-p_{b}^{\mu }\frac{q\cdot p_{a}}{p_{a}\cdot p_{b}}.
\end{equation}
 In the case of massless initial and final hadrons the hadronic and partonic
vectors $ q_{t}^{\mu } $ coincide,
\begin{equation}
\wh{q}_{t}^{\mu }=q^{\mu }_{t}.
\label{qThqT}
\end{equation}

\section{\label{sec:NLOXsec}The structure of the SIDIS cross-section}

The knowledge of five Lorentz scalars $ S_{eA},\, \, Q,\, \, q_{T},\, \, x,\, \, z $
and the lepton azimuthal angle $ \varphi  $ in the hadron frame is sufficient
to specify unambiguously the kinematics of the semi-inclusive scattering event
$ e+A\rightarrow e+B+X $. In the following, I will discuss the hadron cross-section
$ d\sigma _{BA} $, which is related to the parton cross-section $ d\widehat{\sigma }_{ba} $
by\begin{equation}
\label{hadcs}
\frac{d\sigma _{BA}}{dxdzdQ^{2}dq_{T}^{2}d\varphi }=\sum _{a,b}\int _{z}^{1}\frac{d\xi _{b}}{\xi _{b}}D_{B/b}(\xi _{b},\mu _{D})\int _{x}^{1}\frac{d\xi _{a}}{\xi _{a}}F_{a/A}(\xi _{a},\mu _{F})\frac{d\widehat{\sigma }_{ba}(\mu _{F},\mu _{D})}{d\widehat{x}d\widehat{z}dQ^{2}dq_{T}^{2}d\varphi }.
\end{equation}
 Here $ F_{a/A}(\xi _{a},\mu _{F}) $ denotes the distribution function (PDF)
of the parton of a type $ a $ in the hadron $ A $, and $ D_{B/b}(\xi _{b},\mu _{D}) $
is the fragmentation function (FF) for a parton type $ b $ and the final
hadron $ B $. The sum over the labels $ a,b $ includes contributions from
all parton types, \textit{i.e.,} $ g,u,\bar{u},d,\bar{d},\dots \, \,  $.
In the following, a sum over the indices $ i,j $ will include contributions
from active flavors of quarks and antiquarks only, \textit{i.e.,} it will not
include a gluonic contribution. The parameters $ \mu _{F} $ and $ \mu _{D} $
are the factorization scales for the PDFs and FFs. To simplify the following
discussion and calculations, I assume that the factorization scales $ \mu _{F},\, \mu _{D} $
and the renormalization scale $ \mu  $ are the same:
\begin{equation}
\mu _{F}=\mu _{D}= \mu.
\end{equation}

The analysis of semi-inclusive DIS can be conveniently organized by separating
the dependence of the parton and hadron cross-sections on the leptonic angle
$ \varphi  $ and the boost parameter $ \psi  $ from the other kinematical
variables $ x,\, \, z,\, \, Q $ and $ q_{T} $ \cite{phipsi}. This separation
does not depend on the details of the hadronic dynamics.
Following
\cite{Meng1}, I express the hadron (or parton) cross-section as a sum over
products of functions of these lepton angles in the hadron frame $ A_{\rho }(\psi ,\varphi ) $,
and structure functions $ \, ^{\rho }V_{BA}(x,z,Q^{2},q_{T}^{2}) $ (or $ \prescr{\rho }\widehat{V}_{ba}(\widehat{x},\widehat{z},Q^{2},q_{T}^{2},\mu ) $,
respectively):\begin{equation}
\label{ang1}
\frac{d\sigma _{BA}}{dxdzdQ^{2}dq_{T}^{2}d\varphi }=\sum _{\rho =1}^{4}\prescr{\rho }V_{BA}(x,z,Q^{2},q_{T}^{2})A_{\rho }(\psi ,\varphi ),
\end{equation}
\begin{equation}
\label{ang2}
\frac{d\widehat{\sigma }_{ba}(\mu )}{d\widehat{x}d\widehat{z}dQ^{2}dq_{T}^{2}d\varphi }=\sum _{\rho =1}^{4}\prescr{\rho }\widehat{V}_{ba}(\widehat{x},\widehat{z},Q^{2},q_{T}^{2},\mu )A_{\rho }(\psi ,\varphi ).
\end{equation}
The coefficients $ \prescr{\rho }V_{BA} $ (or $ \prescr{\rho }\Vh _{ba} $)
of the angular functions $ A_{\rho }(\psi ,\varphi ) $ are independent of
one another. 

At the energy of HERA, hadroproduction via parity-violating $ Z $-boson exchanges
can be neglected, and only four out of the nine angular functions listed in
\cite{Meng1} contribute to the cross-sections (\ref{ang1}-\ref{ang2}). They
are\begin{eqnarray}
A_{1}=1+\cosh ^{2}\psi , &  & A_{3}=-\cos \varphi \sinh 2\psi ,\nonumber \\
A_{2}=-2, &  & A_{4}=\cos 2\varphi \sinh ^{2}\psi .
\end{eqnarray}

\begin{figure}[H]
{\par\centering \resizebox*{1\textwidth}{!}{\includegraphics{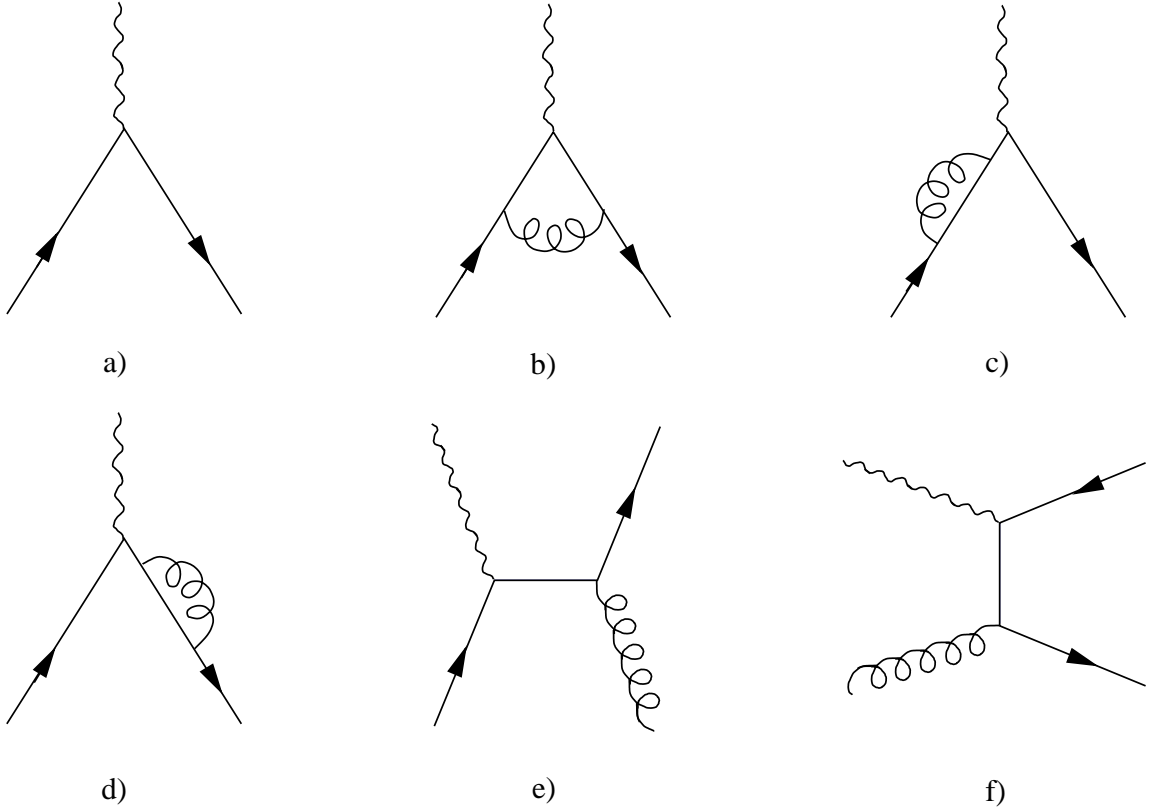}} \par}

\caption{\label{fig:diags} Feynman diagrams for semi-inclusive DIS: (a) LO; (b-d) NLO
virtual diagrams; (e-f) NLO real emission diagrams}
\end{figure}

Out of the four structure functions, $ \prescr{1}\widehat{V}_{ba} $ for the
angular function $ A_{1}=1+\cosh ^{2}\psi  $ has a special status, since
only $ \prescr{1}\widehat{V}_{ba} $ receives contributions from the lowest
order of PQCD (Figure \ref{fig:diags}a). At $ \Oas  $, only the contribution
to the $ \prescr{1}\widehat{V}_{ba} $ structure function diverges in the
limit $ q_{T}\rightarrow 0 $.

\section{Leading-order cross section}
Consider first the ${\cal O}(\alpha_S^0) $
process of the quark-photon scattering 
(Fig.\,\ref{fig:diags}a). This process contributes to the total rate of SIDIS
at the leading order (LO). 
There is no LO contribution from gluons. 
Due to the conservation of
the 4-momentum in the parton-level diagram, at this order\begin{equation}
\label{LOmomenta}
p^{\nu }_{b}=p^{\nu }_{a}+q^{\nu }.
\end{equation}
 This condition and Eqs.\,(\ref{qTmup},\ref{qThqT}) imply that\begin{equation}
q_{T}^{2}=-q_{t}\cdot q_{t}=0.
\end{equation}
Also the longitudinal variables are\footnote{%
To obtain Eqs.\,(\ref{xiaxxibz}), consider, for instance, Eq.\,(\ref{LOmomenta})
in the Breit frame for $ \nu =0 $ and $ 3 $. By using explicit expressions
(\ref{qh}-\ref{pBh},\ref{pa},\ref{pb}) for the parton and hadron momenta
at $ q_{T}=0 $, we find \begin{eqnarray*}
\frac{Q}{2}\left( \frac{1}{\widehat{x}}-\widehat{z}\right)  & = & 0,\\
\frac{Q}{2}\left( \frac{1}{\widehat{x}}-2+\widehat{z}\right)  & = & 0.
\end{eqnarray*}
Eqs.\,(\ref{xiaxxibz}) are solutions for this system of equations.
}\begin{eqnarray}
\xi _{a}=x, & \quad  & \xi _{b}=z,\nonumber \\
\widehat{x}= & \widehat{z} & =1,\label{xiaxxibz} 
\end{eqnarray}
 so that the momentum of the final-state hadron $ B $ is\begin{equation}
p_{B}^{\mu }=z\, \left( xp_{A}^{\mu }+q^{\mu }\right) .
\end{equation}

Since both quarks and electrons are spin-1/2 particles, the LO cross section
is proportional to $ 1+\cosh ^{2}\psi \equiv A_{1}(\psi ,\varphi ) $ (Callan-Gross
relation \cite{CallanGrossRelation}). Hence the LO cross section is\begin{equation}
\label{LO}
\left. \frac{d\wh{\sigma}_{ij}}{dxdzdQ^{2}dq_{T}^{2}d\varphi }\right|
_{LO}=\sFs \delta (\vec{q}_{T})\frac{A_{1}(\psi ,\varphi )}{2}
\delta_{ij} e_{j}^{2}\delta(1-\xh)\delta(1-\zh),
\end{equation}
where $ \vec{q}_{T}\equiv (0,q_{T},0,0) $ in the hadron frame. In Eq.\,(\ref{LO})
the parameter $ \sigma _{0} $ collects various constant factors coming from
the hadronic side of the matrix element,\begin{equation}
\sigma _{0}\equiv \frac{Q^{2}}{4\pi S_{eA}x^{2}}\Bigl (\frac{e^{2}}{2}\Bigr ).
\end{equation}
 The factor $ F_{l} $ that comes from the leptonic side is defined by\begin{equation}
F_{l}=\frac{e^{2}}{2}\frac{1}{Q^{2}}.
\end{equation}
 $e_{j} $ denotes the fractional electric charge of a participating quark
or antiquark of the flavor $ j $; $ e_{j}=2/3 $ for up quarks and $ -1/3 $
for down quarks. 

The LO cross section (\ref{LO}) does not explicitly depend on $ Q^{2} $,
but rather on $ x $ and $ z $. This phenomenon is completely analogous
to the Bjorken scaling in completely inclusive DIS \cite{Bjorken69}, \emph{i.e.,}
independence of DIS structure functions from the photon's virtuality $ Q^{2} $.
Just as in the case of inclusive DIS, the scaling of the LO SIDIS cross section
is approximate due to the dependence of PDFs and FFs 
on the factorization scale $ \mu  $. This scale
is naturally chosen of order $ Q $, the only momentum scale in the
LO kinematics. When $ \mu \approx Q $ varies, the PDFs and FFs
change according to Dokshitser-Gribov-Lipatov-Altarelli-Parisi (DGLAP) differential
equations (\ref{DGLAPEqsF},\ref{DGLAPEqsD}). By solving the DGLAP equations,
one sums dominant contributions from the collinear radiation along
the directions of the hadrons $A$ and $B$ through \emph{all}
orders of PQCD. Formally, the scale dependence of the PDFs and FFs is 
an $\Oas$ effect, so that it is on the same footing in the LO calculation
as other neglected higher-order QCD corrections.  
By observing the dependence of the LO cross section on $\mu$, 
we can  qualitatively test the importance of such neglected corrections.  
One finds that this dependence 
is substantial, so that a calculation of $\Oas$ corrections
is needed to reduce theoretical uncertainties. 
Let us now turn to this more elaborate calculation.

\section{The higher-order radiative corrections}
The complete set of $ \Oas  $ corrections to the SIDIS cross section
is shown in \linebreak Figs.\,\ref{fig:diags}b-f. These corrections 
contribute to the total rate at the next-to-leading order (NLO).

At this order, one has to account for the virtual corrections to the LO subprocess
$ \stackrel{(-)}{q}\gamma ^{*}\rightarrow \stackrel{(-)}{q} $ (Figs.\,\ref{fig:diags}b-d),
as well as for the diagrams describing the real emission
subprocesses $ \stackrel{(-)}{q}\gamma ^{*}\rightarrow \stackrel{(-)}{q}g $
and $ g\gamma ^{*}\rightarrow q\bar{q} $, with the subsequent fragmentation
of the final-state quark, antiquark or gluon (Figs.~\ref{fig:diags}e-f). The
explicit expression for the $ \Oas  $ cross section is given in Appendix
\ref{AppendixPertXSec}.

Due to the momentum conservation, the momentum of the unobserved final-state
partons (\emph{e.g.} the gluon in Fig.~\ref{fig:diags}e) can be expressed
in terms of $ q^{\mu },\, p^{\mu }_{a},\, p_{b}^{\mu } $:\begin{equation}
\label{pg}
p_{x}^{\mu }=q^{\mu }+p_{a}^{\mu }-p_{b}^{\mu }.
\end{equation}
 When there is no QCD radiation ($ p_{x}^{\mu }=0 $), the momentum of $ b $
satisfies the leading-order relationship $ p_{b}^{\mu }=p_{a}^{\mu }+q^{\mu } $,
so that $ q_{T}=0 $. 
If $ q_{T}/Q\ll 1 $, the perturbative parton-level
cross section is dominated by the term with $ \rho =1 $. In the limit $ q_{T}^{2}/Q^{2}\rightarrow 0 $,
but $ q_{T}\neq 0 $, $ \prescr{1}{\Vh _{ba}} $ behaves as $ 1/q_{T}^{2} $
times a series in powers of $ \alpha _{S} $ and logarithms $ \ln (q_{T}^{2}/Q^{2} $),
\begin{equation}
\label{NLL}
\prescr{1}{\Vh }_{ba}\approx \frac{\sigma _{0}F_{l}}{2\pi S_{eA}}
\frac{1}{q_{T}^{2}}\sum _{k=1}^{\infty }\Biggl (\frac{\alpha _{S}}{\pi}
\Biggr )^{k}\sum _{m=0}^{2k-1}
\widehat{v}^{(km)}_{ba}(\widehat{x},\widehat{z})\ln ^{m}
\Biggl (\frac{q_{T}^{2}}{Q^{2}}\Biggr ),
\end{equation}
where the coefficients $ \widehat{v}^{(km)}_{ba}(\widehat{x},\widehat{z}) $
are generalized functions of the variables $ \widehat{x} $ and $ \widehat{z} $.
Obviously, the coefficient of the order $ \alpha _{S}^{k} $ in Eq.\,(\ref{NLL})
coincides with the most divergent part of the $ {\cal O}(\alpha _{S}^{k}) $
correction to the SIDIS cross section from the real emission subprocesses. This
coefficient will be called the \emph{asymptotic} part of the real emission correction to $\prescr{1}{\Vh}_{ba}$ 
at $ {\cal O}(\alpha _{S}^{k}) $.  

Convergence of the series in (\ref{NLL}) deteriorates rapidly as $ q_{T}/Q\rightarrow 0 $
because of the growth of the terms $ (q_{T}^{-2})\ln ^{m}(q_{T}^{2}/Q^{2}) $.
Ultimately the structure function $ \prescr{1}{\Vh _{ba}} $ has a non-integrable
singularity at $ q_{T}=0 $. Its asymptotic behavior is very different from
that of the structure functions $ \prescr{2,3,4}{\Vh _{ba}} $, which are
less singular and, in fact, integrable at $ q_{T}=0 $. This singular behavior
of $ \prescr{1}{\Vh _{ba}} $ is generated by infrared singularities of the
perturbative cross section that are located at $ q_{T}=0 $. Indeed, according
to the discussion in Section\,\ref{sec:InfraredSafety}, 
the diagrams with the emission
of massless particles generate singularities when the momentum 
$ p_{1}^{\beta } $ of one of the particles 
is soft ($ p_{1}^{\beta }\rightarrow 0 $) or collinear
to the momentum $ p_{2}^{\beta } $ of another participating particle ($ p_{1\beta }p_{2}^{\beta }=0 $).
The soft singularities in the real emission corrections cancel with
the soft singularities in the virtual corrections. For instance, at $ \Oas  $ 
the soft singularities
of the diagrams shown in Fig.\,\ref{fig:diags}e-f cancel
 with the soft singularities
of the diagrams shown in Fig.\,\ref{fig:diags}b-d. The remaining
collinear singularities are included in the PDFs and FFs, so that
they should be subtracted from $ \prescr{1}{\Vh _{ba}} $. 

There exist two qualitatively different approaches for handling such singularities.
The first approach deals with the singularities order by order in perturbation
theory; the second approach identifies and sums the most singular terms in all
orders of the perturbative expansion. In the next two Subsections, I discuss
regularization of infrared singularities in each of these two approaches.

\subsection{Factorization of collinear singularities at ${\cal O}(\alpha_S)$
\label{sub:NLOFixedOrder}}

Let us begin by considering the first approach, in which singularities are regularized
independently at each order of the series in $ \alpha _{S} $. 
The singularity in the $ {\mathcal{O}}(\alpha _{S}) $ part of the asymptotic
expansion (\ref{NLL}) can be regularized by introducing a ``separation scale''
$ q_{T}^{S} $ and considering the fixed-order cross section separately
in the regions $ 0\leq q_{T}\leq q_{T}^{S} $ and $ q_{T}>q_{T}^{S} $.
The value of $ q_{T}^{S} $ should be small enough for the approximation
(\ref{NLL}) to be valid over the whole range $ q_{T}\leq q_{T}^{S} $.

The quantity $ q_{T}^{S} $ plays the role of a phase space slicing parameter.
In the region $ 0\leq q_{T}\leq q_{T}^{S} $, we can apply the modified
minimal subtraction ($ \MSbar  $) factorization scheme \cite{MSBar} to take
care of the singularities at $ q_{T}=0 $. In the $ \MSbar  $ scheme, the
regularization is done through continuation of the parton-level cross section
to $ n=4-2\eps ,\, \, \eps >0 $ dimensions \cite{HV}. The $ n $-dimensional
expression for the ${\cal O}(\alpha_S)$ part of the 
asymptotic expansion (\ref{NLL}) 
of $ \prescr{1}{\Vh }_{ba}(\widehat{x},\widehat{z},Q^{2},q_{T}^{2}) $
is 
\begin{eqnarray}
 &  & \left. \prescr{1}{\Vh }_{ba}\right| _{\displaystyle 1/q_{T}^{2},\Oas }=
\left(\frac{2\pi\mu_n}{\wh z}\right)^{4-n}
\frac{\sigma _{0}F_{l}}{2\pi S_{eA}}\alpi \frac{1}{2q_{T}^{2}}\sum _{j}\delta _{bj}\delta _{ja}e_{j}^{2}\times \nonumber \\
 && 
\Biggl [
\delta (1-\wh{z})\left\{ P^{(1)}_{qq}(\wh{x})+P^{(1)}_{qg}(\wh{x})\right\} +\left\{ P^{(1)}_{qq}(\wh{z})+P^{(1)}_{gq}(\wh{z})\right\} \delta (1-\wh{x})\nonumber \\
 & + & 2\delta (1-\wh{z})\delta (1-\wh{x})\left[ C_{F}\log \frac{Q^{2}}{q_{T}^{2}}-\frac{3}{2}C_{F}\right] +{\mathcal{O}}\Bigl (\frac{\alpha _{S}}{\pi },q_{T}^{2}\Bigr )\Biggr ].\label{NLL2} 
\end{eqnarray}
Here the color factor $ C_{F}=(N_{c}^{2}-1)/(2N_{c})=4/3, $ $ N_{c}=3 $
is the number of quark colors in QCD. The functions $ P^{(1)}_{ij}(\xi ) $
entering the convolution integrals in (\ref{NLL2}) are the unpolarized $ \Oas  $
splitting kernels \cite{DGLAP}:
\begin{eqnarray}
\label{Pqq}
P^{(1)}_{qq}(\xi ) &=& C_{F}\left[ \frac{1+\xi ^{2}}{1-\xi }\right] _{+},\\
P^{(1)}_{qg}(\xi ) &=& \frac{1}{2}\left( 1-2\xi +2\xi ^{2}\right) ,\\
P^{(1)}_{gq}(\xi ) &=& C_{F}\frac{1+(1-\xi )^{2}}{\xi }.
\end{eqnarray}
The ``+''-prescription in $ P^{(1)}_{qq}(\xi ) $ regularizes $ P^{(1)}_{qq}(\xi ) $
at $ \xi =1 $; it is defined as\[
\int _{0}^{1}d\xi \, \left[ f(\xi )\right] _{+}g(\xi )\equiv \int _{0}^{1}d\xi \, f(\xi )\, \left( g(\xi )-g(1)\right) .\]

The scale parameter $ \mu _{n} $ in (\ref{NLL2}) is introduced to restore
the correct dimensionality of the parton-level cross section $ d\widehat{\sigma }_{ba}/(d\widehat{x}d\widehat{z}dQ^{2}dq_{T}^{n-2}d\varphi ) $
for $ n\neq 4 $. The soft and collinear singularities appear as terms proportional
to $ 1/\eps ^{2} $ and $ 1/\eps  $ when $ n\rightarrow 4 $. The soft
singularity in the real emission corrections cancels with the soft singularity
in the virtual corrections. At $ \Oas , $ the virtual corrections (Fig. \,\ref{fig:diags}b-d)
evaluate to 
\begin{eqnarray}
\left. \frac{d\wh{\sigma }_{ba}}{dxdzdQ^{2}dq_{T}^{2}d\varphi }
\right|_{\displaystyle virt,\Oas } & = & -\frac{\alpha _{S}}{2\pi }C_{F}\left( \frac{4\pi \mu _{n}^{2}}{Q^{2}}\right) ^{\eps }\frac{1}{\Gamma (1-\eps )}\left( \frac{2}{\eps ^{2}}+\frac{3}{\eps }+8\right) \times \nonumber \\
 & \times  & \left. \frac{d\wh{\sigma }_{ba}}{dxdzdQ^{2}dq_{T}^{2}d\varphi }\right| _{LO},
\end{eqnarray}
where the LO cross section is given in Eq.\,(\ref{LO}). 

The remaining collinear singularities are absorbed into the partonic PDFs and
FFs.
When the partonic PDFs and FFs are subtracted from the partonic cross section
$ d\widehat{\sigma } $, the remainder is finite and independent of the types
of the external hadrons. We denote this finite remainder as $ (d\widehat{\sigma })_{hard} $.
The convolution of $ (d\widehat{\sigma })_{hard} $ with the hadronic PDFs
and FFs yields a cross section for the external hadronic states $ A $ and
$ B $. The ``hard'' part depends on an arbitrary factorization
scale $ \mu  $ through terms like $ P^{(1)}_{ab}\ln (\mu /K) $, where
$ P^{(1)}_{ab}(\xi ) $ are splitting functions, and $ K $
is some momentum scale in the process. The scales $ \mu  $ and $ \mu _{n} $
are related as 
\[
\mu ^{2}=4\pi e^{-\gamma _{E}}\mu _{n}^{2}.
\]
The dependence on the factorization scale in the hard part is compensated, up
to higher-order terms in $ \alpha _{S} $, by scale dependence of the long-distance
hadronic functions.

After the cancellation of soft singularities and factorization of collinear
singularities, one can calculate analytically the integral of $ (\prescr{1}{\Vh _{ba}})_{hard} $
over the region \linebreak 
$ 0\nlb \leq \nlb q_{T}\nlb \leq \nlb q_{T}^{S} $. 
At $ {\mathcal{O}}(\alpha _{S}) $
this integral is given by 
\pagebreak
\begin{equation}
\label{NLOcs}
\int _{0}^{(q_{T}^{S})^{2}}dq_{T}^{2}\left( \prescr{1}{\Vh _{ba}}\right) _{hard}=\frac{\sigma _{0}F_{l}}{2\pi S_{eA}}\sum _{j}e_{j}^{2}\Biggl \{\prescr{1}{\Vh ^{LO}_{ba,j}}+\frac{\alpha _{S}}{\pi }\prescr{1}{\Vh ^{NLO}_{ba,j}}\Biggr \}.
\end{equation}
The LO and NLO structure functions
are \begin{eqnarray}
\prescr{1}{\Vh ^{LO}_{ba,j}} &=& 
\delta (1-\widehat{z})\delta (1-\widehat{x})\delta _{bj}\delta _{ja},\\
\prescr{1}{\Vh ^{NLO}_{ba,j}} & = & -\frac{1}{2}\Biggl [\Biggl (C_{F}\ln ^{2}{\frac{Q^{2}}{(q_{T}^{S})^{2}}}-3C_{F}\ln {\frac{Q^{2}}{(q_{T}^{S})^{2}}}\Biggr )\delta (1-\widehat{z})\delta (1-\widehat{x})\delta _{bj}\delta _{ja}\nonumber \\
 & + & \ln {\frac{\mu ^{2}}{(\widehat{z}\, q_{T}^{S})^{2}}}\biggl (\delta (1-\widehat{z})\delta _{bj}P^{(1)}_{ja}(\widehat{x})+P^{(1)}_{bj}(\widehat{z})\delta (1-\widehat{x})\delta _{ja}\biggr )\Biggr ]\nonumber \\
 & + & \delta (1-\widehat{z})\delta _{bj}{c}^{in(1)}_{ja}(\widehat{x})+{c}^{out(1)}_{bj}(\widehat{z})\delta (1-\widehat{x})\delta _{ja}.\label{V1NLO} 
\end{eqnarray}
 The coefficient functions $ {c}^{in,out(1)}_{ba}(\xi ) $ that appear in
$ \prescr{1}{\Vh ^{NLO}_{ba,j}} $ are given by \begin{eqnarray}
{c}^{(1)in}_{ji}(\xi )={c}^{out(1)}_{ij}(\xi ) & = & \delta
_{ij}C_{F}\Biggl [\frac{1}{2}(1-\xi )-2\delta (1-\xi )\Biggr
],\label{Csdis0}
\\ 
{c}^{(1)in}_{jg}(\xi ) &=& \frac{1}{2}\xi (1-\xi ),\\
{c}^{(1)out}_{gj}(\xi ) & = & \frac{C_{F}}{2}\xi .\label{Csdis} 
\end{eqnarray}

Now consider the kinematical region $ q_{T}>q_{T}^{S} $, where the approximation
(\ref{NLL2}) no longer holds. In this region, $ (\prescr{1}{\Vh _{ba}})_{hard} $
should be obtained from the exact NLO result. With this prescription, the integral
over $ q_{T}^{2} $ can be calculated as\begin{eqnarray}
\int _{0}^{\max q_{T}^{2}}dq_{T}^{2}\frac{d\widehat{\sigma }_{ba}}{d\widehat{x}d\widehat{z}dQ^{2}dq_{T}^{2}d\varphi } & = & \nonumber \\
A_{1}(\psi ,\varphi )\Biggl \{\int _{0}^{(q_{T}^{S})^{2}}dq_{T}^{2}\left( \prescr{1}{\Vh _{ba}}\right) _{hard} & + & \int _{(q_{T}^{S})^{2}}^{\max q_{T}^{2}}dq_{T}^{2}\left( \prescr{1}{\Vh _{ba}}\right) _{hard}\Biggr \}\nonumber \\
+\sum _{\rho =2}^{4}A_{\rho }(\psi ,\varphi )\int _{0}^{\max q_{T}^{2}}dq_{T}^{2}\left( \prescr{\rho }{\Vh _{ba}}\right) ,
\label{pert}
\end{eqnarray}
 where $ \max q_{T}^{2} $ is the maximal value of $ q_{T}^{2} $ allowed
by kinematics. The first integral on the right-hand side is calculated analytically,
using the approximation (\ref{NLOcs}); the second and third integrals are calculated
numerically, using the complete perturbative result of the order $ {\mathcal{O}}(\alpha _{S}) $.
The numerical calculation is done with the help of a Monte Carlo integration
package written in the style of the programs Legacy and 
ResBos used earlier for resummation
in vector boson production at hadron-hadron colliders \cite{BY}.

\subsection{All-order resummation of large logarithmic terms\label{sub:NLOresum}}

A significant failure of the computational procedure in (\ref{pert}) is that
it cannot be applied to the description of the $ q_{T} $-dependent differential
cross sections. Indeed, the cancellation of the infrared singularities is achieved by integration of the cross section over the region 
$0\leq q_T \leq q_T^{S}$. However the shape of the $q_T$ distribution
is arbitrary and depends on the choice of the parameter $q_T^{S}$ that
specifies the lowest $q_T$ bin $0\leq q_T \leq q_T^{S}$. 
The fundamental problem is that
the terms in (\ref{NLL}) with small powers of $ \alpha _{S} $ do not reliably
approximate the complete sum in the region $ q_{T}\ll Q $.

This problem justifies the second approach
to the regularization of the singularities at $ q_{T}=0 $, in which large
logarithms in (\ref{NLL}) and virtual corrections at $ q_{T}=0 $ are summed
to all orders. 
A better approximation for $ \prescr{1}{\Vh }_{ba} $ at $ q_{T}/Q\ll 1 $
is provided by the Fourier transform of a $ \vec{b} $-space function $ \wh{\widetilde{W}}_{ba}(b,Q,\widehat{x},\widehat{z},\mu ) $,
which sums the dominant terms in (\ref{NLL}) and virtual corrections through
all orders of $ \alpha _{S} $:\begin{equation}
\label{V1resum}
\left. \prescr{1}{\Vh }_{ba}(\wh{x},\wh{z},Q^{2},q_{T}^{2},\mu )\right|_{\wt{W}}
=\frac{\sigma _{0}F_{l}}{2S_{eA}}\int \frac{d^{2}b}{(2\pi )^{2}}
e^{i\vec{q}_{T}\cdot \vec{b}}\whwtW _{ba}(b,Q,\widehat{x},\widehat{z},\mu ).
\end{equation}
 Here $ \vec{b} $ is a vector conjugate to $ \vec{q}_{T} $, and $ b $
denotes the magnitude of $ \vec{b} $. Hence $ \prescr{1}\Vh _{ba} $ at
\emph{all} values of $ q_{T} $ can be approximated by
\begin{equation}
\label{V1aY}
\prescr{1}\widehat{V}_{ba}(\widehat{x},\widehat{z},Q^{2},q_{T}^{2},\mu )=\left. \prescr{1}{\Vh }_{ba}(\wh{x},\wh{z},Q^{2},q_{T}^{2},\mu )\right| _{\wt{W}}+\prescr{1}\widehat{Y}_{ba}(\widehat{x},\widehat{z},Q^{2},q_{T}^{2}),
\end{equation}
 where $ \prescr{1}\widehat{Y}_{ba} $ is the difference between the 
$\Oas$ expression for $ \prescr{1}\Vh _{ba} $
(cf. Appendix\,\ref{AppendixPertXSec})
and $\Oas$ asymptotic part
(\ref{NLL2}), taken at $n=4$. This difference
is finite in the limit $ q_{T}\rightarrow 0 $. 

The complete \emph{hadron-level} resummed cross section can be obtained by including
the finite parton structure functions for $ \rho =2,3,4 $ and convolving
the parton-level structure functions with PDFs and FFs
(cf. Eqs.\,(\ref{hadcs}-\ref{ang2})):
\begin{equation}
\label{resum}
\left. \frac{d\sigma _{BA}}{dxdzdQ^{2}dq_{T}^{2}d\varphi }\right| _{resum}=\sFs \frac{A_{1}(\psi ,\varphi )}{2}\int \frac{d^{2}b}{(2\pi )^{2}}e^{i\vec{q}_{T}\cdot \vec{b}}\widetilde{W}_{BA}(b,Q,x,z)+Y_{BA}.
\end{equation}
 In this equation, the hadron-level $ b $-dependent form-factor $ \widetilde{W}_{BA}(b,Q,x,z) $
is the sum of convolutions of parton-level form-factors $ \whwtW _{ba}(b,Q,\wh{x},\wh{z}) $
with the PDFs and FFs:
\begin{equation}
\widetilde{W}_{BA}=\sum _{a,b}D_{B/b}\otimes \whwtW _{ba}\otimes F_{a/A}.
\end{equation}
 $ Y_{BA} $ denotes the complete \textit{finite piece},\begin{equation}
Y_{BA}\equiv \prescr{1}Y_{BA}+\sum _{\rho =2}^{4}\prescr{\rho }V_{BA}A_{\rho }(\psi ,\varphi ),
\end{equation}
where
\begin{equation}
\prescr{1}{Y}_{BA}\equiv\sum _{a,b}D_{B/b}\otimes \prescr{1}{\wh{Y}}_{ba}\otimes F_{a/A}.
\end{equation}
 The explicit expression for $ Y_{BA} $ is 
presented in Appendix \ref{AppendixPertXSec}. \\
\indent At small $ b $ and large $ Q $ (\emph{i.e.}, in the region where
perturbative dynamics is expected to dominate) the general structure of $ \widetilde{W}_{BA}(b,Q,x,z) $
can be found from first principles~\cite{CS81,CSS85}: \begin{equation}
\label{W}
\widetilde{W}_{BA}(b,Q,x,z)=\sum _{j}e_{j}^{2}(D_{B/b}\otimes {\cal C}^{out}_{bj})(z,b)({\cal C}_{ja}^{in}\otimes F_{a/A})(x,b)e^{-S_{BA}(b,Q)}.
\end{equation}
According to the discussion in Section\,\ref{sec:TwoScale}, the form-factor
$\widetilde{W}_{BA}$ is the all-order sum of the large logarithms, which remain
after the cancellation of soft singularities and factorization of
collinear singularities. The soft contributions 
are included in the Sudakov function $ S_{BA}(b,Q) $. 
At small $ b $, $ S_{BA}(b,Q) $ does
not depend on the types of the external hadrons and looks like\begin{equation}
\label{SudP}
S_{BA}(b,Q)=\int _{C_{1}^{2}/b^{2}}^{C_{2}^{2}Q^{2}}\frac{d\ov \mu ^{2}}{\ov \mu ^{2}}\left( \ASud (\alpha _{S}(\ov \mu ),C_{1})\ln \frac{C_{2}^{2}Q^{2}}{\ov \mu ^{2}}+\BSud (\alpha _{S}(\ov \mu ),C_{1},C_{2})\right) \equiv S^{P}(b,Q),
\end{equation}
with
\begin{eqnarray}
\ASud (\alpha _{S}(\ov \mu ),C_{1}) &=& 
\sum _{k=1}^{\infty }\ASud _{k}(C_{1})\Biggl (\frac{\alpha _{S}(\ov \mu )}{\pi }\Biggr )^{k},\\
\BSud (\alpha _{S}(\ov \mu ),C_{1},C_{2}) &=&
\sum _{k=1}^{\infty }\BSud _{k}(C_{1},C_{2})\Biggl (\frac{\alpha _{S}(\ov \mu )}{\pi }\Biggr )^{k}.
\end{eqnarray}

%
%

Contributions from the collinear partons are included
in the functions $ {\cal \mathcal{C}}^{in}(\widehat{x},b,\mu ) $ and 
$ {\cal \mathcal{C}}^{out}(\widehat{z},b,\mu ) $.
These functions can also be expanded in series of 
$ \alpha _{S}/\pi  $, as
\begin{equation}
{\cal \mathcal{C}}^{in}_{ij}(\widehat{x},b,\mu ) = \sum _{k=0}^{\infty }
{\cal \mathcal{C}}^{in(k)}_{ij}(\widehat{x},C_{1},C_{2,}\mu b)
\Biggl (\frac{\alpha _{S}(\mu )}{\pi }\Biggr )^{k},
\end{equation}
\begin{equation}
{\cal \mathcal{C}}^{out}_{ij}(\widehat{z},b,\mu ) = 
\sum _{k=0}^{\infty }{\cal \mathcal{C}}^{out(k)}_{ij}(\widehat{z},C_{1},C_{2},\mu b)\Biggl (\frac{\alpha _{S}(\mu )}{\pi }\Biggr )^{k}.
\end{equation}

According to Eq.\,(\ref{SudP}), the integration in $ S^{P}(b,Q) $
is performed between two scales $ C_1/b $ and $ C_2 Q $,
where $C_1$ and $C_2$ are constants of order 1.  
These constants also appear in terms proportional to $
\delta (1-\wh{x}) $ or $ \delta (1-\wh{z}) $
in the $ {\cal C} $-functions.
The complete factor $ \wt{W}(b,Q) $ is approximately independent
from $ C_{1}$ and $ C_{2}$.
In addition, the $ {\cal \mathcal{C}} $-functions
depend on the factorization scale $ \mu $ that separates singular collinear 
contributions included in the PDFs and FFs from 
the finite collinear contributions included in the ${\cal C}$-functions. 
To suppress certain logarithms in
$ \Oas  $ parts of the $ {\cal C}^{in} $ and $ {\cal C}^{out} $ functions,
it is convenient to choose\begin{eqnarray}
C_{1} & = & 2e^{-\gamma _{E}}\equiv b_{0},\\
C_{2} & = & 1,\\
\mu  & = & b_{0}/b,
\end{eqnarray}
 where $ \gamma _{E}=0.577215... $ is the Euler constant.

Using our NLO results, I find the following expressions for 
the coefficients $ \ASud _{k}(C_{1}) $, $ \BSud _{k}(C_{1},C_{2}) $
and the $ {\cal C} $-functions: 
\begin{eqnarray}
\ASud _{1} & = & C_{F},\label{A1} \\
\BSud _{1} & = & 2C_{F}\log \biggl (\frac{e^{-3/4}C_{1}}{b_{0}C_{2}}\biggr ).\label{B1} 
\end{eqnarray}
 To the same order, the expressions for the $ \mathcal{C} $-functions are 

\begin{itemize}
\item LO:\begin{eqnarray}
{\cal \mathcal{C}}^{in(0)}_{jk}(\widehat{x},\mu b) & = & \delta _{jk}\delta (1-\widehat{x}),\label{LO1} \\
{\cal \mathcal{C}}^{out(0)}_{jk}(\widehat{z},\mu b) & = & \delta _{jk}\delta (1-\widehat{z}),\\
{\cal \mathcal{C}}^{in(0)}_{jg}= & {\cal \mathcal{C}}^{out(0)}_{gj} & =0;
\end{eqnarray}

\item NLO:
\bea
{\cal \mathcal{C}}^{in(1)}_{jk}(\widehat{x},\mu b) & = & \frac{C_{F}}{2}(1-\widehat{x})-P^{(1)}_{qq}(\widehat{x})\log \Bigl (\frac{\mu b}{b_{0}}\Bigr )\nonumber \\
 & - & C_{F}\delta (1-\widehat{x})\biggl (\frac{23}{16}+\log ^{2}\Bigl (\frac{e^{-3/4}C_{1}}{b_{0}C_{2}}\Bigr )\biggr ),\label{C1in} \\
{\cal \mathcal{C}}^{in(1)}_{jg}(\widehat{x},\mu b) & = & \frac{1}{2}\widehat{x}(1-\widehat{x})-P^{(1)}_{qg}(\widehat{x})\log \Bigl (\frac{\mu b}{b_{0}}\Bigr ),\label{C1in2} \\
{\cal \mathcal{C}}^{out(1)}_{jk}(\widehat{z},\mu b) & = & \frac{C_{F}}{2}(1-\widehat{z})-P^{(1)}_{qq}(\widehat{z})\log \Bigl (\frac{\mu b}{\wh{z}b_{0}}\Bigr )\nonumber \\
 & - & C_{F}\delta (1-\widehat{z})\biggl (\frac{23}{16}+\log ^{2}\Bigl
(\frac{e^{-3/4}C_{1}}{b_{0}C_{2}}\Bigr )\biggr ),\label{C1out} 
\\
{\cal \mathcal{C}}^{out(1)}_{gj}(\widehat{z},\mu b) &=&  \frac{C_{F}}{2}\widehat{z}-P^{(1)}_{gq}(\widehat{z})\log \Bigl (\frac{\mu b}{\wh{z}b_{0}}\Bigr ).\label{C1out2} 
\eea

\end{itemize}
In these formulas, the indices $ j $ and $ k $ correspond to quarks and
antiquarks, and $ g $ to gluons. In Appendix \ref{ch:resumOas} I show that
the expansion of the integral over $b$ in Eq.\,(\ref{resum}) 
up to the order $ \Oas  $, with
perturbative coefficients given in Eqs.\,(\ref{A1}-\ref{C1out2}),
reproduces the small-$q_T$ limit of the 
fixed-order $ \Oas  $ cross section discussed in 
Subsection\,\ref{sub:NLOFixedOrder}.

Due to the crossing relations between 
parton-level SIDIS, vector boson production,
and $ e^{+}e^{-} $ hadroproduction,
the $ {\cal \mathcal{C}}^{in} $-functions are essentially the same in SIDIS
and the Drell-Yan process; and the $ {\cal \mathcal{C}}^{out} $-functions
are essentially the same in SIDIS and $ e^{+}e^{-} $ hadroproduction. At
NLO the only difference stems from the fact that the momentum transfer $ q^{2} $
is spacelike in DIS and timelike in the other two processes. Hence the virtual diagrams
Figs.~\ref{fig:diags}b-d differ by $ \pi ^{2} $ for spacelike and timelike
$ q^{2} $. Correspondingly, $ {\cal \mathcal{C}}^{in(1)}_{jk} $ 
for SIDIS does not contain the term $ (\pi ^{2}/3)\delta (1-\widehat{x})$,
which is present in the $ {\cal \mathcal{C}}^{in(1)}_{jk} $-function
for the Drell-Yan process. Similarly,
$ {\cal \mathcal{C}}^{out(1)}_{jk} $
for SIDIS does not contain the term 
$ (\pi ^{2}/3)\delta (1-\widehat{z}) $, which is present in 
the $ {\cal \mathcal{C}}^{out(1)}_{jk} $-function
for $ e^{+}e^{-} $ hadroproduction.

Up to now, I was discussing the behavior of the resummed cross-section
at short distances. The representation (\ref{W}) should be modified at large
values of the variable $ b $ to account for nonperturbative long-distance
dynamics. The authors of Ref.\,\cite{CSS85} suggested the following ansatz for $ \wt{W}_{BA} $
which is valid at all values of $ b $: \begin{equation}
\wt{W}_{BA}(b,Q,x,z)=\sum _{j}e_{j}^{2}(D_{B/b}\otimes {\cal \mathcal{C}}^{out}_{bj})(z,b_{*})({\cal \mathcal{C}}_{ja}^{in}\otimes F_{a/A})(x,b_{*})e^{-S_{BA}}.
\end{equation}
 Here the variable\begin{equation}
\label{bstar}
b_{*}\equiv \frac{b}{\sqrt{1+\Bigl (\frac{b}{b_{max}}\Bigr )^{2}}}
\end{equation}
 serves to reproduce the perturbative solution (\ref{W}) at $ b\ll b_{max} $,
with $ b_{max}\approx 0.5\mbox {\, GeV}^{-1} $, and turn off the perturbative
dynamics for $ b\geq b_{max} $. Furthermore, the Sudakov factor is modified,
being written as the sum of the perturbatively calculable part $ S^{P}(b_{*},Q) $
given by Eq.\,(\ref{SudP}), and a nonperturbative part, 
which is only partially constrained
by the theory:\begin{equation}
\label{SPNP}
S_{BA}(b,Q,x,z)=S^{P}(b_{*},Q,x,z)+S^{NP}_{BA}(b,Q,x,z).
\end{equation}

 An explicit solution for the function $ S_{BA}^{NP}(b,Q,x,z) $ has not been
found yet. 
Nonetheless, the renormalization properties of the theory require that the $ Q $
dependence in the nonperturbative Sudakov term be separated from the dependence
on the other kinematical variables, \textit{i.e}.,\begin{equation}
S^{NP}_{BA}(b,Q,x,z)=g_{BA}^{(1)}(b,x,z)+g_{BA}^{(2)}(b,x,z)\log \frac{Q}{Q_{0}},
\end{equation}
 with $ Q_{0}\approx 1\, \, \mbox {\, GeV} $. The theory does not predict
the functional forms of $ g_{BA}^{(1)}(b,x,z) $ and $ g_{BA}^{(2)}(b,x,z) $,
so these must be determined by fitting experimental data. In addition, if 
$ S^{NP}_{BA}(b,Q,x,z) $
indeed describes long-distance dynamics, it should vanish or be much smaller
than $ S^{P}(b,Q,x,z) $ in the perturbative region $ b<b_{max}. $ In the
analysis of the experimental results, we may find that the fit to the data prefers
a parametrization of $ S^{NP}(b) $ that is not small in comparison to the
perturbative part of $ \widetilde{W}_{BA} $ at $ b<b_{max} $. Such observation
will be an 
evidence in favor of important dynamics that is not included in the $ b $-space
resummation formula with coefficients calculated at the given order of PQCD.
Therefore, this work uses an interpretation of $ S^{NP} $ that is
broader than its original definition in \cite{CSS85}. $ S^{NP} $ will parametrize
not only large-$ b $ physics, but additional contributions to $ \widetilde{W}_{BA} $
at \textit{all} values of $ b $ that are not included in the perturbative
part of $ \widetilde{W}_{BA} $. In the following parts of the thesis I will
test whether the data are consistent with the assumption that these additional
contributions are small in comparison to the perturbative part of $ \wt{W}_{BA} $
when $ b<b_{max} $.

Before ending this section, I would like to comment on a subtle difference between
$ {\cal \mathcal{C}}^{in} $ and $ {\cal \mathcal{C}}^{out} $. While the
initial-state coefficient functions $ {\cal \mathcal{C}}^{in(1)}_{ba}(\widehat{x},C_{1},C_{2},b,\mu ) $
given in Eqs.\,(\ref{C1in}) and (\ref{C1in2}) depend on the factorization
scale $ \mu  $ through a factor $ \ln {[\mu b/b_{0}]} $, the final-state
functions $ {\cal \mathcal{C}}^{out(1)}_{ba}(\widehat{z},C_{1},C_{2},b,\mu ) $
given in Eqs.\,(\ref{C1out}) and (\ref{C1out2}) depend instead on $ \ln {[\mu b/(b_{0}\widehat{z})]} $.
The additional term $ \propto \ln {\widehat{z}} $ in the functions $ {\cal \mathcal{C}}^{out(1)}_{ba}(\widehat{z},C_{1},C_{2},b,\mu ) $
becomes large and negative when $ \widehat{z}\rightarrow 0 $, so that it
can significantly influence the $ \Oas  $ contribution at small values 
of $ \widehat{z}$. As a result, the resummed total rate tends to be lower than
its fixed-order counterpart for $ z\lesssim 0.1 $. This issue is discussed
in more detail in Section\,\ref{sec:Multiplicity}. Similarly, the $ {\cal \mathcal{O}}(\alpha _{S}) $
part of the NLO structure function $ \prescr{1}{\Vh ^{NLO}_{ba,j}} $ in (\ref{V1NLO})
depends on $ \mu  $ through a logarithm $ \ln {[\mu ^{2}/(\widehat{z}q_{T}^{S})^{2}]} $.

The appearance of the additional terms $ \propto \ln {\widehat{z}} $ in the
functions $ {\cal \mathcal{C}}^{out(1)}_{bj} $ and $ \prescr{1}{\Vh ^{NLO}_{ba,j}} $
reflects the specifics of separation of the $ {\cal \mathcal{O}}(\alpha _{S}) $
``hard'' cross section $ (d\widehat{\sigma })_{hard} $ from the collinear
contributions to the FFs in the $ \MSbar  $ factorization scheme. The easiest
way to see the specific origin of the $ \ln \widehat{z} $ terms is to notice
that the dependence on the parameter $ \mu _{n} $ in the $ n $-dimensional
expression (\ref{NLL2}) for $ \prescr{1}\Vh _{ba}(\widehat{x},\widehat{z},Q^{2},q_{T}^{2}) $
comes through a factor $ (2\pi \mu _{n}/\widehat{z})^{4-n} $, rather than
through a more conventional $ (2\pi \mu _{n})^{4-n} $. In its turn, $ \wh{z} $
appears in $ (2\pi \mu _{n}/\widehat{z})^{4-n} $, because the $ \MSbar  $-scheme
prescribes to continue to $ n-2 $ dimensions the transverse momentum $ \wh{\vec{p}}_{T} $
of the outgoing parton, rather than the vector $ \vec{q}_{T}=\vec{\wh{p}_{T}}/\wh{z} $
relevant to the resummation calculation. It is this factor that generates the
$ \mu  $-dependent logarithmic terms $ \ln {[\mu b/(b_{0}\widehat{z})]} $
in the functions $ {\cal \mathcal{C}}^{out(1)}_{bj}(\widehat{z},C_{1},C_{2},b,\mu ) $
and $ \prescr{1}{\Vh ^{NLO}_{ba,j}} $. The $ {\cal \mathcal{C}}^{in(1)}_{ja} $-functions
do not include $ \ln {\widehat{z}} $ because they are evaluated along the
direction $ \widehat{z}=1 $ in the phase space. In contrast, nothing forbids
such a term in the functions $ {\cal \mathcal{C}}^{out(1)}_{bj} $, in which
$ \widehat{z} $ can be anything between $ z $ and 1. Moreover, the $ \ln \wh{z} $
terms are needed to reproduce $ \MSbar  $ expressions for $ \Oas  $ coefficient
functions in completely inclusive DIS \cite{Furmanski82} by integration of
$ d\wh{\sigma }_{ba}/(dQ^{2}d\wh{x}d\wh{z}dq_{T}^{2}) $ over $ q_{T} $
and $ \wh{z} $. 

\newpage
\section{Hadronic multiplicities and energy flows \label{sec:Formalism:ZFlows}}

Knowing the hadronic cross-section, it is possible to calculate the multiplicity
of the process, which is defined as the ratio of this cross-section and the
total inclusive DIS cross-section for the given leptonic cuts:\begin{equation}
\mbox {Multiplicity}\equiv \frac{1}{d\sigma _{tot}/dxdQ^{2}}\frac{d\sigma }{dxdzdQ^{2}dq_{T}^{2}d\varphi}.
\end{equation}
 Both the cross-section and the multiplicity depend on the properties of the
final-state fragmentation. The analysis can be simplified by considering energy
flows which do not have such dependence. A traditional variable used in the
experimental literature is a transverse energy flow $ \langle E_{T}\rangle  $
in one of the coordinate frames, defined as
\begin{equation}
\label{ET}
\langle E_{T}\rangle _{\Phi _{B}}=
\frac{1}{\sigma _{tot}}\sum _{B}\int _{\Phi _{B}}d\Phi _{B}\, \, 
E_{T}\frac{d\sigma (e+A\rightarrow e+B+X)}{d\Phi _{B}}.
\end{equation}
 This definition involves an integration over the available phase space $ \Phi _{B} $
and a summation over all possible species of the final hadrons $ B $. Since
the integration over $ \Phi _{B} $ includes integration over the longitudinal
component of the momentum of $ B $, the dependence of $ \langle E_{T}\rangle  $
on the fragmentation functions drops out due to the normalization condition\begin{equation}
\sum _{B}\int \,  z\, D_{B/b}(z)dz=1.
\end{equation}

Instead of $ \langle E_{T}\rangle  $, I will analyze the flow of
the variable $ z $. This flow is defined as~\cite{zflowdef}
\begin{equation}
\label{zflow}
\frac{d\Sigma _{z}}{dx\, \, dQ^{2}\, \, dq_{T}^{2}d\varphi}=
\sum _{B}\int _{z_{min}}^{1}\, \, z\, \, 
\frac{d\sigma (e+A\rightarrow e+B+X)}{dx\, \, dz\, \, dQ^{2}\, \,
dq_{T}^{2}d\varphi}\,\, dz.
\end{equation}
 I prefer to use $ \Sigma _{z} $ rather than $ \langle E_{T}\rangle  $
because $ \langle E_{T}\rangle  $ is not Lorentz invariant, which complicates
its usage in the theoretical analysis\footnote{%
The $ z $-flow $ \Sigma _{z} $ is related to the energy distribution function
$ \Sigma  $ calculated in \cite{Meng2} as $ \Sigma _{z}=(2xE_{A}/Q^{2})\Sigma  $.
Here $ E_{A} $ is the energy of the initial hadron in the HERA lab frame.
}. Since $ q_{T} $ is related to the pseudorapidity in the hCM   frame
via Eq.\,(\ref{qTetacm}), and the transverse energy of a nearly massless
particle in this frame is given by\begin{equation}
E_{T}\approx p_{T}=zq_{T},
\end{equation}
 the experimental information on $ d\Sigma _{z}/(dx\, \, dQ^{2}\, \, dq_{T}^{2}) $
can be derived from the hCM  frame pseudorapidity ($ \eta _{cm} $) distributions
of $ \langle E_{T}\rangle  $ in bins of $ x $ and $ Q^{2} $. If mass
effects are neglected, we have\begin{equation}
\label{Sz2ET}
\frac{d\langle E_{T}\rangle }{dxdQ^{2}d\eta _{cm}d\varphi}=
2q_{T}^{3}\frac{d\Sigma _{z}}{dxdQ^{2}dq^{2}_{T}d\varphi}.
\end{equation}

By the factorization theorems of QCD, the hadron-level $ z $-flow $ \Sigma _{z} $
can be written as the convolution of a parton-level $ z $-flow $ \widehat{\Sigma }_{z} $
with the PDFs, \begin{equation}
\frac{d\Sigma _{z}}{dxdQ^{2}dq_{T}^{2}d\varphi }=\sum _{a}\int _{x}^{1}\frac{d\xi _{a}}{\xi _{a}}F_{a/A}(\xi _{a},\mu _{F})\frac{d\widehat{\Sigma }_{z}(\mu )}{d\widehat{x}dQ^{2}dq_{T}^{2}d\varphi }.
\end{equation}
 Similarly to the SIDIS cross section, the $ z $-flow can be expanded in
a sum over the leptonic angular functions $ A_{\rho }(\psi ,\varphi ) $:
\begin{equation}
\label{angz}
\frac{d\widehat{\Sigma }_{z}(\mu )}{d\widehat{x}dQ^{2}dq_{T}^{2}d\varphi }=\sum _{\rho =1}^{4}\prescr{\rho }\Vh _{za}(\widehat{x},Q^{2},q_{T}^{2},\mu )A_{\rho }(\psi ,\varphi ),
\end{equation}
 where the structure functions $ \prescr{\rho }\Vh _{za}(\widehat{x},Q^{2},q_{T}^{2},\mu ) $
for the $ z $-flow are related to the structure functions $ \prescr{\rho }\Vh _{ba}(\widehat{x},\widehat{z},Q^{2},q_{T}^{2},\mu ) $
for the SIDIS cross section by \pagebreak[3]
\begin{equation}
\label{Vcsz}
\prescr{\rho }\Vh _{za}(\widehat{x},Q^{2},q_{T}^{2},\mu )=\sum _{b}\int _{0}^{1}\widehat{z}d\widehat{z}\, \prescr{\rho }\Vh _{ba}(\widehat{x},\widehat{z},Q^{2},q_{T}^{2},\mu ).
\end{equation}
The resummed $ z $-flow is calculated as \begin{equation}
\label{resumz}
\frac{d\Sigma _{z}}{dxdQ^{2}dq_{T}^{2}d\varphi }=\sFs \frac{A_{1}(\psi ,\varphi )}{2}\int \frac{d^{2}b}{(2\pi )^{2}}e^{i\vec{q}_{T}\cdot \vec{b}}\widetilde{W}_{z}(b,Q,x)+Y_{z},
\end{equation}
 where \begin{equation}
\label{Wz}
\widetilde{W}_{z}(b,Q,x)=\sum _{j}e_{j}^{2}{\mathcal{C}}^{out}_{z}\, e^{-S_{z}(b,Q,x)}\, ({\mathcal{C}}_{ja}^{in}\otimes F_{a/A})(x,b_{*},\mu ).
\end{equation}
As in the case of the resummation of hadronic cross sections, only the structure
function $ ^{1}V_{zA} $ for the angular function $ A_{1}=1+\cosh ^{2}\psi  $
has to be resummed. 

The functions $ {\mathcal{C}}_{ja}^{in} $ in (\ref{Wz}) are the same as
in (\ref{W}). The coefficient $ {\mathcal{C}}^{out}_{z} $ is \begin{equation}
\label{Coutz}
{\mathcal{C}}^{out}_{z}=1+\frac{\alpha _{S}}{\pi }C_{F}\Bigl (-\frac{7}{16}-\frac{\pi ^{2}}{3}-\ln ^{2}\frac{e^{-3/4}C_{1}}{C_{2}b_{0}}\Bigr ).
\end{equation}
The parameter $ b_{*} $, given by (\ref{bstar}) with
$ b_{max}=0.5\mbox {\, GeV}^{-1} $, is introduced in (\ref{Wz}) to smoothly
turn off the perturbative dynamics when $ b $ exceeds $ b_{max} $. The
term $ Y_{z} $ in (\ref{resumz}) is the difference between the complete
fixed-order expression at $ {\mathcal{O}}(\alpha _{S}) $ for $ d\Sigma _{z}/(dxdQ^{2}dq_{T}^{2}d\varphi ) $
and its most singular part in the limit $ q_{T}\rightarrow 0 $; that is,
\begin{equation}
Y_{z}=\frac{d\Sigma _{z}}{dxdQ^{2}dq_{T}^{2}d\varphi }-\Biggl (\frac{d\Sigma _{z}}{dxdQ^{2}dq_{T}^{2}d\varphi }\Biggr )_{asym}.
\end{equation}
 The asymptotic part calculated to $ {\mathcal{O}}(\alpha _{S}) $ is 
\pagebreak[3]
\begin{eqnarray}
 &  & \Biggl (\frac{d\Sigma _{z}}{dxdQ^{2}dq_{T}^{2}d\varphi }\Biggr )_{asym}=\sFs \alpi \frac{1}{2q_{T}^{2}}\frac{A_{1}(\psi ,\varphi )}{2\pi }\nonumber \\
 & \times  & \sum _{j}e_{j}^{2}\Biggl [\Bigl \{(P^{(1)}_{qq}\otimes F_{j/A})(x,\mu )+(P^{(1)}_{qg}\otimes F_{g/A})(x,\mu )\Bigr \}\nonumber \\
 & + & 2F_{j/A}(x,\mu )\Bigl \{C_{F}\ln \frac{Q^{2}}{q_{T}^{2}}-\frac{3}{2}C_{F}\Bigr \}+{\mathcal{O}}\Bigl (\biggl (\frac{\alpha _{S}}{\pi }\biggr )^{2},q_{T}^{2}\Bigr )\Biggr ].\label{asymz} 
\end{eqnarray}

Similar to (\ref{SPNP}), the $ z $-flow Sudakov factor $ S_{z} $ is a
sum of perturbative and nonperturbative parts,\begin{equation}
\label{SzPNP}
S_{z}(b,Q,x)=S^{P}(b_{*},Q,x)+S^{NP}_{z}(b,Q,x).
\end{equation}
 The NLO perturbative Sudakov factor $ S^{P} $ is given by the universal
$ x $-independent expression (\ref{SudP}). As in the case of SIDIS multiplicities,
the renormalization group invariance requires that the dependence of $ S_{z}^{NP} $on
$ \ln Q $ be separated from the dependence on other variables: \begin{equation}
\label{SzNP}
S^{NP}_{z}(b,Q,x)=g^{(1)}(b,x)+g^{(2)}(b,x)\log \frac{Q}{Q_{0}}.
\end{equation}
 In principle, the $ z $-flow Sudakov factor $ S_{z}(b,Q,x) $ is related
to the Sudakov factors $ S_{BA}(b,Q,x,z) $ of the contributing hadroproduction
processes $ e+A\ra e+B+X $ through the relationship\begin{equation}
\label{SS}
e^{-S_{z}(b,x)}=\frac{1}{{\cal C}_{z}^{out}(b_{*},\mu )}\sum _{B}\int zdze^{-S_{BA}(b,Q,x,z)}(D_{B/b}\otimes {\cal C}_{b,j}^{out})(z,b_{*},\mu ).
\end{equation}
 In practice, the efficient usage of this relationship to constrain the Sudakov
factors is only possible if the fragmentation functions and the hadronic contents
of the final state are accurately known. I do not use the relationship (\ref{SS})
in my calculations.

\section{Relationship between the perturbative and resummed cross-sections. Uncertainties
of the calculation \label{sec:TheoryUncertainties}}

In the numerical calculations, some care is needed to treat the uncertainties
in the definitions of the asymptotic and resummed cross sections, although formally these uncertainties are of order $ {\mathcal{O}}((\alpha _{S}/\pi )^{2},q_{T}^{-1}) $.

\subsection{\label{sub:Matching}Matching}

The generic structure of the resummed cross-section (\ref{resum}), calculated
up to the order $ {\mathcal{O}}((\alpha _{S}/\pi )^{N}) $, is\begin{equation}
\label{r}
\sigma _{resum}^{(N)}=\sigma _{\wt{W}}+Y^{(N)}.
\end{equation}
 In (\ref{r}), the $ \wt{W} $-piece receives all-order contributions from
large logarithmic terms
\begin{equation}
\frac{1}{q_{T}^{2}}\sum ^{\infty }_{k=1}\left( \frac{\alpha _{S}}{\pi }\right) ^{k}\sum _{m=0}^{2k-1}v^{(km)}\log ^{m}{(q_{T}^{2}/Q^{2})}.
\end{equation}
The $ Y $-piece is the difference of the fixed-order perturbative
and asymptotic cross-sections:
\begin{equation}
\label{Y2}
Y^{(N)}=\sigma _{pert}^{(N)}-\sigma _{asym}^{(N)}.
\end{equation}

In the small-$ q_{T} $ region, we expect cancellation up to terms of order
$ {\mathcal{O}}(\alpha _{S}^{N+1}/\pi ^{N+1}) $ between the perturbative
and asymptotic pieces in (\ref{Y2}), so that the $ \wt{W} $-piece dominates
the resummed cross-section (\ref{r}). On the other hand, the expression for
the asymptotic piece coincides with the expansion of the $ \wt{W} $-piece
up to the order $ {\mathcal{O}}(\alpha _{S}^{N}/\pi ^{N}) $, so that
at large $ q_{T} $ the resummed cross-section (\ref{r})
is formally equal to the perturbative cross-section up to corrections of order
$ {\mathcal{O}}(\alpha _{S}^{N+1}/\pi ^{N+1}) $.

In principle, due to the cancellation between the perturbative and asymptotic
pieces at small $ q_{T} $, and between the resummed and asymptotic piece
at large $ q_{T} $, the resummed formula $ \sigma _{resum} $ is at least
as good an approximation of the physical cross-section as the perturbative cross-section
$ \sigma _{pert} $ of the same order. However, in the NLO calculation at
$ q_{T}\gg Q $ it is safer to use the fixed order cross-section instead of
the resummed expression. At the NLO order of $ \alpha _{S} $, the difference
between the $ \wt{W} $-piece and the asymptotic piece at large $ q_{T} $
may still be non-negligible in comparison to the perturbative piece. In particular,
due to the fast rise of the PDFs at small $ x $, the resummed and asymptotic
pieces receive large contributions from the small-$ x $ region, while the
perturbative piece does not (see the next Subsection for details). Therefore,
the resummed cross-section $ \sigma _{resum} $ may differ significantly from
the NLO cross-section $ \sigma _{pert} $. This difference does not mean that
the resummed cross-section agrees with the data better than the fixed-order
one. At $ q_{T}\geq Q $, the NLO cross-section is no longer dominated by
the logarithms that are resummed in Eq.\,(\ref{r}). In other words, the resummed
cross-section (\ref{r}) does not include some terms in the NLO cross-section
that become important at $ q_{T}\approx Q $. For this reason, at $ q_{T}>Q $
the resummed cross-section may show unphysical behavior; for example, it can
be significantly higher the NLO cross section or even oscillate if the $ \wt{W} $-term
changes rapidly near the boundary between the perturbative and nonperturbative
regions. 

As the order of the perturbative calculation increases, the agreement between
the resummed and the fixed-order perturbative cross-sections is expected to
improve. Indeed, such improvement was shown in the case of vector boson production
\cite{BY}, where one observes a smoother transition from the resummed to the
fixed-order perturbative cross-section if the calculation is done at the next-to-next-to-leading
order. Also, at the NNLO the switching occurs at larger values of the transverse
momentum of the vector boson than in the case of the NLO. 

Since the fixed-order result is more reliable at $ q_{T}\gtrsim Q $, the
switching from the resummed to the fixed-order perturbative cross-section should
occur at $ q_{T}\approx Q $. However, there is no unique point at which this
switching happens. Similarly, it is not possible to say beforehand which of
the two cross sections agrees better with the data in the region $ q_{T}\approx Q $.
In SIDIS at small $ x $, the NLO $ z $-flow underestimates
the data at $ q_{T}\gtrsim Q $, while the resummed $ z $-flow is in better
agreement. Therefore, it makes more sense to use the resummed $ z $-flow
in this region, without switching to the fixed-order piece. On the other hand,
in the charged particle production one \emph{has} to switch to the NLO cross
section at $ q_{T}\approx Q $ in order to reproduce the measured $ p_{T} $-distributions.

\subsection{\label{sub:KinCorr}Kinematical corrections at \protect$ q_{T}\approx Q\protect $}

In this Subsection I discuss the differences between the kinematics implemented
in the definitions of the asymptotic and resummed cross-sections, 
and the kinematics of the perturbative piece at non-zero
values of $ q_{T} $.

\begin{figure}[H]
\resizebox*{1\textwidth}{!}{\includegraphics{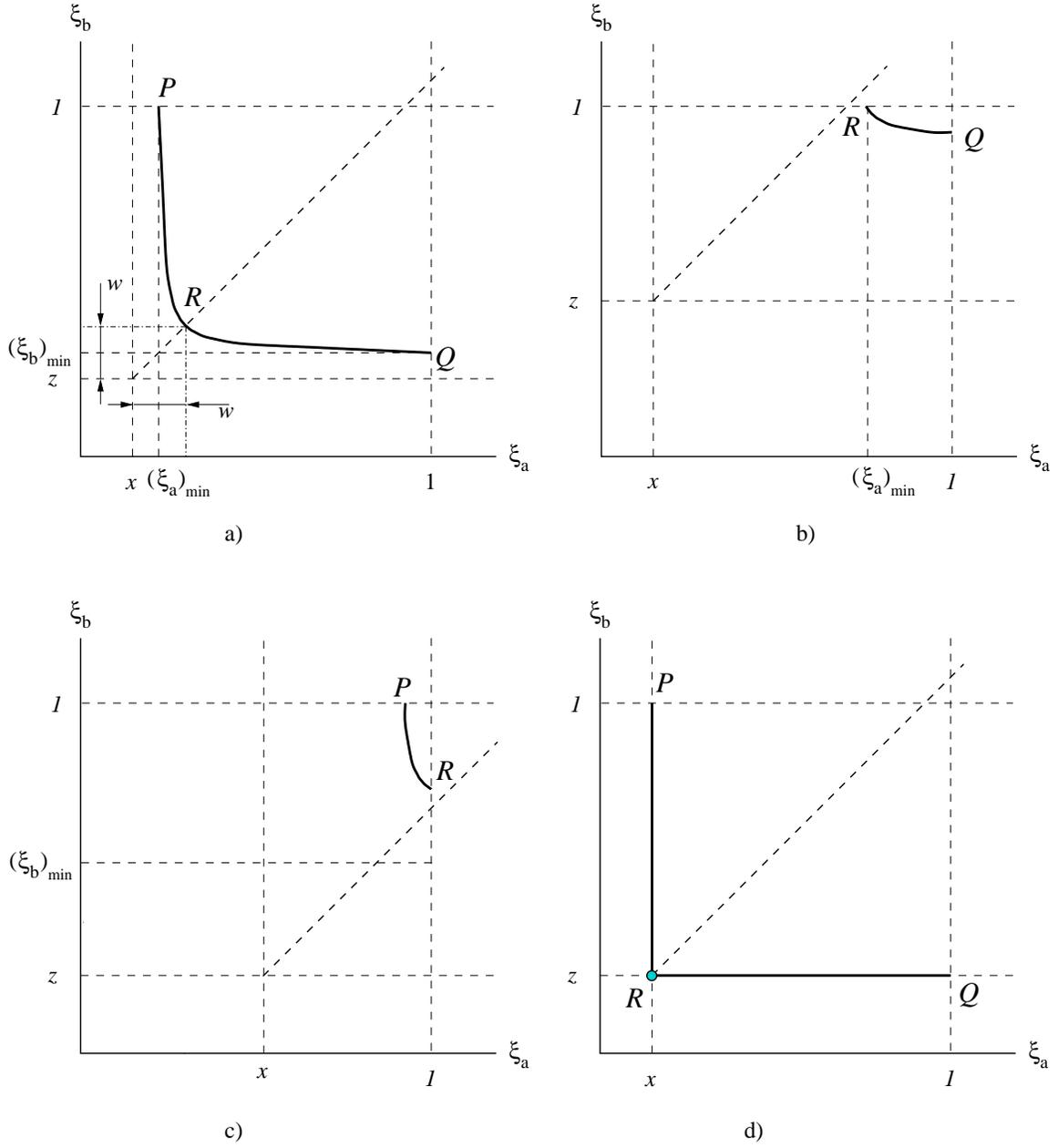}}

\caption{\label{fig:xiaxib} The contours of the integration over \protect$ \xi _{a}\protect $,
\protect$ \xi _{b}\protect $ for (a,b,c) the perturbative cross-section;
(d) the asymptotic and resummed cross-sections}
\end{figure}

Let us first discuss the NLO approximation to the hadronic cross
section (\ref{hadcs}).
The integrand
of the NLO cross section contains the delta-function 
\begin{equation}
\label{cond}
\delta \Biggl [\frac{q_{T}^{2}}{Q^{2}}-\biggl (\frac{1}{\widehat{x}}-1\biggr )
\biggl (\frac{1}{\widehat{z}}-1\biggr )\Biggr ]=
xz\delta \Biggl [(\xi _{a}-x)(\xi _{b}-z)-xz\frac{q_{T}^{2}}{Q^{2}}\Biggr ],
\end{equation}
 which comes from the parton-level cross-section (\ref{sighat}). Depending
on the values of $ x,z,Q^{2},q_{T}^{2} $, the contour of the integration
over $ \xi _{a} $ and $ \xi _{b} $ determined by (\ref{cond}) can have
one of three shapes shown in Fig.~\ref{fig:xiaxib}a,b,c. For $ q_{T}\ll Q $
the integration proceeds along the contour in Fig.~\ref{fig:xiaxib}a, and
the integral in (\ref{hadcs}) can be written in either of two alternative
forms
\begin{eqnarray}
\frac{d\sigma _{BA}}{dxdzdQ^{2}dq_{T}^{2}d\varphi } & = & 
\int _{(\xi _{a})_{min}}^{1}\frac{d\xi _{a}}{\xi _{a}-x}{\cal M}_{BA}
(\xi _{a},\xi _{b};\widehat{x},\widehat{z},Q^{2},q_{T}^{2},\varphi )\nonumber \\
 &  & \nonumber \\
 & = & \int _{(\xi _{b})_{min}}^{1}\frac{d\xi _{b}}{\xi _{b}-z}
{\cal M}_{BA}(\xi _{a},\xi _{b};\widehat{x},\widehat{z},Q^{2},q_{T}^{2},\varphi ),
\end{eqnarray}
 where
\bea
&&{\cal M}_{BA}(\xi _{a},\xi _{b};\widehat{x},\widehat{z},Q^{2},q_{T}^{2},\varphi )=\nonumber\\
&&\frac{\sigma _{0}F_{l}}{4\pi S_{eA}}\frac{\alpha _{S}}{\pi }\widehat{x}\widehat{z}\sum _{a,b}D_{B/b}(\xi _{b})F_{a/A}(\xi _{a})\sum _{\rho =1}^{4}\prescr{\rho }f_{ba}(\widehat{x},\widehat{z},Q^{2},q_{T}^{2})A_{\rho }(\psi ,\varphi ).
\eea
 The lower bounds of the integrals are
\bea
(\xi _{a})_{min} &=& \frac{w^{2}}{1-z}+x,\\
(\xi _{b})_{min} &=& \frac{w^{2}}{1-x}+z,
\eea
 with\begin{equation}
w\equiv \frac{q_{T}}{Q}\sqrt{xz}.
\end{equation}

Alternatively, the cross-section can be written in a form symmetric with respect
to $ x $ and $ z $,
\begin{equation}
\frac{d\sigma _{BA}}{dxdzdQ^{2}dq_{T}^{2}d\varphi } =  
\int _{x+w}^{1}\frac{d\xi _{a}}{\xi _{a}-x}{\cal M}_{BA}
+ \int _{z+w}^{1}\frac{d\xi _{b}}{\xi _{b}-z}{\cal M}_{BA},\label{sym} 
\end{equation}
where the integrals are calculated along the branches $ RP $ and $ RQ $
in Fig.~\ref{fig:xiaxib}a, respectively. As $ q_{T}\rightarrow 0 $,\[
(\xi _{a})_{min}\rightarrow x,\, \, (\xi _{b})_{min}\rightarrow z,\]
 and the contour $ PRQ $ approaches the contour of integration of the asymptotic
cross-section shown in Fig.~\ref{fig:xiaxib}d. The horizontal
(or vertical) branch contributes to the convolutions with splitting functions
in Eq.\,(\ref{NLL2}) arising from the initial (or final) state collinear singularities,
while the soft singularities of Eq.\,(\ref{NLL2}) are located at the point $ \xi _{a}=x,\, \, \xi _{b}=z $.

On the other hand, as $ q_{T} $ increases up to values around $ Q $, the
difference between the contours of integration of the perturbative and asymptotic
cross-sections may become significant. First, as can be seen from (\ref{sym}),
in the perturbative piece $ \xi _{a} $ and $ \xi _{b} $ are always higher
than $ x+w $ or $ z+w $ , while in the asymptotic piece they vary between
$ x $ or $ z $ and~unity. At small $ x $ (or small $ z $) the difference
between the phase spaces of the perturbative and asymptotic pieces may become
important due to the steep rise of the PDFs (or FFs) in this region. Indeed,
for illustration consider a semi-inclusive DIS experiment at small $ x $.
Let $ q_{T}/Q=0.5 $, $ z=0.5 $, and $ x=10^{-4} $; then $ x+w=3.2\cdot 10^{-3}\gg x=10^{-4} $.
In combination with the fast rise of the PDFs at small $ x $, this will enhance
the difference between the perturbative and resummed cross-sections.

Second, for $ x $ or $ z $ near unity, it could happen that $ x+w\geq 1 $
or $ z+w\geq 1 $, which would lead to the disappearance of one or two branches
of the integration of the perturbative piece (Fig.~\ref{fig:xiaxib}b,c). In
this situation the phase space for nearly collinear radiation along the direction
of the initial or final parton is suppressed. Again, this may degrade the consistency
between the perturbative and asymptotic piece, since the latter includes contributions
from both branches of the collinear radiation. Fortunately, the $ x-z $ asymmetry
of the phase space in SIDIS is not important in the analysis of the existing
data from HERA, since it covers the small-$ x $ region and is less sensitive
to the contributions from the large $ z $ region, where the rate of the hadroproduction
is small. 

The numerical analysis below includes a correction that imitates the
phase space contraction in the low-$ x $ region.
This correction is incorporated by replacing $ x $ in Eqs.\,(\ref{NLL2}, \ref{resum})
by the rescaled variable\begin{equation}
\label{tildex}
\tilde{x}=\frac{Q^{2}+q_{T}^{2}}{Q^{2}}x.
\end{equation}
 This substitution simulates the phase space contraction of the perturbative
piece. At small $ q_{T} $, the rescaling reproduces the exact asymptotic
and resummed pieces, but at larger $ q_{T} $
it excludes the unphysical integration region of $ \xi _{a}\approx x $. \newpage

\newpage

\chapter{Resummation in semi-inclusive DIS: numerical results\label{ch:Phenomenology}}

Despite the abundance of experimental publications on SIDIS, none of them presents
dependence of SIDIS observables on the variable $ q_{T} $. Hence $ q_{T} $-distributions,
which are sensitive to the multiple parton radiation, 
have to be derived from the
published data on less direct distributions. The $ q_{T} $-distributions
for some of the HERA data were reconstructed for the first time in \cite{nsy1999,nsy2000}.
In this Chapter, I concentrate on the analysis of the $ q_{T} $-distributions
for the $ z- $flow (\ref{zflow}) \[
\frac{1}{\sigma _{tot}}\frac{d\Sigma _{z}}{dxdQ^{2}dq_{T}^{2}},\]
which can be derived from published pseudorapidity distributions for the transverse
energy flow in the hCM frame \cite{H1z1,H1z2}. I will also discuss several
observables, including the average value of $ q_{T}^{2} $, that were measured
in the production of light charged hadrons \cite{E665,ZEUSchgd96}. 

\begin{figure}[H]
{\par\centering \resizebox*{1\textwidth}{!}{\includegraphics{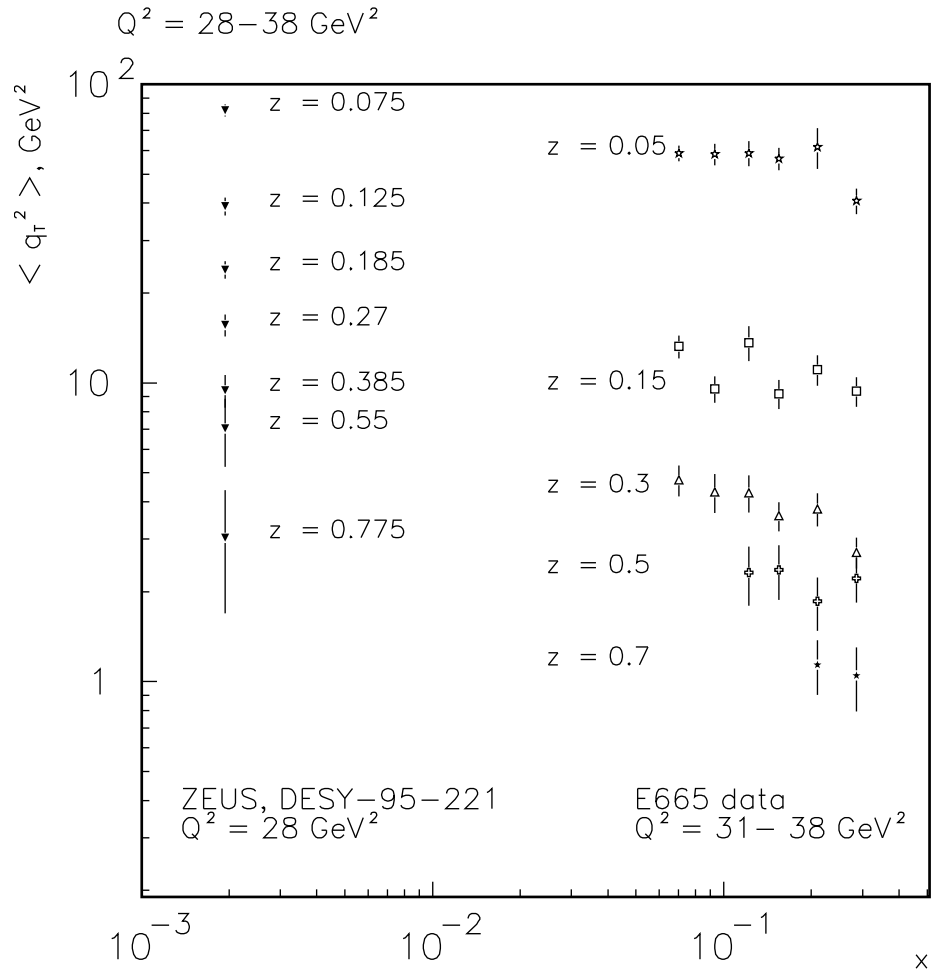}} \par}

\caption{\label{fig:qt2avchgd}The average \protect$ q_{T}^{2}\protect $ as a function
of \protect$ x\protect $ and \protect$ z\protect $ in the charged particle
production at \protect$ Q^{2}=28-38\mbox {\, GeV}^{2}\protect $. The data
points are extracted from published distributions \protect$ \langle p_{T}^{2}\rangle \protect $
vs. \protect$ x_{F}\protect $ \cite{E665,ZEUSchgd96} using the method described
in Section\,\ref{sec:Multiplicity}.}
\end{figure}

\begin{figure}[H]
{\par\centering \resizebox*{1\textwidth}{!}{\includegraphics{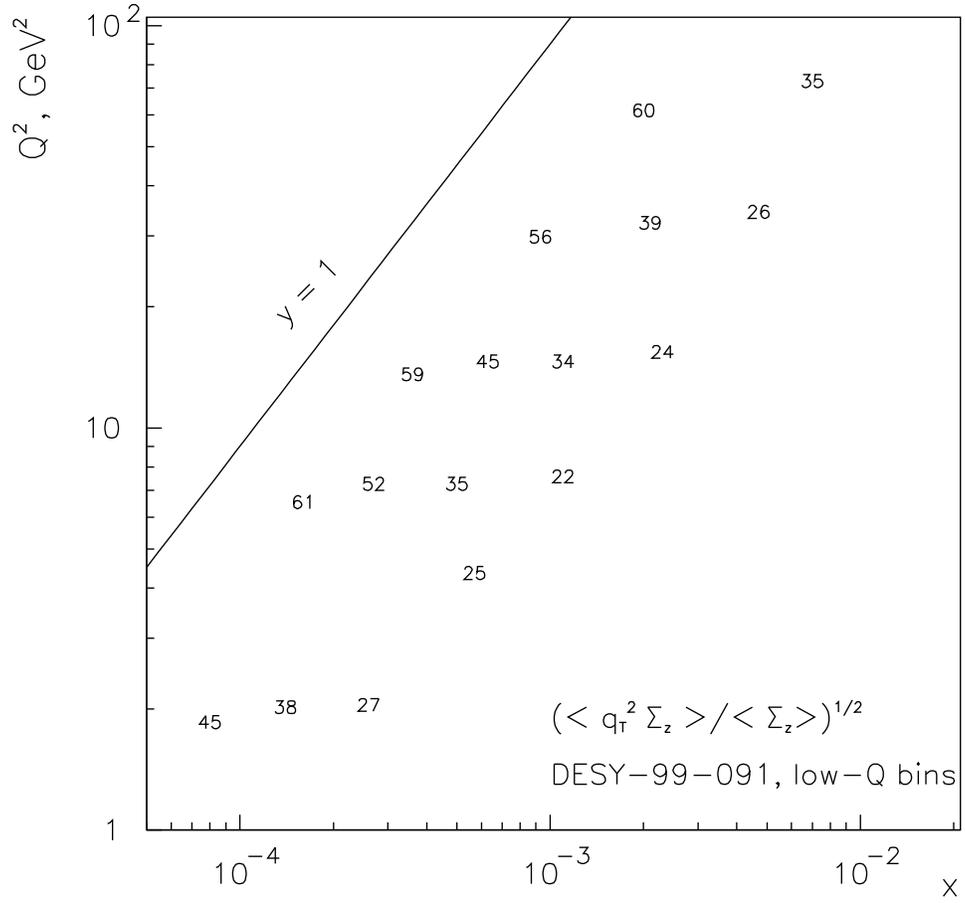}} \par}

\caption{\label{fig:qtavzflow}\protect$ \sqrt{\left\langle q_{T}^{2}\Sigma _{z}\right\rangle /\left\langle \Sigma _{z}\right\rangle }\protect $
reconstructed from distributions \protect$ d\left\langle E_{T}\right\rangle /d\eta_{cm}\protect $
in bins of \protect$ x\protect $ and \protect$ Q^{2}\protect $ \cite{H1z2}. }
\end{figure}

Reconstruction of the $ q_{T} $-dependence reveals an interesting trend in
the data: namely, the average $ q_{T} $ (or average $ q_{T}^{2} $) increases
rapidly when either $ x $ or $ z $ decreases. This trend is illustrated
in Figs.\,\ref{fig:qt2avchgd} and \ref{fig:qtavzflow}. Figure \ref{fig:qt2avchgd}
shows the average $ q_{T}^{2} $ in the charged particle production for several
bins of $ x $ and $ z $ at $ 28\leq Q^{2}\leq 38\mbox {\, GeV}^{2} $.
The procedure of reconstruction of $ \left\langle q_{T}^{2}\right\rangle  $
is described in detail in Section\,\ref{sec:Multiplicity}. As can be seen from
Figure \ref{fig:qt2avchgd}, $ \left\langle q_{T}^{2}\right\rangle  $ in
the ZEUS data sample ($ \left\langle x\right\rangle =1.94\cdot 10^{-3} $)
is several times higher than $ \left\langle q_{T}^{2}\right\rangle  $ at
the same value of $ \left\langle z\right\rangle  $ and larger values 
of $ \left\langle x\right\rangle  $
in the E665 data sample ($ \left\langle x\right\rangle =0.07-0.29 $). 
$ \left\langle q_{T}^{2}\right\rangle  $
increases even faster when $ \left\langle z\right\rangle  $ decreases and
$ \left\langle x\right\rangle  $ is fixed. For instance, at 
$ \left\langle x\right\rangle =1.94\cdot 10^{-3} $
$ \left\langle q_{T}^{2}\right\rangle  $ increases from $ 3\mbox {\, GeV}^{2} $
at $ \left\langle z\right\rangle =0.775 $ to $ 82\mbox {\, GeV}^{2} $
at $ \left\langle z\right\rangle =0.075 $. 

A similar trend is apparently 
present in the behavior of 
the quantity $ \sqrt{\left\langle q_{T}^{2}\Sigma _{z}\right\rangle /
\left\langle \Sigma _{z}\right\rangle }, $
which was derived from the data for the distributions 
$ d\left\langle E_{T}\right\rangle /d\eta_{cm} $
published in \cite{H1z2}. This quantity is shown in Figure \ref{fig:qtavzflow}
as a function of $ Q^{2} $ and $ x $.\footnote{%
The distributions 
$ \sqrt{\left\langle q_{T}^{2}\Sigma _{z}\right\rangle /
\left\langle \Sigma _{z}\right\rangle } $
were derived by converting distributions $ d\left\langle E_{T}\right\rangle /d\eta_{cm} $
in $ d\Sigma _{z}/dq_{T}^{2} $ with the help of Eq.~(\ref{Sz2ET}) and then
averaging $ \left\langle q_{T}^{2}\Sigma _{z}\right\rangle  $ and $ \left\langle \Sigma _{z}\right\rangle  $
over the experimental bins of $ q_{T}. $ In each bin of $ q_{T} $, central
values of $ \Sigma _{z} $ and $ q_{T}^{2} $ were used. This procedure
provides a reasonable estimate 
for $ \sqrt{\left\langle q_{T}^{2}\Sigma _{z}\right\rangle /\left\langle \Sigma _{z}\right\rangle } $
if the experimental $ q_{T} $-bins cover all available
range of $ q_{T}$. Figure \ref{fig:qtavzflow} shows $ \sqrt{\left\langle q_{T}^{2}\Sigma _{z}\right\rangle /\left\langle \Sigma _{z}\right\rangle } $
for the ``low-$ Q $'' data set of from \cite{H1z2}, which satisfies
this criterion.
} At each value of $ Q^{2} $, $ \sqrt{\left\langle q_{T}^{2}\Sigma _{z}\right\rangle /\left\langle \Sigma _{z}\right\rangle } $
becomes larger when $ x $ decreases. Also, $ \sqrt{\left\langle q_{T}^{2}\Sigma _{z}\right\rangle /\left\langle \Sigma _{z}\right\rangle } $
is roughly constant along the lines of constant $ y=Q^{2}/xS_{eA} $ (\emph{i.e.},
the lines parallel to the kinematical boundary $ y=1 $). Larger values of $ \sqrt{\left\langle q_{T}^{2}\Sigma _{z}\right\rangle /\left\langle \Sigma _{z}\right\rangle } $
at smaller $ x $ are the evidence of ``broader'' 
distributions $ d\Sigma _{z}/dq_{T} $. In the subsequent Sections, I discuss
this phenomenon in the context of the $ q_{T} $-resummation
formalism.

In this Chapter I assume that the angle $ \varphi  $ is not monitored in
the experiment, so that it will be integrated out in the following discussion.
Correspondingly, the numerical results for $ d\sigma _{BA}/(dxdzdQ^{2}dq_{T}^{2}) $
will not depend on terms in Eqs.\,(\ref{ang1}\nlb[4]-\nlb[4]\ref{ang2}) proportional to the angular
functions $ A_{3} $ and $ A_{4} $, which integrate to zero. The dependence
on the azimuthal angle $ \varphi  $ is discussed in more detail in Chapter
\ref{ch:AzimuthalAsymmetries}.

\section{Energy flows\label{sec:ZFlows}}

\subsection{General remarks}

As was discussed at length in Subsection\,\ref{sub:NLOresum}, knowledge of the
resummed SIDIS cross section can be used to predict the pseudorapidity spectrum
of the transverse energy flow in the hadron Breit frame or the hCM frame. It
is advantageous to study the energy flows, because they are less dependent on
the specifics of the final-state fragmentation of the scattered partons into the
observed hadrons. I therefore start the presentation of the numerical results
with the comparison of the resummation formalism to the experimentally measured
pseudorapidity distributions for the transverse energy flow in the hCM frame. 

I consider the data on $ d\left\langle E_{T}\right\rangle /d\eta_{cm} $
which has been published in \cite{H1z1,H1z2}. I consider seven bins of $ x $
and $ Q $ from \cite{H1z1} ($ 10\leq \left\langle Q^{2}\right\rangle \leq 50\, \, \mbox {\, GeV}^{2} $,
$3.7\cdot 10^{-4}\leq \left\langle x\right\rangle \leq 4.9\cdot 10^{-3} $) 
and two sets of bins of $ x $ and $ Q $ from \cite{H1z2} (``low-$ Q^{2} $''
set covering $ 13.1<\langle Q^{2}\rangle <70.2\mbox {\, GeV}^{2} $, $ 8\times 10^{-5}<\langle x\rangle <7\times 10^{-3} $
and ``high-$ Q^{2} $'' set covering $ 175<\langle Q^{2}\rangle <2200\mbox {\, GeV}^{2} $
and $ 0.0043<\langle x\rangle <0.11 $). 

The experimental
distributions $ d\left\langle E_{T}\right\rangle /d\eta_{cm} $ at a fixed
value of $ W^{2}=Q^{2}(1-x)/x $ can be converted into the distributions $ d\Sigma _{z}/dq_{T} $
using Eqs.\,(\ref{qTetacm},\ref{Sz2ET}):
\begin{equation}
q_{T}=We^{-\eta _{cm}}
\end{equation}
and
\begin{equation}
\label{Sz2ET2}
\frac{d\Sigma _{z}}{dxdQ^{2}dq_{T}}=\frac{1}{q_{T}^{2}}\frac{d\langle E_{T}\rangle }{dxdQ^{2}d\eta _{cm}}.
\end{equation}
The ``derived'' data for $ d\Sigma _{z}/(dxdQ^{2}dq^{2}_{T}) $ can
be compared with the resummed $ z $-flow (\ref{resumz}), which is calculated
as 
\begin{equation}
\frac{d\Sigma _{z}}{dxdQ^{2}dq^{2}_{T}}=\frac{\pi }{S_{eA}}\sigma _{0}F_{l}\left( 1+\cosh ^{2}\psi \right) \int \frac{d^{2}b}{(2\pi )^{2}}e^{i\vec{q}_{T}\cdot \vec{b}}\widetilde{W}_{z}(b,Q,x)+Y_{z},
\end{equation}
 where 
\begin{equation}
\label{Wz2}
\widetilde{W}_{z}(b,Q,x)=\sum _{j}e_{j}^{2}{\mathcal{C}}^{out}_{z}\, e^{-S_{z}(b,Q,x)}\, ({\mathcal{C}}_{ja}^{in}\otimes F_{a/A})(x,b_{*},\mu ).
\end{equation}
The Sudakov factor $ S_{z} $ in Eq.\,(\ref{Wz2}) is \[
S_{z}(b,Q,x)=S^{P}(b_{*},Q,x)+S^{NP}_{z}(b,Q,x),\]
where the perturbative part $ S^{P} $ is given by Eq.\,(\ref{SudP}),
and a realistic parametrization of the nonperturbative part $ S_{z}^{NP}(b,Q,x) $
can be obtained by comparison with experimental data at low and intermediate
values of $ Q $, especially with the measured pseudorapidity distributions
at $ Q\approx 3-20 $ GeV. At high $ Q $, we expect the data to be dominated
by the perturbatively calculable parton radiation and be less sensitive to 
the nonperturbative
effects incorporated in $ S_{z}^{NP}(b,Q,x) $. According to the renormalization group
invariance argument, $ S_{z}^{NP} $ includes a part that is proportional
to $ \ln Q $:\begin{equation}
S^{NP}_{z}(b,Q,x)=g^{(1)}(b,x)+g^{(2)}(b,x)\log \frac{Q}{Q_{0}},
\end{equation}
where the parameter $ Q_{0}\approx 1\mbox {\, GeV}^{-1} $ prevents $ \ln {Q/Q_{0}} $
from being negative in the region of validity of PQCD. In the following analysis,
I use two parametrizations of $ S_{z}^{NP}(b,Q,x), $ which I will call parametrizations
1 and 2. 

\begin{itemize}
\item \textbf{Parametrization 1} \, was \, proposed \, in our paper \cite{nsy1999} 
with D. Stump and C.\nolinebreak[4]-\nolinebreak[4]P.~Yuan based on the analysis
of the data in Ref.~\cite{H1z1}: \begin{equation}
\label{SNPzpar1}
S^{NP}_{z}(b,Q,x)=b^{2}\left\{ g^{(1)}(x)+\frac{1}{2}\Bigl (\left. g^{(2)}(b,Q)\right| _{DY}+\left. g^{(2)}\right| _{e^{+}e^{-}}(b,Q)\Bigr )\right\} ,
\end{equation}
where $ \left. g^{(2)}(b,Q)\right| _{DY} $ and $ \left. g^{(2)}(b,Q)\right| _{e^{+}e^{-}} $
are $ Q $-dependent terms in the nonperturbative Sudakov factors in 
Drell-Yan process and $ e^{+}e^{-} $
hadroproduction.
The parametrization of the function $ g^{(2)}(b,Q) $ in Eq.\,(\ref{SNPzpar1})
is suggested by the crossing symmetry between SIDIS, the Drell-Yan process and
$ e^{+}e^{-} $ hadroproduction. Due to this symmetry, the functions $ g^{(2)}(b,Q) $
in these processes may be related as~\cite{Meng2}\begin{equation}
\label{g2}
\left. g^{(2)}(b,Q)\right| _{SIDIS}=\frac{1}{2}\Bigl (\left. g^{(2)}(b,Q)\right| _{DY}+\left. g^{(2)}(b,Q)\right| _{e^{+}e^{-}}\Bigr ).
\end{equation}
 If the relationship (\ref{g2}) is true, then the function $ g^{(2)}(b,Q) $
in SIDIS is completely known once parametrizations for the functions $ g^{(2)}(b,Q) $
in the Drell-Yan and $ g^{(2)}(b,Q) $ in $ e^{+}e^{-} $ hadroproduction
processes are available. In practice, the only known parametrization of the
nonperturbative Sudakov factor in the $ e^{+}e^{-} $ 
hadroproduction was obtained
in Ref.\,\cite{CS85} by fitting the resummation formula to the data
at $ Q=27 $ GeV. Most of the $ \left\langle E_{T}\right\rangle  $ data
from HERA correspond to significantly smaller values of $ Q $, where the
usage of the parametrization \cite{CS85} is questionable. In addition,
the known parametrizations of the nonperturbative Sudakov factors for the Drell-Yan
\cite{DWS,LY,BY,BLLY,Landry2000} and $ e^{+}e^{-} $ hadroproduction \cite{CS85} processes
correspond to slightly different scale choices:\begin{equation}
\label{C121}
C_{1}=b_{0},\quad C_{2}=1
\end{equation}
 and\begin{equation}
\label{C122}
C_{1}=b_{0},\quad C_{2}=e^{-3/4},
\end{equation}
 respectively. Therefore, the known functions $ \left. g^{(2)}\right| _{DY}(b) $
and $ \left. g^{(2)}\right| _{e^{+}e^{-}}(b) $ are not 100\% compatible
and in principle should not be combined to obtain $ g^{(2)}(b) $ for SIDIS.
In the numerical calculation, I have used the functions $ \left. g^{(2)}(b)\right| _{DY} $
from \cite{DWS} and $ \left. g^{(2)}(b)\right| _{e^{+}e^{-}} $ from \cite{CS81},
despite the fact that $ \left. g^{(2)}(b)\right| _{DY} $ was fitted to Drell-Yan
data using a different $ C_{2} $ value than in $ \left. g^{(2)}(b)\right| _{e^{+}e^{-}} $.
Explicitly, the $ Q $-dependent part $ g^{(2)}(b,Q) $ in Eq.\,(\ref{SNPzpar1})
is\begin{equation}
\label{g2_2}
g^{(2)}(b,Q)=\frac{1}{2}b^{2}\Biggl (0.48\log (\frac{Q}{2Q_{0}})+5.32C_{F}\log \Bigl (\frac{b}{b_{*}}\Bigr )\log \Bigl (\frac{C_{2}Q}{C_{1}Q_{0}}\Bigr )\Biggr ).
\end{equation}
 In Eq.\,(\ref{g2_2}), the constants are 
$ C_{1}=2e^{-\gamma_E },\, \, C_{2}=e^{-3/4},Q_{0}=1\, \, \mbox {\, GeV} $.
The variable $ b_{*} $ is given by Eq.\,(\ref{bstar}), with 
$ b_{max}=0.5\, \, \mbox {\, GeV}^{-1} $.

The functional form of $ g^{(1)}(b,x) $ in terms of $ b $ and $ x $
was parametrized as\begin{equation}
g^{(1)}(b,x)=(-4.58+\frac{0.58}{\sqrt{x}})b^{2},
\end{equation}
 where the numerical coefficients were determined by fitting the experimental
data. These data cover a limited region of $ x $ and $ Q^{2} $
($ 10\leq Q^{2}\leq 50\, \, \mbox {\, GeV}^{2} $, $ 3.7\cdot 10^{-4}\leq x\leq 4.9\cdot 10^{-3} $
), so that the parametrization~1 should not be used away from this region. Also, the
dependence of $ S_{z}^{NP}(b,Q,x) $ on $ Q $ cannot be determined reliably
using exclusively the data from Ref.\,\cite{H1z1}, since all pseudorapidity
distributions in this publication are presented in a small range of $ Q\approx 2-6 $
GeV. This circumstance motivated us to model the $ Q $-dependent terms in
the parametrization~1 by using the crossing relationship (\ref{g2}) instead
of trying to find these terms from the comparison with the data.
\item \textbf{Parametrization 2} overcomes several shortcomings of the parametrization\,1.
The parametrization\,2 was proposed in \cite{nsy2000}, where the analysis
of Ref.\,\cite{nsy1999} was repeated using the latest and more comprehensive
data on the transverse energy flow \cite{H1z2}. From our analysis, we found
that the data from Refs.\,\cite{H1z1,H1z2} are consistent with the following
representative parametrization of the nonperturbative Sudakov factor: \begin{equation}
\label{SNPzpar2}
S^{NP}_{z}(b,Q,x)=b^{2}\, \left( 0.013\frac{(1-x)^{3}}{x}+0.19\ln {\frac{Q}{Q_{0}}}+C\right) ,
\end{equation}
 where the parameter $ Q_{0} $ is fixed to be 2 GeV to prevent $ \ln {Q/Q_{0}} $
from being negative in the region of validity of PQCD, and where we set $ C=0 $
for reasons explained later. 

The H1 Collaboration presented pseudorapidity distributions
of the transverse energy flow for $ Q^{2} $ up to $ 2200\mbox {\, GeV}^{2} $.
However, the data at such high $ Q^{2} $ is rather
insensitive to the nonperturbative dynamics because of the poor
resolution of the H1 detector in the region of large $Q^2$ and $\eta_{cm}$. 
Thus the H1 data at very high $ Q^{2} $ is not informative
about the $ Q^{2} $-dependence of $ S_{z}^{NP}(b,Q,x) $ either. 
Fortunately, the H1 Collaboration presented distributions in two bins
at intermediate values of $ \langle Q^{2}\rangle  $, namely $ \langle Q^{2}\rangle =59.4\mbox {\, GeV}^{2} $
and $ \langle Q^{2}\rangle =70.2\mbox {\, GeV}^{2} $. Together with the data
from Refs.\,\cite{H1z1,H1z2} at lower values of $ Q $, these distributions
provide the first direct tests of the $ Q^{2} $-dependence of $ S_{z}^{NP}(b,Q,x) $.
Therefore the parametrization~2 of $ S_{z}^{NP}(b,Q,x) $ includes a numerical
value for the coefficient of $ \ln Q/Q_{0} $, which yields reasonable agreement
with all of the analyzed data. The resulting value for this coefficient differs
noticeably from its model expression in the parametrization~1. However,
we should not draw too strong a conclusion from this difference, because it
might be caused by ambiguities in the separation of $ Q^{2} $ dependence
and $ x $ dependence in the existing data. To draw a strong conclusion about
the crossing symmetry model, experimental pseudorapidity distributions in a
larger range of $ x $ at intermediate values of $ Q^{2} $, as well as
improvements in the knowledge of the nonperturbative Sudakov factor in the $ e^{+}e^{-} $-hadroproduction
will be needed.
\end{itemize}

\subsection{Comparison with the data}

\begin{figure}[H]
{\par\centering \resizebox*{0.8\textwidth}{!}{\includegraphics{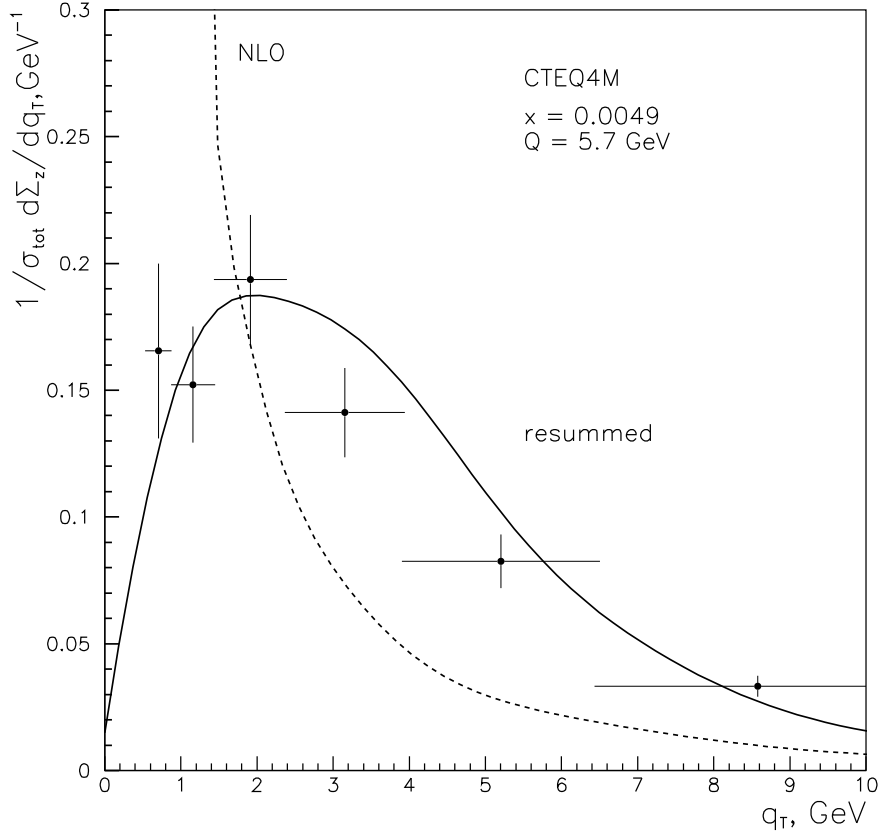}} \par}

\caption{\label{fig:nloandresum} Comparison of the NLO perturbative and
resummed expressions for the \protect$ z\protect $-flow distribution
with the existing experimental data from HERA \protect\cite{H1z1}. The 
data is for \protect
$ \langle x\rangle =0.0049,\, \, \langle Q^{2}\rangle =32.6\mbox{\, GeV}^{2}\protect $.
The resummed curve is calculated using the parametrization~1 of \protect$ S_{z}^{NP}.\protect $
CTEQ4M PDFs \cite{CTEQ4} were used.}
\end{figure}

\begin{figure}[H]
\resizebox*{1\textwidth}{!}{\includegraphics{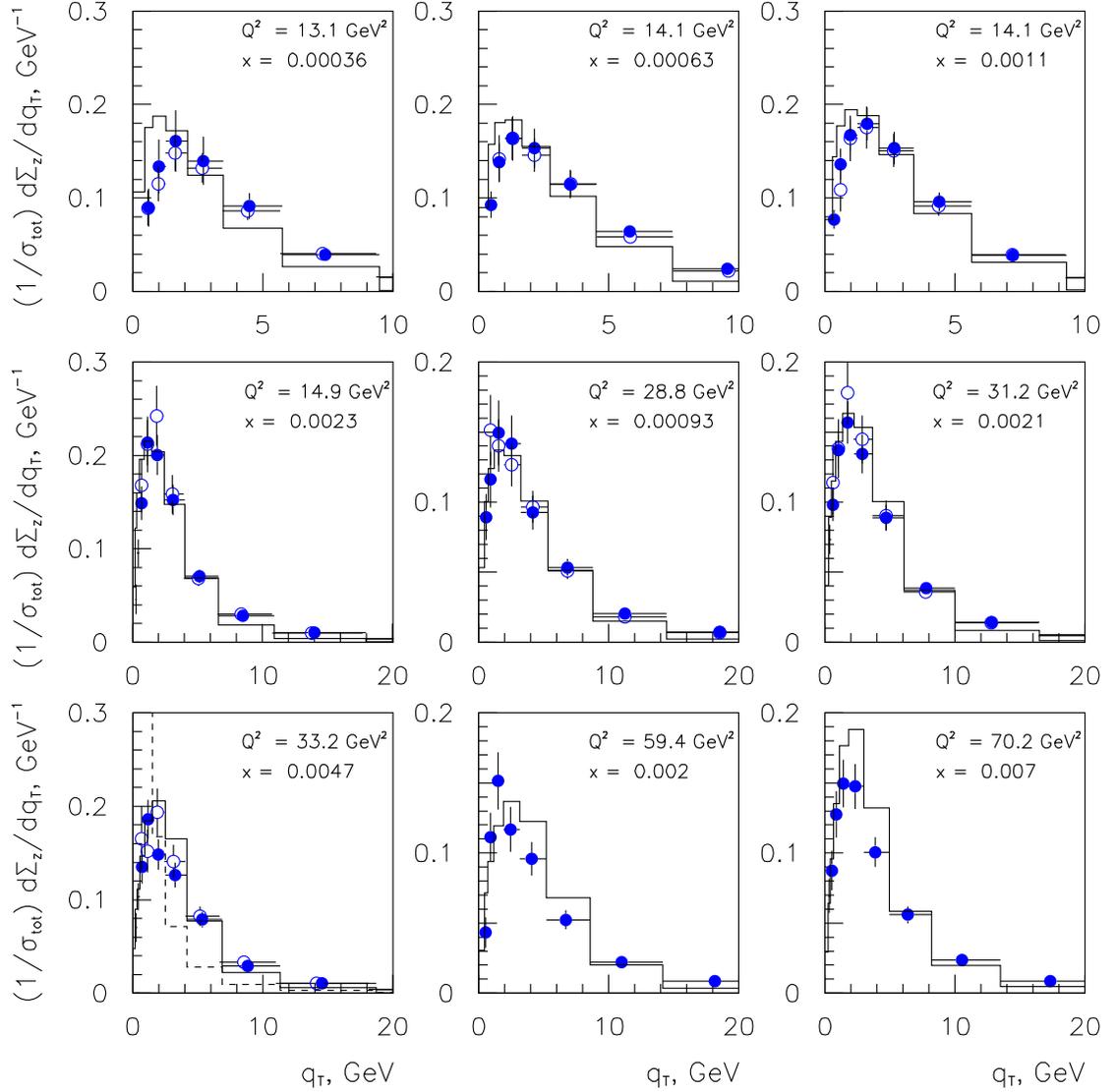}} \vspace{-2cm}

\caption{Comparison of the resummed \protect$ z\protect $-flow 
(solid curve) in the current region
of the hCM frame with the data in the low-\protect$ Q^{2}\protect $ bins
from Refs.\,\protect\cite{H1z2} (filled circles) and \protect\cite{H1z1} (empty
circles). For the bin with \protect$ \langle Q^{2}\rangle =33.2\mbox {\, GeV}^{2}\protect $
and \protect$ \langle x\rangle =0.0047\protect $, the fixed-order
\protect$ {\mathcal{O}}(\alpha _{S})\protect $
contribution for the factorization scale 
\protect$ \mu=Q\protect $ is shown as the dashed curve.
\label{fig:b17} }
\end{figure}

\begin{figure}[H]
\resizebox*{1\textwidth}{!}{\includegraphics{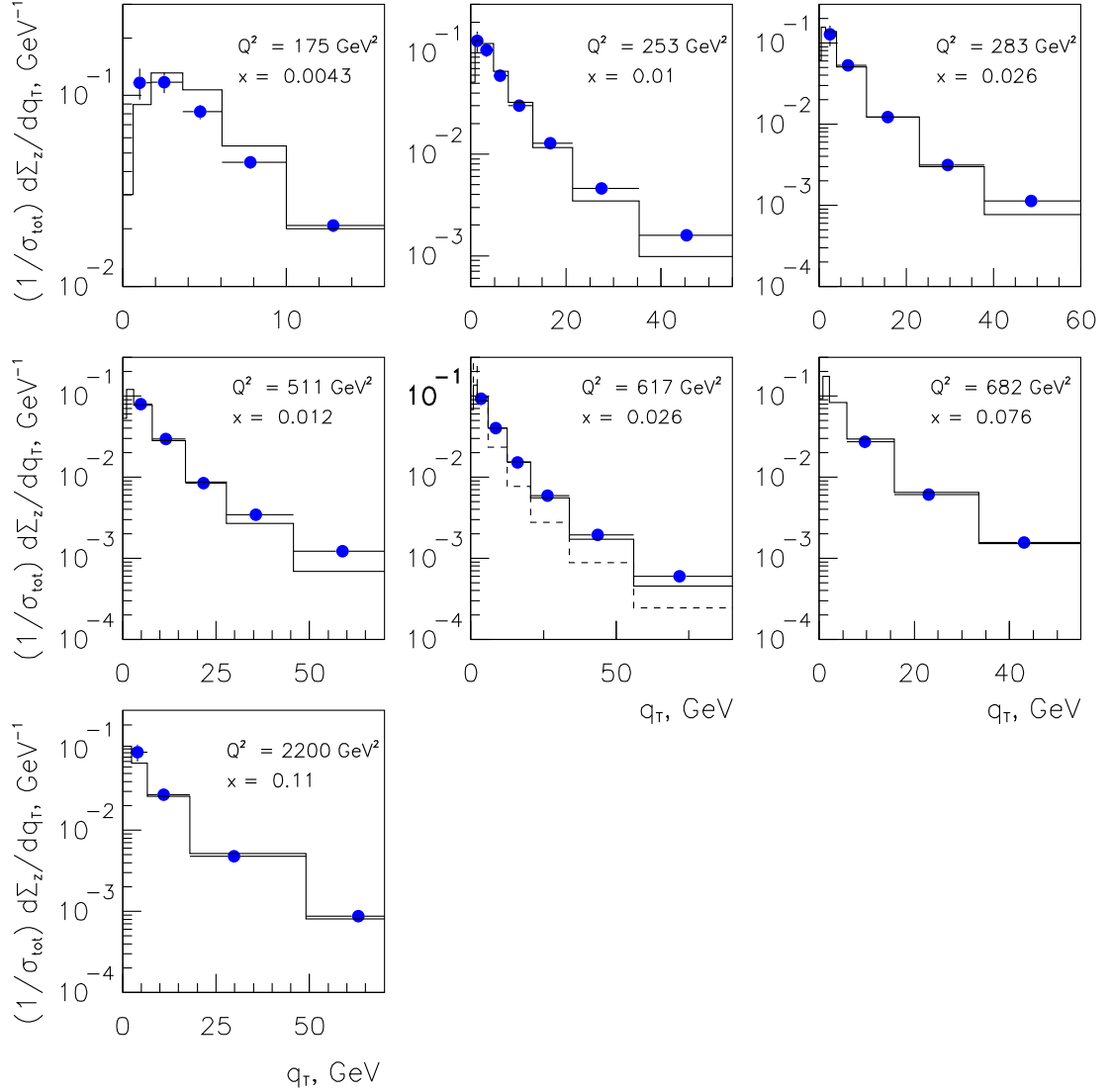}} \vspace{-2cm}

\caption{Comparison of the resummed \protect$ z\protect $-flow in the current region
of the hCM frame with the data in the high-\protect$ Q^{2}\protect $ bins
from Ref.\,\protect\cite{H1z2}. For the bin with \protect$ \langle Q^{2}\rangle =617\mbox {\, GeV}^{2}\protect $
and \protect$ \langle x\rangle =0.026\protect $, the \protect$ {\mathcal{O}}(\alpha _{S})\protect $
contribution for \protect$ \mu _{F}=Q\protect $ is shown as a dashed curve.
\label{fig:b47} }
\end{figure}

The numerical results below were obtained using the parameters of the HERA electron-proton
collider. The energies of the proton and electron beams are taken to be equal
to 820 and 27.5 \mbox{GeV}, respectively. All calculations were performed using
CTEQ5M1 parton distribution functions \cite{CTEQ5} and the parametrization\,2
of the nonperturbative Sudakov function $ S_{z}^{NP} $ (Eq.\,(\ref{SNPzpar2})),
unless stated otherwise. 
The theoretical results in Figs.\,\ref{fig:nloandresum}-\ref{fig:b47} were
obtained using the kinematical correction to the asymptotic and resummed cross-sections
at non-zero $ q_{T} $, which was discussed in Subsection\,\ref{sub:KinCorr}.
The factorization and renormalization scales
of the perturbative and asymptotic pieces are all set equal to $ \mu =Q $.
The resummed piece was calculated using $ C_{1}=b_{0},\, C_{2}=1,\,
\mu =b_{0}/b $, where $b_0 \equiv 2 e^{-\gamma_E}$.

In Fig.\,\ref{fig:nloandresum}, I present the comparison of the existing data
from \cite{H1z1} in one of the bins of $ x $ and $ Q^{2} $ ($ \langle x\rangle =0.0049,\, \, \langle Q^{2}\rangle =32.6\, \, \mbox {\, GeV}^{2} $)
with the NLO perturbative and resummed $ z $-flows. 
Figure~\ref{fig:nloandresum} demonstrates
two important aspects of the NLO $ q_{T} $ distribution (dashed curve):
 namely, the
NLO $z$-flow exceeds the data at small $ q_{T} $ and is below the data at $ q_{T}\geq Q $.
In fact, the deficit of the NLO prediction in comparison
with the data at medium and large $ q_{T} $ ($ q_{T}\geq 5\, \, \mbox {\, GeV} $)
is present in the entire region of $ x $ and $ Q^{2} $ that was studied.

As I discussed in Section\,\ref{sec:TheoryUncertainties}, one can trust the
resummed calculation only for reasonably small values of $ q_{T}/Q $. For
large values of $ q_{T} $, the fixed-order perturbative result is more reliable.
This means that the NLO resummation formalism will not give an accurate description
of the data for $ q_{T}\gg Q $ 
due to the small magnitude of the NLO perturbative
$ z $-flow in this region.

\vspace{1cm}

\newpage

\newpage

The excess of the data over the NLO calculation at large $q_T$
(cf. Figs.\,\ref{fig:nloandresum}-\ref{fig:b47})
can be interpreted as a signature
of other intensive hadroproduction mechanisms at hCM   pseudorapidities 
$ \eta_{cm}\leq 2 $.
A discussion of the cross-sections in this pseudorapidity region is beyond the
scope of this thesis. There exist several possible explanations of the data in
this region, for instance, the enhancement of the cross-section due to BFKL
showering \cite{BFKL} or resolved photon contributions \cite{Kramer,Jung}.
It is clear, however, that better agreement between
the data and the theory,
in a wider range of $ \eta _{cm} $, will be achieved when next-to-next-to-leading
order contributions, like the ones contributing to (2+1) jet production \cite{Catani},
are taken into account.

On the other hand, Figs.\,\ref{fig:nloandresum}-\ref{fig:b47} 
illustrate that the resummed
$ z $-flow is in better agreement with the data, over a wide range of $ q_{T}/Q $,
but also lies below the data if $ q_{T}/Q $ significantly exceeds unity.
The better consistency between the resummed $ z $-flow and the data suggests
that the resummed $ z $-flow should be used up to values of $ q_{T}/Q\sim 1-4 $,
\emph{i.e.}, without switching to the fixed-order expression. This
procedure was followed in the derivation of our numerical results.

Let us discuss the features of the data presented in Figs.\,\ref{fig:b17} and
\ref{fig:b47}. First, the data in the low-$ Q^{2} $ bins is significantly
influenced by nonperturbative effects and therefore is sensitive to the details
of the parametrization of $ S^{NP}_{z}(b,Q,x) $.
This feature can be seen from the abundance of data points around the maximum
of the $ q_{T} $-distribution, where the shape is mainly determined by 
$ S^{NP}_{z}(b,Q,x) $. Also, the low-$ Q^{2} $ data from HERA is
characterized by small values of $ x $, between $ 10^{-4} $ and $ 10^{-2} $.
For the theory to be consistent with the data from Ref.\,\cite{H1z1}
in this range of $ x $, the nonperturbative Sudakov factor must increase
rapidly as $ x\rightarrow 0 $, at least as $ 1/\sqrt{x} $. Such $ x $-dependence
is implemented in the parametrization\,1 of $ S_{z}^{NP}. $ In our
newer analysis, we found that growth of $ S^{NP}_{z}(b,Q,x) $ as $ 1/x $
at small $ x $ is in better agreement with the more recent data from \cite{H1z2}.

Second, the data in the high-$ Q^{2} $ bins of Fig.\,\ref{fig:b47} shows
a behavior that is qualitatively different from Fig.\,\ref{fig:b17}. In the
region covered by the experimental data points, the $ q_{T} $ distribution
is a monotonically decreasing function of $ q_{T} $, which shows good agreement
with the resummed $ z $-flow over a significant range\footnote{%
I point out once again that both the $ {\mathcal{O}}(\alpha _{S}) $ and
resummed $ z $-flow lie below the data at very large $ q_{T} $, in all
bins of $ x $ and $ Q^{2} $ in Figures\,\ref{fig:b17} and \ref{fig:b47}.
} of $ q_{T} $. In the region $ q_{T}<10 $ GeV, \textit{i.e.}, where the
maximum of the $ q_{T} $ distribution is located and where nonperturbative
effects are important, the experimental $ q_{T} $-bins are too large to provide
any information about the shape of $ d\Sigma _{z}/dq_{T} $. Thus, as mentioned
earlier, the published high-$ Q^{2} $ $ z $-flow data from Ref.\,\cite{H1z2}
is not sensitive to the dynamics described by the nonperturbative Sudakov factor
$ S^{NP}_{z}(b,Q,x) $.

%
%

A third comment is that most of the high-$ Q^{2} $ data points in Fig.\,\ref{fig:b47}
correspond to $ \langle x\rangle >10^{-2} $. If the resolution of the H1
measurements at high $ Q^{2} $ were better in the small-$ q_{T} $ region,
then the high-$ Q^{2} $ data would also reveal the behavior of $ S^{NP}_{z}(b,Q,x) $
at large $ x $. But, as mentioned before, the published data in the high-$ Q^{2} $
bins are not very sensitive to the shape of the $ z $-flow at small $ q_{T} $.
Therefore it is not possible to impose any constraints on $ S^{NP}_{z}(b,Q,x) $
at large values of $ x $, except that it should be positive, $ S^{NP}_{z}(b,Q,x)>0 $.
For this reason we  have chosen the $ x $-dependent part of $ S^{NP}_{z}(b,Q,x) $
in the parametrization\,2 such that $ S^{NP}_{z}(b,Q,x) $ grows approximately
as $ 1/x $ as $ x\rightarrow 0 $ and is positive for all $ x $. For
the same reason, we chose $ C=0 $ in the parametrization\,2. Although
the most general parametrization of $ S_{z}^{NP}(b,Q,x) $ can have $ C\neq 0 $,
the current data cannot distinguish between the parametrization\,2 with
$ C=0 $ and $ C\neq 0 $, as long as the value of $ C $ is not very
large.

\begin{figure}[H]
\resizebox*{\textwidth}{!}{\includegraphics{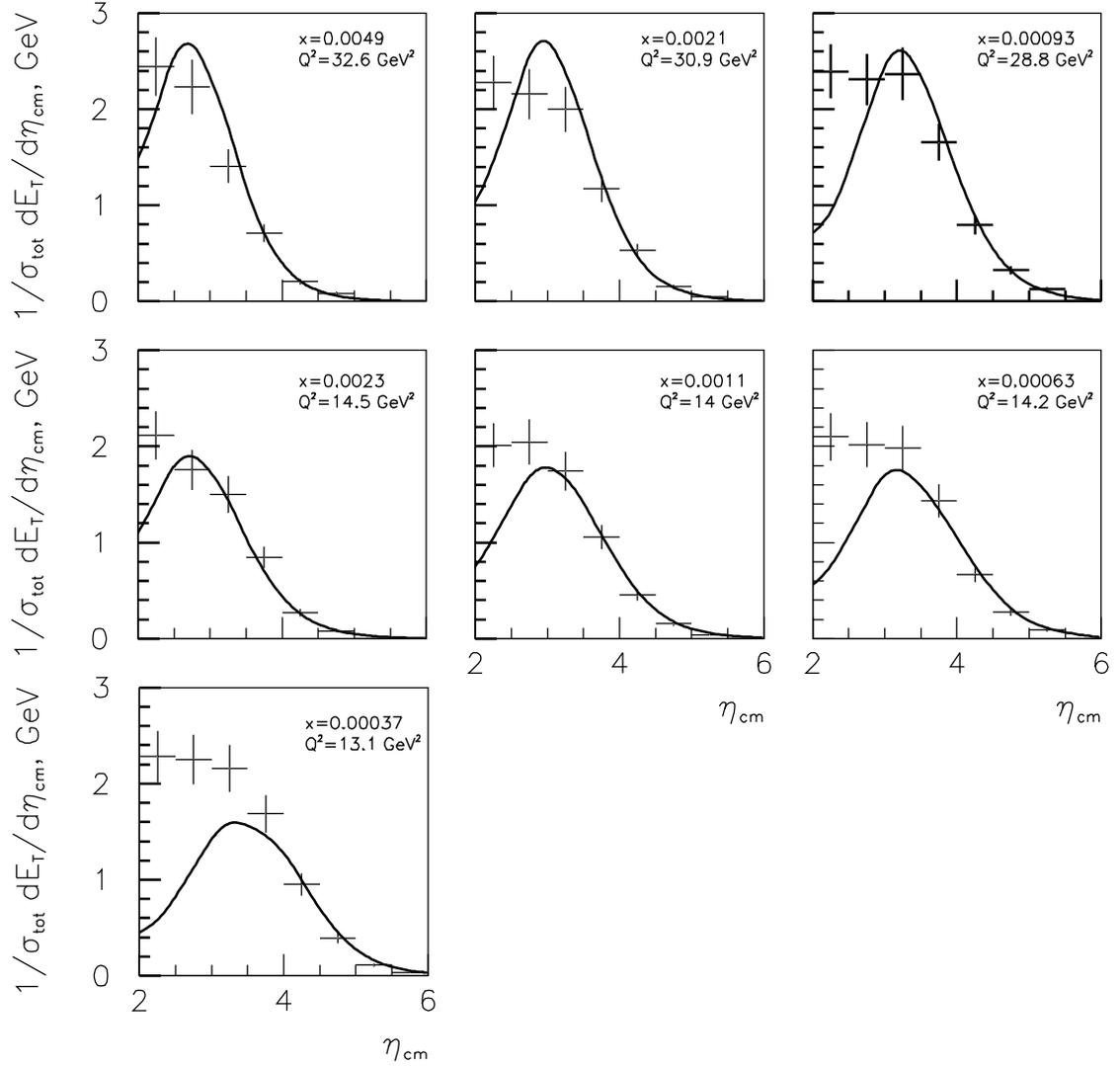}}

\caption{\label{fig:ET} The hCM pseudorapidity distributions of the transverse energy
flow in the current fragmentation region. The data are from \protect\cite{H1z1}. CTEQ4M PDFs
and the parametrization\protect$ \, \protect $1 of \protect$ S_{z}^{NP}\protect $
were used.}
\end{figure}

Finally, Fig.\,\ref{fig:ET} shows the results of our calculation presented
as the hCM    pseudorapidity distributions of the transverse energy flow $ \langle E_{T}\rangle  $.
This quantity is obtained by the transformation (\ref{Sz2ET}). The small-$ q_{T} $
region, where the resummation formalism is valid, corresponds to large pseudorapidities.
In this region, the agreement between our calculation and the data is good.
At smaller pseudorapidities (larger $ q_{T} $), one sees the above-mentioned
excess of the data over the perturbative NLO calculation. In the 
$ \langle E_{T}\rangle$~vs.~$ \eta _{cm} $ plot, this excess is magnified because of the factor $ q_{T}^{2} $
in the transformation (\ref{Sz2ET2}).

\subsection{How trustworthy is the resummed \protect$ z\protect $-flow at large \protect$ q_{T}\protect $?}

As noted earlier, the $ {\mathcal{O}}(\alpha _{S}) $ fixed-order $ z $-flow
is much larger than the data in the region $ q_{T}/Q\ll 1 $ and smaller
than the data in the region $ q_{T}/Q\gtrsim 1 $. In the small-$ q_{T} $
region, the resummed $ z $-flow is, by its construction, more reliable than
the fixed-order result. In the large-$ q_{T} $ region, the resummed $ z $-flow,
with the kinematical correction (\ref{tildex}) included, is also in better
agreement with the data than the fixed-order calculation. But theoretically,
the resummed $ z $-flow at large $ q_{T}/Q $ is not absolutely trustworthy,
because it does not include those parts of the fixed-order $ z $-flow that
are subleading in the limit $ q_{T}\rightarrow 0 $, but which might be important
at large $ q_{T} $. If the NLO result were in a good agreement with the data
at large $ q_{T} $, it would be justified to consider it a more reliable
prediction in this region. But since the $ {\mathcal{O}}(\alpha _{S}) $ contribution
is systematically smaller than the data, higher-order corrections are presumably
necessary in order to describe the region $ q_{T}\gtrsim Q $ reliably.

A systematic approach for improving the theoretical description of the large-$ q_{T} $
region would require inclusion of the complete $ {\mathcal{O}}(\alpha _{S}^{2}) $
terms in both the fixed-order and resummed $ z $-flows. But because such
a calculation is not available, it might be beneficial to use the resummed $ z $-flow
as a better theoretical prediction both in the region $ q_{T}/Q\ll 1 $, where
application of the resummation formalism is fully justified, and for $ q_{T} $
up to several units of $ Q $, where the resummed $ z $-flow agrees with
the data better than the fixed-order one. Then the use of the resummed $ q_{T} $-distributions
of the $ z $-flow will provide more reliable predictions for other observables
relevant to the SIDIS process.\\
\indent As an example, resummation can improve the reliability of the theoretical
prediction for the azimuthal asymmetry of the $ z $-flow. The $ b $-space
resummation formalism affects only the coefficient $ \prescr{1}V_{zA} $ of
the angular function $ A_{1}(\psi ,\varphi ) $. This coefficient is the one
that dominates the $ \varphi  $-integrated $ z $-flow in the small-$ q_{T} $
region, where the energy flow is the most intense. On the other hand, the main
goal of the measurement of angular asymmetries is to study structure functions
other than $ \prescr{1}V_{zA} $, \textit{e.g.,} those corresponding to the
angular functions $ A_{3}(\psi ,\varphi )=-\cos \varphi \sinh 2\psi  $ and
$ A_{4}(\psi ,\varphi )=\cos 2\varphi \sinh ^{2}\psi  $. By using a better
approximation for the coefficient $ \prescr{1}{V}_{zA} $, it is possible
to measure the coefficients $ \prescr{3,4}V_{zA} $ more reliably. Conversely,
by knowing that the all-order resummation effects are important in the region
of small $ q_{T} $ and by concentrating on the region where $ q_{T} $
is of the order $ Q $ or larger, one may find angular asymmetries that are
well approximated in the lowest orders of PQCD. The impact of resummation
on the angular asymmetries is discussed 
in more detail in Chapter\,\ref{ch:AzimuthalAsymmetries}.

\newpage

\section{Normalized distributions of charged particle production \label{sec:Multiplicity}}

Let us now turn to the discussion of particle multiplicities. Although the resummation
formalism, as outlined in Chapter \ref{ch:Formalism}, can describe the cross
section for any massless final-state particle (provided that the fragmentation
functions for this particle are known), in this Section I will concentrate
on distributions of the charged particle multiplicity, defined as \begin{equation}
\frac{1}{\sigma _{tot}}\frac{d\sigma (A+e\rightarrow h^{\pm }+e+X)}{d\Theta }.
\end{equation}
 Here $ \Theta  $ is some kinematical variable, such as the variable $ q_{T}^{2} $
in Eq.\,(\ref{qT2}), the transverse momentum $ p_{T} $ of the final-state
charged particle in the hCM frame, or the Feynman variable $ x_{F} $, \begin{equation}
x_{F}\equiv \frac{2p_{B,cm}^{z}}{W}=z\left( 1-\frac{q_{T}^{2}}{W^{2}}\right) .
\end{equation}

Our calculation assumes that the charged particles registered in the detector
are mostly charged pions, kaons and protons. Therefore the cross section for
charged particle production can be calculated using (\ref{hadcs}) with the
replacement of the fragmentation functions $ D_{B/b}(\xi _{b},\mu ) $ by\begin{equation}
D_{h^{\pm }/b}(\xi _{b},\mu )\equiv\sum _{B=\pi ^{\pm },K^{\pm },p,\bar{p}}D_{B/b}(\xi _{b},\mu ).
\end{equation}
The fragmentation functions $ D_{B/b}(\xi _{b},\mu ) $ are known reasonably
well only for $ \xi _{b}\gtrsim 0.05-0.1 $ \cite{BKK,Kretzer,Bourhis}. Thus,
the formalism presented here is applicable to the production of charged particles
with sufficiently large energies, \textit{i.e.}, for $ z\gtrsim 0.05 $.

Certain experimental distributions are readily available from the literature
\cite{H1chgd97,ZEUSchgd96,EMC,E665}, such as $ d\sigma /dp_{T} $, $ d\sigma /dx_{F} $,
as well as distributions for the average transverse momentum $ \langle p_{T}^{2}\rangle  $.
However, the ``experimental'' $ q_{T} $ distributions must currently
be derived from pseudorapidity distributions by using Eq.\,(\ref{qTetacm}).
Although the distributions $ d\sigma /dp_{T} $ and $ \langle p_{T}^{2}\rangle  $
are quite sensitive to resummation effects, they cannot be interpreted as easily
as the distributions $ d\sigma /dq_{T} $, primarily because the distributions
$ d\sigma /dp_{T} $ and $ \langle p_{T}^{2}\rangle  $ mix resummation
effects at small values of $ q_{T} $ with perturbative contributions from
the region $ q_{T}/Q\gtrsim 1 $. The most straightforward way to study the
effects of multiple parton radiation would be to consider the $ q_{T} $ (or
pseudorapidity) distributions that satisfy the additional requirement $ z>0.05-0.1 $
and that are organized in small bins of $ Q^{2} $ and $ x $. Unfortunately,
such distributions have not been published yet. Although Ref.\,\cite{H1chgd97}
presents distributions $ d\sigma (p+e\rightarrow h^{\pm }+e+X)/d\eta_{cm} $
for some values of $ x $ and $ Q^{2} $, these distributions are integrated
over the full range of $ z $. Therefore, they are sensitive to the uncertainties
in fragmentation functions, mass effects,\footnote{%
Our calculation assumes that all participating particles, including the
final-state hadrons, are massless. Because of this assumption, the production
of final-state hadrons with $ z=0 $ is allowed. However, in realistic SIDIS
experiments there is a non-zero minimal value of $ z $ determined by the
finite mass of the observed hadron. It follows from the definition (\ref{z})
of $ z $ and Eqs.\,(\ref{pAc}, \ref{pBc}) for the initial and final
hadron momenta in the $ \gamma ^{*}p $ \cms frame, that\begin{equation}
\label{zmin}
z=\frac{p_{B,cm}^{+}}{W}\geq \frac{m_{B}}{W},
\end{equation}
 where\[
p_{B,cm}^{+}=E_{B,cm}+p_{B,cm}^{z}.\]
According to Eq.\,(\ref{zmin}), the mass of the final-state hadron
should be included if $ z\sim m_{B}/W\sim \Lambda _{QCD}/W $. Hence, our
massless calculation is not suited for the analysis of the distributions $ d\sigma (A+e\rightarrow h^{\pm }+e+X)/d\eta_{cm} $
from \cite{H1chgd97}, which are sensitive to such small values of $ z $.
} and contributions from diffractive scattering. 

Because the experimental $ q_{T} $ distributions are unavailable, we have
decided to undertake a simpler analysis than the one presented for the energy
flow. Our goal here is to understand how the multiple parton radiation \textit{could}
affect various aspects of charged particle production. For this purpose we focused
our attention on data from the ZEUS Collaboration \cite{ZEUSchgd96}, which
presents the charged particle multiplicity in a phase-space region characterized
by the mean values $ \langle W\rangle =120\mbox {\, GeV} $, $ \langle Q^{2}\rangle =28\mbox {\, GeV}^{2} $,
and the additional constraint $ z>0.05 $. These values of $ \langle W\rangle  $
and $ \langle Q^{2}\rangle  $ translate into an average value of $ x=1.94\times 10^{-3} $.
A simple model for nonperturbative effects at small
$ q_{T} $ will demonstrate that resummation describes qualitative features
of this set of experimental data better than the fixed-order calculation.

\begin{figure}[H]
{\par\centering \resizebox*{!}{9.5cm}{\includegraphics{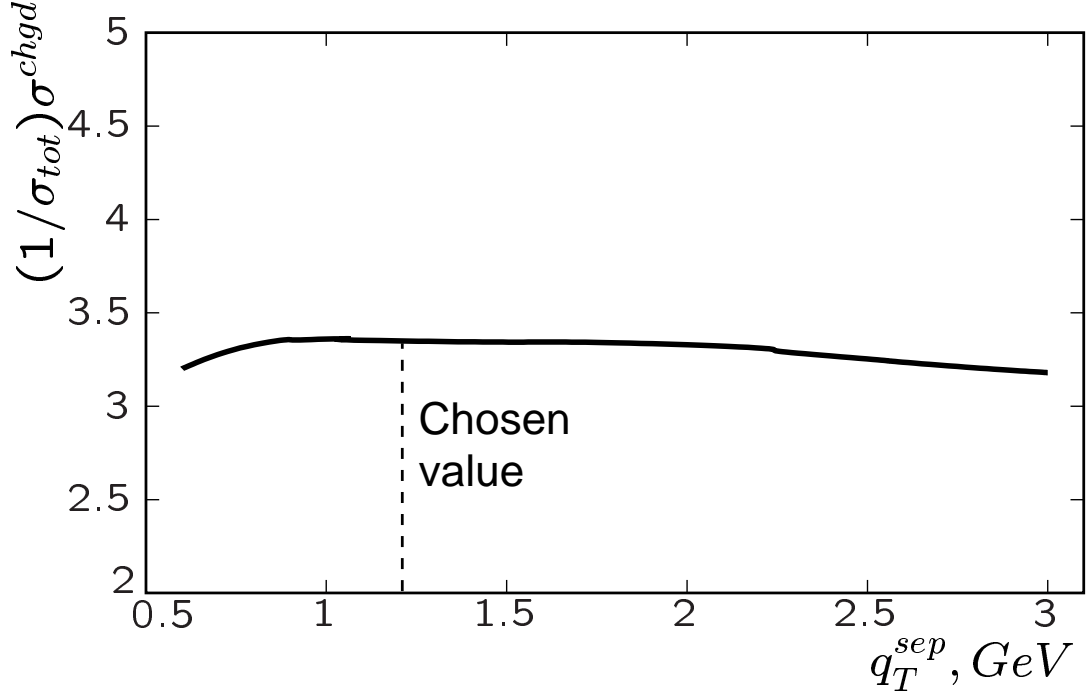}} \par}

\caption{\label{fig:sig_tot_vs_qt_sep} The dependence of the \protect$ {\mathcal{O}}(\alpha _{S})\protect $
prediction for the total charged particle multiplicity on the value of the separation
scale \protect$ q_{T}^{S}\protect $. The calculation is done for \protect$ \langle W\rangle =120\mbox {\, GeV},\, \, \langle Q^{2}\rangle =28\mbox {\, GeV}^{2}\protect $.}
\end{figure}

In all of the cases presented, the strategy is to compare the resummed multiplicity
to that from the next-to-leading order calculation. In the numerical analysis,
the multiplicity was calculated using the CTEQ5M1 PDFs \cite{CTEQ5} and the
FFs from \cite{Kretzer}. For the resummed multiplicity, the ``canonical''
combination $ C_{1}=b_{0},\, C_{2}=1,\, \mu =b_{0}/b $ was used. The NLO
cross section was calculated according to Eq.\,(\ref{pert}), for the factorization
scale $ \mu =Q $. As explained in detail in Section\,\ref{sub:NLOFixedOrder},
the integration of the NLO term over $ q_{T} $ is done separately over the
regions $ 0\leq q_{T}\leq q_{T}^{S} $ and $ q_{T}>q_{T}^{S} $, where
$ q_{T}^{S} $ is a particular type of a phase space slicing parameter.
The final results should not depend on the exact value of $ q_{T}^{S} $
provided that it is chosen in the region where the $ {\mathcal{O}}(\alpha _{S}) $
part of the next-to-leading-logarithmic expansion (\ref{NLL}) approximates
well the exact NLO cross section. In practical calculations, $ q_{T}^{S} $
cannot be chosen to be too small, because the numerical calculation becomes
unstable due to large cancellations between the integrals over the regions $ 0\leq q_{T}\leq q_{T}^{S} $
and $ q_{T}\geq q_{T}^{S} $. The NLO prediction for the integrated charged
particle multiplicity $ \sigma _{chgd}/\sigma _{tot} $ at $ \langle W\rangle =120\mbox {\, GeV} $,
$ \langle Q^{2}\rangle =28\mbox {\, GeV}^{2} $ is practically independent
of $ q_{T}^{S} $ in the region $ 1\lesssim q_{T}^{S}\lesssim 2.5 $
GeV (\textit{\emph{cf.}} Figure\,\ref{fig:sig_tot_vs_qt_sep}). The NLO distributions
shown in the subsequent Figures were calculated for $ q_{T}^{S}=1.2 $~GeV,
which lies within the range of stability of $ \sigma ^{chgd}/\sigma _{tot} $.
As in the case of the $ z $-flow, the resummed charged particle multiplicity
may suffer from matching ambiguities at $ q_{T}/Q\sim 1 $.

In Section\,\ref{sec:ZFlows}, we found that the resummed $ z $-flow
is in better agreement with the experimental distributions than the NLO $ z $-flow,
for the whole range $ q_{T}/Q\lesssim 2-4 $. That result suggests that it
might be preferable to use the resummed $ z $-flow in the whole range $ q_{T}/Q\lesssim 2-4 $
as a better theoretical prediction, until the $ {\mathcal{O}}(\alpha _{S}^{2}) $
prediction for the $ z $-flow in the region $ q_{T}/Q\gtrsim 1 $ becomes
available. In the case of the charged particle multiplicity, the resummed cross
section, which is calculated according to the formula \begin{equation}
\frac{d\sigma _{BA}}{dxdzdQ^{2}dq_{T}^{2}}=\frac{\sigma _{0}F_{l}}{\pi S_{eA}}\int \frac{d^{2}b}{(2\pi )^{2}}e^{i\vec{q}_{T}\cdot \vec{b}}\widetilde{W}_{BA}(b,x,z,Q)+Y_{BA},
\end{equation}
 overestimates the experimentally measured rate for the production of charged
particles with $ p_{T}>2 $ GeV. This discrepancy indicates that the resummed
cross section in the region $ q_{T}/Q\gtrsim 1 $ is too high, so that switching
to the perturbative cross section in this region is in fact required. Therefore,
we have chosen to use the resummed cross section for $ q_{T}\leq 5 $ GeV
and switch to the next-to-leading order cross section for $ q_{T}\geq 5 $
GeV. 

\begin{figure}[tb]
{\par\centering \resizebox*{0.495\textwidth}{!}{\includegraphics{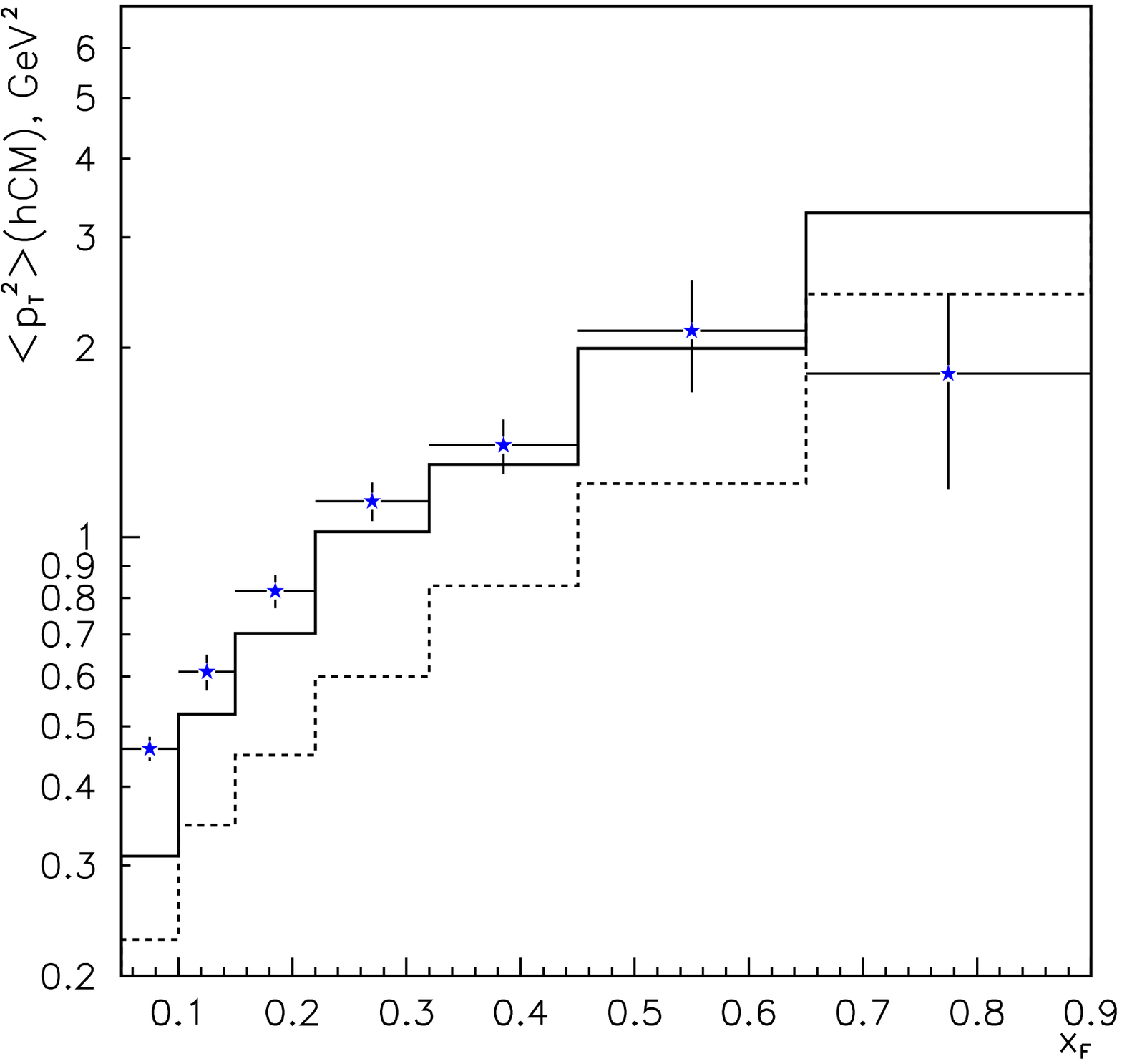}} \resizebox*{0.495\textwidth}{!}{\includegraphics{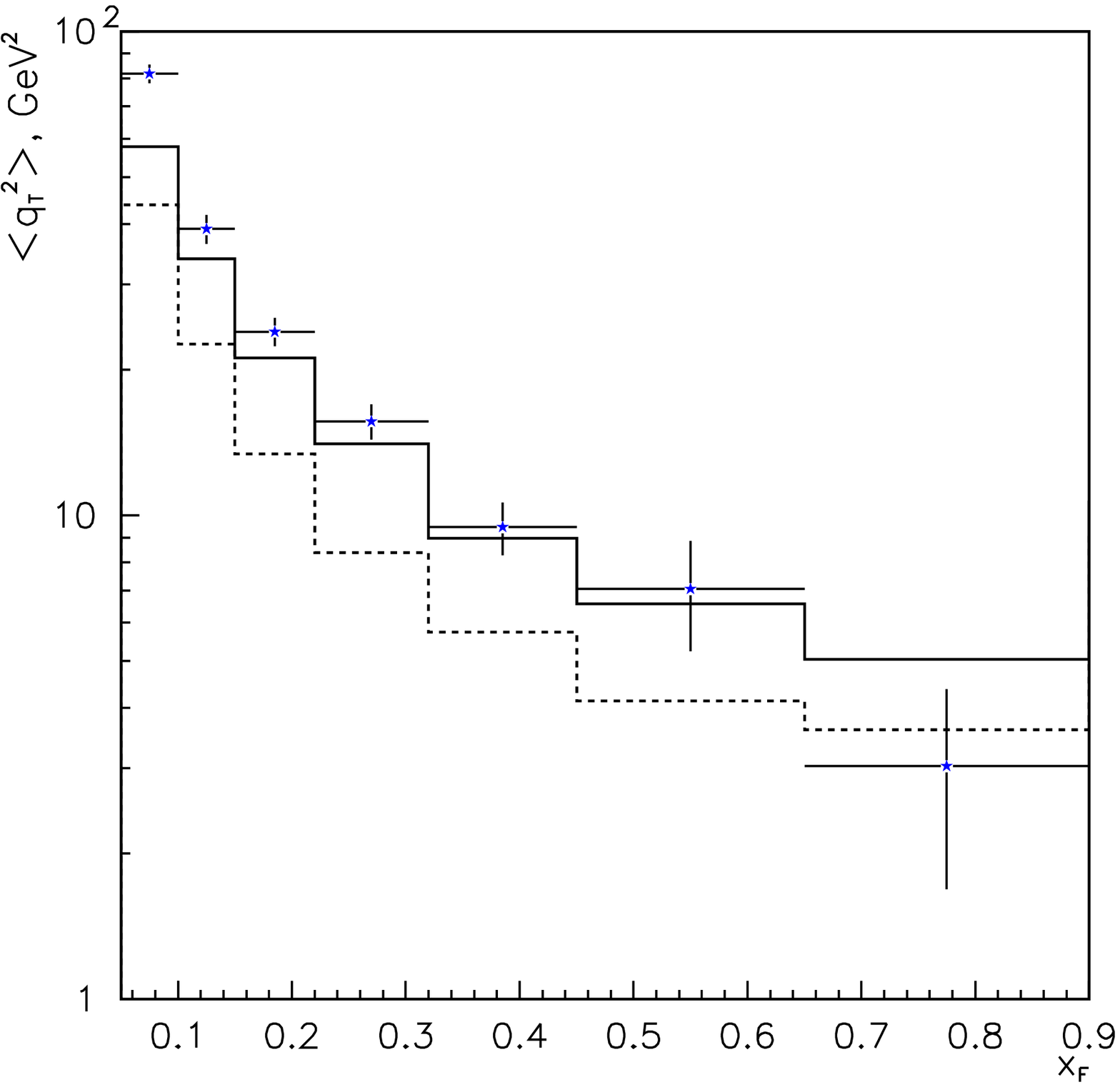}} \\ a)
\hspace{7cm} b) \par}

\caption{\label{fig:seagull} The distributions (a) \protect$ \langle p_{T}^{2}\rangle \protect $~vs.~\protect$ x_{F}\protect $
and (b) \protect$ \langle q_{T}^{2}\rangle \protect $~vs.~\protect$ x_{F}\protect $
for the charged particle multiplicity at \protect$ \langle W\rangle =120\mbox {\, GeV},\, \, \langle Q^{2}\rangle =28\mbox {\, GeV}^{2}\protect $.
The experimental points for the distribution \protect$ \langle p_{T}^{2}\rangle \protect $~vs.~\protect$ x_{F}\protect $
are from Fig.\ 3c of Ref.\,\protect\cite{ZEUSchgd96}. The ``experimental''
points for the distribution \protect$ \langle q_{T}^{2}\rangle \protect $~vs.~\protect$ x_{F}\protect $
are derived using Eq.\,(\ref{pT2toqT2}). The solid and dashed curves correspond
to the resummed and NLO (\protect$ \mu =Q\protect $)
multiplicity, respectively.}
\end{figure}

As in the case of the $ z $-flow, the shape of the $ q_{T} $ distribution
for the charged particle multiplicity at small values of $ q_{T} $ depends
strongly on the unknown nonperturbative Sudakov factor $ S^{NP}(b,Q,x,z) $.
For the purposes of this study, we introduced a preliminary representative parametrization
of the nonperturbative Sudakov factor for the \textit{fixed} values of $ x=1.94\times 10^{-3} $
and $ Q^{2}=28\mbox {\, GeV}^{2} $, \textit{i.e.,} the values that coincide
with the average values of $ x $ and $ Q^{2} $ in \cite{ZEUSchgd96}.
This \textit{$ z $-dependent} parametrization is \begin{equation}
\label{SNPchgd}
S^{NP}\left( b,Q^{2}=28\mbox {\, GeV}^{2},x=1.94\times 10^{-3},z\right) =b^{2}\left( 0.18+0.8\frac{(1-z)^{3}}{z^{1.4}}\right) .
\end{equation}
 Since the ZEUS Collaboration did not publish pseudorapidity distributions for
the charged particle multiplicity $ (1/\sigma _{tot})d\sigma /d\eta_{cm} $
in bins of varying $ z $, we had to deduce information about the $ z $-dependence
of $ S^{NP} $ from the less direct distribution of $ \langle p_{T}^{2}\rangle  $
vs.~$ x_{F} $ presented in Fig.\,3c of \cite{ZEUSchgd96}. This distribution,
known as a ``seagull'' for its characteristic shape (Fig.\,\ref{fig:seagull}a),
can be converted into the more illustrative distribution of $ \langle q_{T}^{2}\rangle  $
vs.~$ x_{F} $ (Fig.\,\ref{fig:seagull}b). Since the major portion of the
registered events comes from the region $ q_{T}^{2}/W^{2}\ll 1 $, or $ x_{F}\approx z $,
a first estimate of the experimental data points for the distribution of $ \langle q_{T}^{2}\rangle  $
vs.~$ x_{F} $ can be obtained by assuming that \begin{equation}
\label{pT2toqT2}
\langle q_{T}^{2}\rangle \approx \frac{\langle p_{T}^{2}\rangle }{\langle z\rangle ^{2}}\approx \frac{\langle p_{T}^{2}\rangle }{\langle x_{F}\rangle ^{2}},
\end{equation}
 where $ \langle x_{F}\rangle  $ denotes central values of $ x_{F} $ in
each bin in Fig.\,\ref{fig:seagull}a.\footnote{%
In principle, a more accurate experimental distribution $ \langle q_{T}^{2}\rangle  $
vs.~$ x_{F} $ can be determined by its direct measurement.
} We refer to the resulting values as ``derived data''.

Note that the shapes of $ \langle p_{T}^{2}\rangle  $ vs. $ x_{F} $ and
$ \langle q_{T}^{2}\rangle  $ vs. $ x_{F} $ are quite different. The transformation
from Fig.\,\ref{fig:seagull}a to Fig.\,\ref{fig:seagull}b shows immediately
that the wing-like shape of the distribution of $ \langle p_{T}^{2}\rangle  $~vs.~$ x_{F} $
should be attributed to a purely kinematical effect, namely an extra factor
$ 1/z^{2} $ which is absent in the distribution of $ \langle q_{T}^{2}\rangle  $
vs.~$ x_{F} $. Once this extra factor is removed, we see from Fig.\,\ref{fig:seagull}b
that $ \langle q_{T}^{2}\rangle  $ increases monotonically and rapidly as
$ z $ approaches zero. In other words, the $ q_{T} $ distribution broadens
rapidly when $ z $ decreases. This behavior is approximately realized by
the simple $ z $-dependent nonperturbative Sudakov factor $ S^{NP}(b,Q,x,z) $
given in Eq.\,(\ref{SNPchgd}).

\begin{figure}[H]
{\par\centering \resizebox*{0.9\textwidth}{!}{\includegraphics{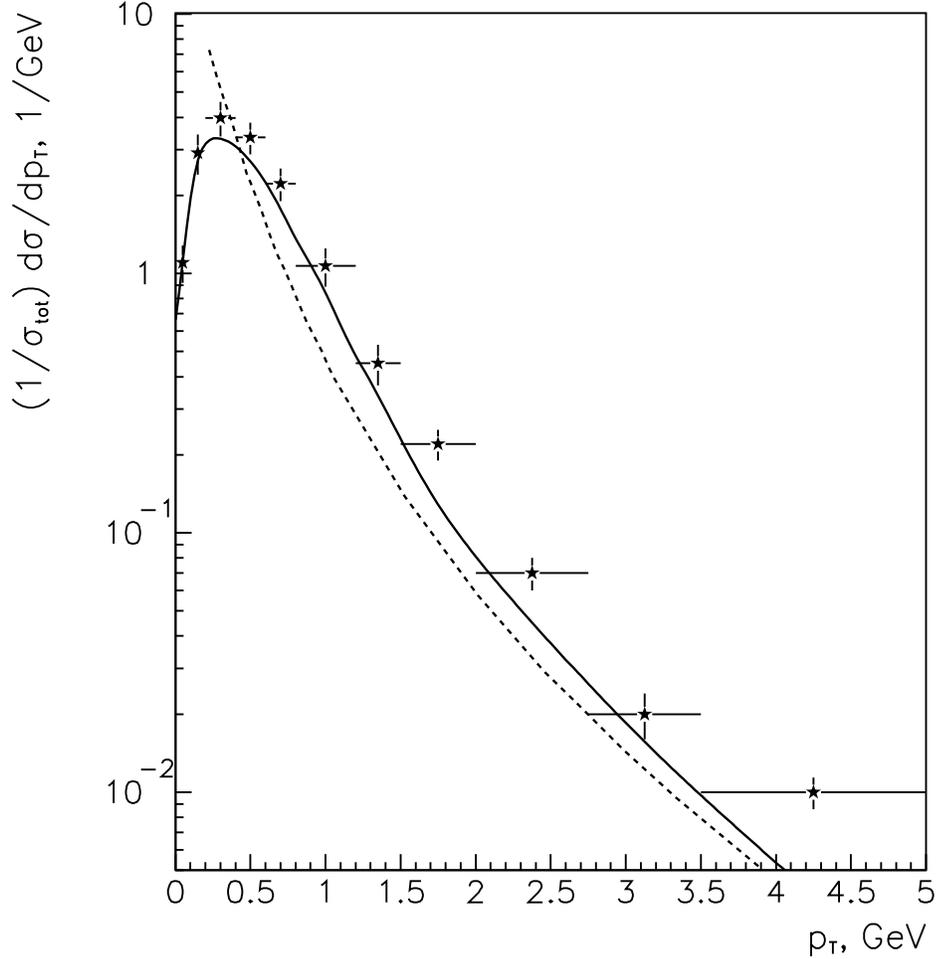}} \par}

\caption{\label{fig:dsigdpt} The dependence of the charged particle multiplicity on
the transverse momentum \protect$ p_{T}\protect $ of the observed particles
in the hCM frame. The data points are from \protect\cite{ZEUSchgd96}. The solid
and dashed curves correspond to the resummed and NLO multiplicities, respectively.}
\end{figure}

The parametrization of $ S^{NP}(b,Q,x,z) $ was chosen to maximize the agreement
between the resummed distribution of $ \langle q_{T}^{2}\rangle  $~vs.~$ x_{F} $
and the ``derived data'' (\textit{\emph{cf.}} Fig.\,\ref{fig:seagull}b).
Figure\,\ref{fig:seagull}b shows that the resummed calculation is in better
agreement with the data points than the NLO expression. We have found it difficult
to reproduce the rapid growth of $ \langle q_{T}^{2}\rangle  $ as $ x_{F}\rightarrow 0 $
in either approach. In the future, it will be interesting to see how a more
precise theoretical study will be able to explain adequately this rapid growth
of $ \langle q_{T}^{2}\rangle  $ in the region $ x_{F}\rightarrow 0 $,
assuming that the actual experimental data for the $ \langle q_{T}^{2}\rangle  $~vs.~$ x_{F} $
distribution resemble the ``derived data'' discussed above.

The resummation also significantly affects the $ p_{T} $ dependence of the
charged particle multiplicity. In Fig.\,\ref{fig:dsigdpt} we present the distribution
$ (1/\sigma _{tot})d\sigma /dp_{T} $. We see that resummation effects must
be included to describe the shape of this distribution at $ p_{T}\leq 1 $
GeV. Furthermore, resummation also improves the agreement between the theory
and the experiment in the whole range of $ p_{T} $. Through Eq.\,(\ref{pTB}),
the improved description of the $ q_{T} $ distribution in the small-$ q_{T} $
region translates into a better agreement with the $ p_{T} $ distribution
in the whole range of $ p_{T} $. Just as in the case of the $ z $-flow,
the fixed-order calculation gives a rate that is too small compared to the data,
which implies that higher-order corrections are important. Until the complete
$ {\mathcal{O}}(\alpha _{S}^{2}) $ corrections are available, the resummation
formalism, which already accounts for the most important contributions in the
region of the phase space with the highest rate (\textit{i.e.,} at small $ q_{T} $),
serves as a better theoretical prediction in the whole range of $ p_{T} $.

\begin{figure}[H]
{\par\centering \resizebox*{1\textwidth}{!}{\includegraphics{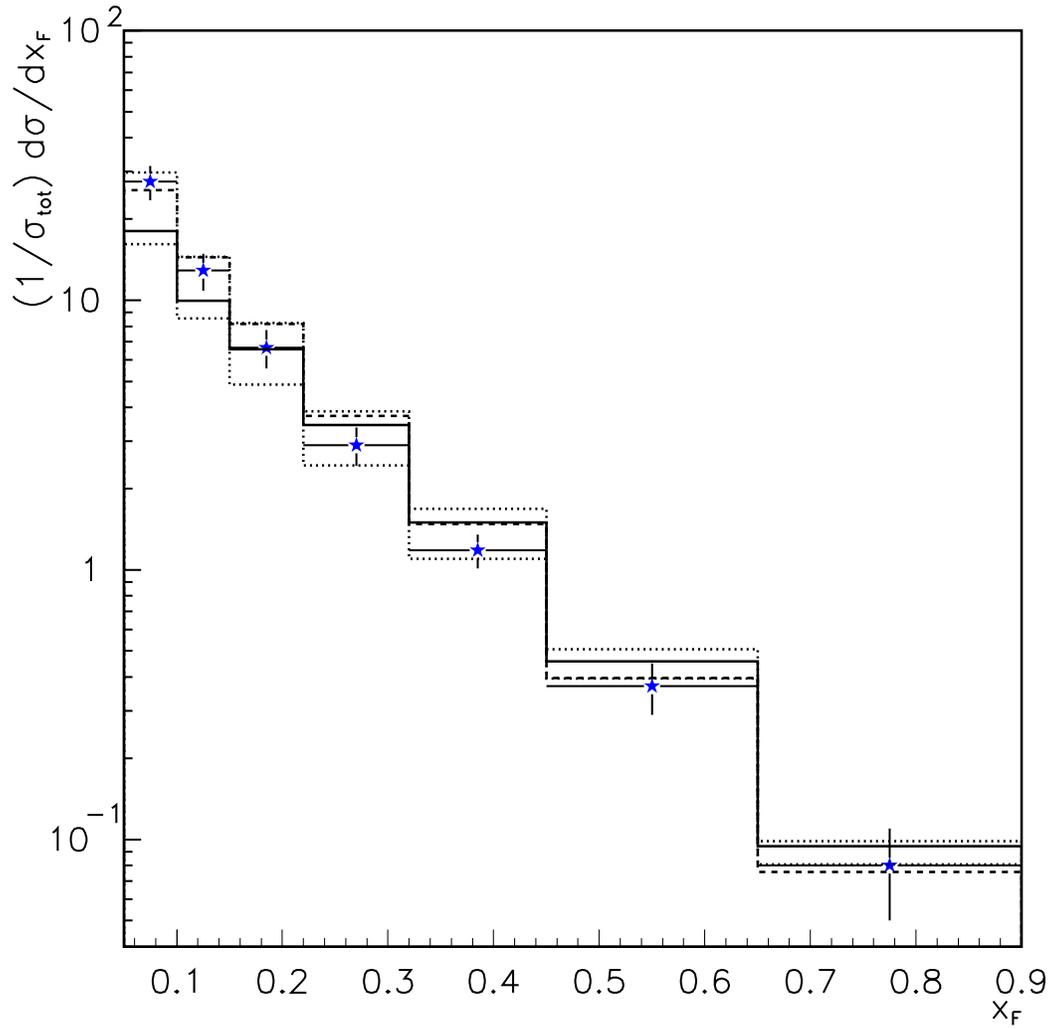}} \par}

\caption{\label{fig:dsigdxf} The dependence of the charged particle multiplicity on
the Feynman variable \protect$ x_{F}\protect $ in the hCM frame. The solid
curve corresponds to the resummed multiplicity. The dashed, lower dotted and
upper dotted curves correspond to the NLO multiplicity calculated for \protect$ \mu =Q,\, \, 0.5Q\protect $
and \protect$ 2Q\protect $, respectively.}
\end{figure}

Finally, Fig.\,\ref{fig:dsigdxf} shows the $ x_{F} $-distribution for the
charged particle multiplicity $ (1/\sigma _{tot})d\sigma /dx_{F} $. We see
that both the resummed and fixed-order distributions are in reasonable agreement
with the data and with earlier published theoretical results for the $ {\mathcal{O}}(\alpha _{S}) $
$ x_{F} $-distributions \cite{GraudenzPLB}. For the fixed-order multiplicity,
we present two additional curves corresponding to different choices of the factorization
scale $ \mu  $ in (\ref{hadcs}); the lower and upper dotted curves correspond
to $ \mu =0.5Q $ and $ 2Q $, respectively. Note that the scale dependence
of the NLO multiplicity increases when $ z\rightarrow 0 $. Also note that
the resummed multiplicity is significantly lower than the data in the two lowest
bins of $ x_{F} $ ($ \langle x_{F}\rangle =0.075 $ and $ 0.125 $),
but consistent with the NLO multiplicity within the uncertainty due to the scale
dependence. Such behavior of the resummed multiplicity results from the dependence
of the $ {\mathcal{O}}(\alpha _{S}) $ coefficient functions $ {\mathcal{C}}^{out(1)}_{ba}(\widehat{z},C_{1},C_{2},b_{*},\mu ) $
on the additional term $ \ln \widehat{z} $ which was given in Eqs.\,(\ref{C1out})
and (\ref{C1out2}) and discussed at the end of Subsection\,\ref{sub:NLOresum}.
This negative logarithm dominates the $ {\mathcal{C}}^{out(1)} $-functions
at very small values of $ \widehat{z} $. Similarly, the integral (\ref{NLOcs})
of the NLO cross section over the lowest bin $ 0\leq q_{T}^{2}\leq (q_{T}^{S})^{2} $
depends on $ \ln {\widehat{z}} $ through the terms \[
-\frac{\alpha _{S}}{2\pi }\ln {\frac{\mu _{F}^{2}}{(\widehat{z}\, q_{T}^{S})^{2}}}\biggl (\delta (1-\widehat{z})\delta _{bj}P_{ja}(\widehat{x})+P_{bj}(\widehat{z})\delta (1-\widehat{x})\delta _{ja}\biggr ),\]
 as given in (\ref{V1NLO}). Numerically, this dependence is less pronounced
than in the resummed cross section. For $ z\lesssim 0.1 $, the growing scale
dependence of the multiplicity in the $ {\mathcal{O}}(\alpha _{S}) $ calculation
indicates that unaccounted higher-order effects become important and are needed
to improve the theory predictions. For example, including the $ {\mathcal{O}}(\alpha ^{2}_{S}) $
coefficient $ {\mathcal{C}}_{ba}^{(2)} $ in the resummed calculation will
be necessary to improve the description of the charged particle multiplicity
in the small-$ z $ region.

\section{Discussion and conclusions\label{sub:DiscussionResults}}

The results in this Chapter demonstrate that multiple radiation of soft and
collinear partons influence a large class of observables and can be described
with the help of the CSS resummation formalism \cite{Collins93,Meng2}. Multiple parton
radiation affects hadroproduction in the current region of deep-inelastic scattering,
\textit{i.e.,} for large pseudorapidity of the final-state hadrons in the photon-proton
c.m.~frame. 

Although the resummation formalism needs further development, in particular
in the procedure for matching the resummed curve to the perturbative result
in the transition region, it already improves the agreement between the theory
and the data and provides interesting insights about qualitative features of
SIDIS. The formalism describes well the behavior of the transverse energy flows
measured at HERA \cite{H1z1,H1z2} in the region of large hCM~pseudorapidity
$ \eta_{cm} \geq 3.0 $. At smaller pseudorapidities the
NLO rate falls below  the existing data. Evidently, this is a signature of the
importance of the NNLO corrections, which were not studied in this paper. The
resummation formalism describes the pseudorapidity distributions of the transverse
energy flow more accurately than the NLO calculation; this formalism also has
good potential to improve the description of various distributions of particle
multiplicity.

The presented analysis shows that the experimentally measured $ q_{T} $ distributions
for the energy flow broaden rapidly as $ x\rightarrow 0 $. This rapid broadening
of the $ q_{T} $ distributions can be realized if the nonperturbative Sudakov
factor in the resummed energy flow increases as $ 1/x $. Similarly, the $ q_{T} $
distribution for the charged particle multiplicity broadens rapidly when $ z\rightarrow 0 $,
which is consistent with the nonperturbative Sudakov factor increasing as $ z^{-1.4} $.
The SIDIS nonperturbative Sudakov factors at small values of $ x $ and $ z $
are therefore qualitatively different from the known nonperturbative Sudakov
factors for vector boson production and $ e^{+}e^{-} $ hadroproduction, which
do not depend on the longitudinal variables at all. The rapid growth of the
nonperturbative Sudakov factor in SIDIS might indicate that the $ ep $ collider
HERA tests the resummation formalism in a new dynamical regime, which was not
yet studied at colliders of other types. In particular, the CSS formula adopted
here assumes the usual DGLAP physics for the evolution of initial and final
state partons \cite{DGLAP}, in which the radiation of unobserved collinear
partons is $ k_{T} $-ordered. The broadening of the $ q_{T} $ distributions
may be a result of the increasing importance of $ k_{T} $-unordered radiation
in the limit $ x\rightarrow 0 $. The growth of the nonperturbative Sudakov
factor $ S^{NP} $ as $ x $ decreases may be caused by the increase of
the intrinsic transverse momentum of the probed partons due to such radiation.

There are several theoretical aspects of the resummation formalism that can
be clarified when more experimental data are published. Perhaps the largest
uncertainty in the predictions of the resummation formalism comes from the unknown
nonperturbative contributions, which in the $ b $-space formalism are included
in the nonperturbative Sudakov factor $ S^{NP}(b) $. I have presented simple
parametrizations of $ S^{NP}(b) $ for the transverse energy flow (cf. Eqs.\,(\ref{SNPzpar1},\ref{SNPzpar2}))
and charged particle multiplicity (cf. Eq.\,(\ref{SNPchgd})). These parametrizations
were found by fitting the resummed energy flow and charged particle multiplicity
to the data from Refs.\,\cite{H1z1,H1z2} and Ref.\,\cite{ZEUSchgd96}, respectively.
Experimental measurements outside the range of those data will make it possible
to further improve these parametrizations and, hence, the accuracy of the resummation
formalism. 

The most straightforward way to study $ S^{NP}(b) $ is by measuring the variation
of the $ q_{T} $ spectra of physical quantities due to variations of one
kinematical variable, with other variables fixed or varying only in a small
range. For the energy flow, it would be beneficial to obtain more data at $ x>10^{-2} $,
where the predictions of the resummation formalism can be tested more reliably,
without potential uncertainties due to the small-$ x $ physics. Another interesting
question is the dependence of the nonperturbative Sudakov factor on the virtuality
$ Q $ of the vector boson. This dependence can be tested by studying the
$ q_{T} $ spectra in a range of $ Q $ with sufficient experimental resolution
in the current fragmentation region. Finally, to study effects of multiple parton
radiation on semi-inclusive production of individual hadrons, it will be interesting
to see the $ q_{T} $ spectra for particle production multiplicities with
the additional constraint $ z>0.05\sim 0.1 $, \textit{i.e.,} in the kinematical
region where the parametrizations of the fragmentation functions are known reasonably
well.

\chapter{Azimuthal asymmetries of SIDIS observables\label{ch:AzimuthalAsymmetries}}

In a recent publication \cite{ZEUSasym} the ZEUS Collaboration at DESY-HERA
has presented data on asymmetries of charged particle ($ h^{\pm } $) production
in the process $ e+p\stackrel{\gamma ^{*}}{\longrightarrow }e+h^{\pm }+X $,
with respect to the angle $ \varphi  $ defined as the angle between the lepton
scattering plane and the hadron production plane (of $ h^{\pm } $ and the
exchanged virtual photon). This angle is shown in Figure\,\ref{fig:hCM}.
The azimuthal asymmetries, $ \langle \cos \varphi \rangle  $
and $ \langle \cos {2\varphi }\rangle  $, as functions of the minimal transverse
momentum $ p_{c} $ of the observed charged hadron $ h^{\pm } $ in the
hadron-photon center-of-mass (hCM) frame, are defined as \begin{equation}
\label{chgdasym}
\langle \cos n\varphi \rangle (p_{c})=\frac{\int d\Phi \int _{0}^{2\pi }d\varphi \cos n\varphi \frac{d\sigma }{dxdzdQ^{2}dp_{T}d\varphi }}{\int d\Phi \int _{0}^{2\pi }d\varphi \frac{d\sigma }{dxdzdQ^{2}dp_{T}d\varphi }},
\end{equation}
 with $ n=1,2 $. In terms of the momenta of the initial proton $ p_{A}^{\mu } $,
the final-state hadron $ p_{B}^{\mu } $, and the exchanged photon $ q^{\mu } $,
the variables in (\ref{chgdasym}) are $ Q^{2}=-q_{\mu }q^{\mu } $, $ x=Q^{2}/2(p_{A}\cdot q) $,
and $ z=(p_{A}\cdot p_{B})/(p_{A}\cdot q) $. $ \int d\Phi  $ denotes the
integral over $ x,z,Q^{2},p_{T} $ within the region defined by $ 0.01<x<0.1 $,
$ 180\mbox {\, GeV}^{2}<Q^{2}<7220\mbox {\, GeV}^{2} $, $ 0.2<z<1,\, \mbox {and\, }p_{T}>p_{c} $.
Nonzero $ \langle {\cos {2\varphi }}\rangle  $ comes from interference of
the helicity $ +1 $ and $ -1 $ amplitudes of the transverse photon polarization;
and nonzero $ \langle {\cos \varphi }\rangle  $ comes from interference of
transverse and longitudinal photon polarization.

More than 20 years ago it was proposed to test QCD by comparing measured azimuthal
asymmetries to the perturbative predictions \cite{GeorgiPolitzer}. However,
it was also realized that nonperturbative contributions and higher-twist effects
may affect the comparison \cite{LeveltMulders94,KKJK,ChgdNP,BoerMulders98}.
For example, intrinsic $ k_{T} $ might be used to parametrize the nonperturbative
effects \cite{KKJK}, and indeed ZEUS did apply this idea to their analysis
of the data. The relative importance of the nonperturbative effects is expected
to decrease as $ p_{T} $ increases. Thus, the azimuthal asymmetries in semi-inclusive
deep-inelastic scattering (SIDIS) events with large $ p_{T} $ should be dominated
by perturbative dynamics. 

By comparing their data to the PQCD calculation at the leading order in $ \alpha _{S} $
\cite{Gehrmann,KoppMendez,Mendez78}, the ZEUS Collaboration concluded that the magnitude
of the measured asymmetry $ \langle \cos \varphi \rangle  $ exceeds the theoretical
prediction for $ p_{c}<1\mbox {\, GeV} $, and $ \langle \cos 2\varphi \rangle  $
is systematically above the theoretical prediction for $ p_{c}>1.25 $ GeV.
ZEUS also estimated the possible nonperturbative contribution, by introducing
a transverse momentum $ k_{T} $ of the initial-state parton in the proton,
and similarly of the final-state hadron due to nonperturbative fragmentation.
It was found that this nonperturbative contribution is negligible for $ \langle \cos 2{\varphi }\rangle  $.
For $ \langle \cos {\varphi }\rangle  $, the nonperturbative contribution
can be sizable (up to 20\%), but it is not large enough to account for the difference
between the data and the $ {\mathcal{O}}(\alpha _{S}) $ calculation at low
$ p_{c} $. Hence, it was suggested that the discrepancy in $ \langle \cos {\varphi }\rangle  $
may be caused by large higher-order corrections.

From the comparison to the PQCD calculation at the leading order in $ \alpha _{S} $
\cite{Gehrmann,KoppMendez}, the ZEUS Collaboration concluded that the data
on the azimuthal asymmetries at large values of $ p_{c} $, although not well
described by the QCD predictions, do provide clear evidence for a PQCD contribution
to the azimuthal asymmetries. In this Chapter, the ZEUS data is discussed in
a framework of QCD resummation formalism \cite{CS81,CSS85,Collins93,Meng2,nsy1999,nsy2000,nsyasym}
that takes into account the effects of multiple soft parton emission. The discussion
targets two objectives. First, it is shown that the analysis of $ \langle \cos \varphi \rangle  $
and $ \langle \cos {2\varphi }\rangle  $ based on fixed-order QCD is unsatisfactory
because it ignores large logarithmic corrections due to soft parton emission.
In addition, perturbative and nonperturbative contributions are mixed in the
transverse momentum distributions, so that the presented data does not clarify
the dynamical mechanism that generates the observed asymmetries. Second, I
make two suggestions for improvement of the analysis of the ZEUS data. I show
that perturbative and nonperturbative contributions can be separated more clearly
in asymmetries depending on a variable $ q_{T} $ related to the pseudorapidity
of the final hadron in the hCM frame. I also suggest to measure the asymmetries
of the \emph{transverse energy flow} that are simpler and may be calculated
reliably. I present numerical predictions for the asymmetries of transverse
energy flow. These predictions are the most important result in this Chapter.

\section{Large logarithmic corrections and resummation}

The resummation formalism applied here was discussed in Chapter \ref{ch:Formalism}.
It describes production of nearly massless hadrons in the current fragmentation
region, where the production rate is the highest. In this region, transverse
momentum distributions are affected by large logarithmic QCD corrections due
to radiation of soft and collinear partons. The leading logarithmic contributions
can be summed through all orders of PQCD \cite{Collins93,Meng2,nsy1999,nsy2000}
by applying a method originally proposed for jet production in $ e^{+}e^{-} $
annihilation \cite{CS81} and vector boson production at hadron-hadron colliders
\cite{CSS85}. 

According to Eq.\,(\ref{ang1}), the spin-averaged cross section for
SIDIS in a parity-conserving channel, \textit{e.g.,} $ \gamma ^{*} $ exchange,
can be decomposed into a sum of independent contributions from four basis functions
$ A_{\rho }(\psi ,\varphi ) $ of the leptonic angular parameters $ \psi ,\varphi  $
\cite{Meng1}: \[
\frac{d\sigma }{dxdzdQ^{2}dq_{T}^{2}d\varphi }=\sum _{\rho =1}^{4}\prescr{\rho }{V}(x,z,Q^{2},q_{T}^{2})A_{\rho }(\psi ,\varphi ).\]
 Here $ \psi  $ is the angle of a hyperbolic rotation (a boost) in Minkowski
space; it is related to the conventional DIS variable $ y $, by $ y=Q^{2}/xS_{eA}=2/(1+\cosh \psi ) $.
The angular basis functions are $ A_{1}=1+\cosh ^{2}\psi  $, $ A_{2}=-2 $,
$ A_{3}=-\cos \varphi \sinh 2\psi  $, $ A_{4}=\cos 2\varphi \sinh ^{2}\psi  $.
Of the four structure functions $ \prescr{\rho }V $, only $ \prescr{1}V $
and $ \prescr{2}V $ contribute to the denominator of (\ref{chgdasym}), \textit{i.e.,}
the $ \varphi  $-integrated cross section. Of these two terms, $ \prescr{1}V $
is more singular, and it dominates the rate. According to the
discussion in Chapter\,\ref{ch:Formalism}, the singular contributions
in $ \prescr{1}V $ can be conveniently explored by  introducing a
scale $ q_{T} $ related to the \textit{polar} angle ($ \theta
_{B,cm}$) of the direction of  the final hadron (B) in the
hCM frame:
\begin{equation}
q_{T}=Q\sqrt{1/x-1}\, \exp {(-\eta_{cm})},
\end{equation}
 where $ \eta_{cm} $ is the pseudorapidity of the charged hadron in the
hCM frame. In the limit $ q_{T}\rightarrow 0 $, the structure
function $ \prescr{1}V $ is dominated by large logarithmic terms; it has
the form $ q_{T}^{-2}\sum _{k=1}^{\infty }(\alpha _{S}/\pi )^{k}\sum _{m=0}^{2k-1}v^{(km)}\ln ^{m}(q_{T}^{2}/Q^{2}) $,
where $ v^{(km)} $ are some generalized functions. To obtain a stable theoretical
prediction, these large terms must be resummed through all orders of PQCD. The
other structure functions $ ^{2,3,4}V $ are finite at this order; they will
be approximated by fixed-order $ {\cal O}(\alpha _{S}) $ expressions. 

In Eq.\,(\ref{chgdasym}), the numerator of $ \langle \cos {\varphi }\rangle  $
or $ \langle \cos {2\varphi }\rangle  $ depends only on the structure function
$ \prescr{3}V $ or $ \prescr{4}V $, respectively. The measurement of $ \langle \cos {\varphi }\rangle  $
or $ \langle \cos {2\varphi }\rangle  $ must be combined with good knowledge
of the $ \varphi  $-integrated cross section, \textit{i.e.,} the denominator
of (\ref{chgdasym}), to provide experimental information on the structure function
$ \prescr{3}V $ or $ \prescr{4}V $. Thus it is crucial to check whether
the theory can reproduce the $ \varphi  $-integrated cross section as a function
of $ p_{T} $ before comparing the prediction for (\ref{chgdasym}) to the
data. But Figure~\ref{fig:dsigdpt} shows that the $ {\mathcal{O}}(\alpha _{S}) $
fixed-order cross section is significantly lower than the data from \cite{ZEUSchgd96}
in the range of $ p_{T} $ relevant to the ZEUS measurements. This difference
signals the importance of higher-order corrections and undermines the validity
of the $ {\mathcal{O}}(\alpha _{S}) $ result as a reliable approximation
for the numerator of Eq.\,(\ref{chgdasym}). 

On the other hand, the resummation calculation with a proper choice of the nonperturbative
function yields a much better agreement with the experimental data for the $ p_{T} $-distribution
from \cite{ZEUSchgd96}. One might try to improve the theoretical description
of the ZEUS data using resummation for the denominator of Eq.\,(\ref{chgdasym}).
However, the resummation calculation for $ d\sigma /(dxdzdQ^{2}dq_{T}^{2}d\varphi ) $
in the phase space region relevant to the ZEUS data is currently not possible,
largely because of the uncertainty in the parametrization of the nonperturbative
contributions in this region. The impact parameter ($ b $-space) resummation
formalism \cite{Collins93,Meng2} includes a nonperturbative Sudakov factor
, which contains the effects of the intrinsic transverse momentum of the initial-state
parton and the nonperturbative fragmentation contributions to the transverse
momentum of the final-state hadron (cf. Eq.\,(\ref{SPNP})). Without first determining
this nonperturbative factor, \textit{e.g.,} from other measurements, it is not
possible to make a trustworthy theoretical prediction for the denominator of
Eq.\,(\ref{chgdasym}) and, hence, these azimuthal asymmetries.

\begin{figure}[H]
{\par\centering \resizebox*{!}{9cm}{\includegraphics{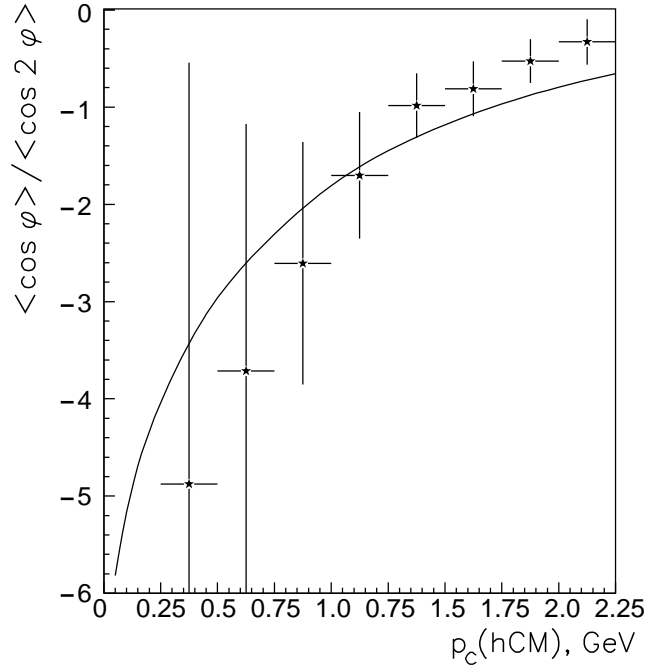}} \par}

\caption{\label{fig:cosphi2cos2phi}Comparison of the \protect$ {\mathcal{O}}(\alpha _{S})\protect $
prediction for the ratio \protect$ \langle \cos \varphi \rangle /\langle \cos {2\varphi }\rangle \protect $
with the ratio of experimentally measured values of \protect$ \langle \cos \varphi \rangle \protect $
and \protect$ \langle \cos {2\varphi }\rangle \protect $ from \protect\cite{ZEUSasym}.
The error bars are calculated by adding the statistical errors of \protect$ \langle \cos \varphi \rangle \protect $
and \protect$ \langle \cos {2\varphi }\rangle \protect $ in quadrature. Systematic
errors are not included. The theoretical curve is calculated for \protect$ \langle x\rangle =0.022\protect $,
\protect$ \langle Q^{2}\rangle =750\mbox {\, GeV}^{2}\protect $, using the
CTEQ5M1 parton distribution functions \protect\cite{CTEQ5} and fragmentation
functions by S.~Kretzer from \protect\cite{Kretzer}.}
\end{figure}

The azimuthal asymmetries measured by ZEUS may also be sensitive to uncertainties
in the fragmentation to $ h^{\pm } $ in the final state. Indeed, the cross
section in Eq.\,(\ref{chgdasym}) includes convolutions of hard scattering cross
sections with fragmentation functions (FFs), integrated over the range $ 0.2<z<1 $.
Although the knowledge of FFs is steadily improving \cite{BKK,Kretzer,Bourhis},
there is still some uncertainty about their $ z $-dependence and flavor structure
for the range of $ Q $ relevant to the ZEUS measurement. Therefore the most
reliable tests of the theory would use observables that are not sensitive to
the final-state fragmentation. The asymmetries $ \langle \cos n\varphi \rangle  $
would be insensitive to FFs if the dependence on the partonic variable $ \widehat{z} $
were similar in the hard parts of the numerator and denominator of Eq.\,(\ref{chgdasym}),
so that the dependence on the FFs would approximately cancel. It is shown in
Appendix~\ref{AppendixPertXSec} that the partonic structure function $ \prescr{1}\Vh  $,
which dominates the denominator of (\ref{chgdasym}), contains terms proportional
to $ 1/\wh{z}^{2} $ that increase rapidly as $ \wh{z} $ decreases. However,
the most singular terms in the partonic structure functions $ \prescr{3,4}\Vh  $
are proportional to $ 1/\wh{z} $. Therefore, the dependence on the FFs does
not cancel in the azimuthal asymmetries.

A curious fact appears to support the suggestion that the theoretical predictions
for $ \langle \cos n\varphi \rangle  $ depend significantly on the fragmentation
functions. While each of the measured asymmetries, $ \langle \cos \varphi \rangle  $
and $ \langle \cos 2\varphi \rangle  $, deviates from the $ {\mathcal{O}}(\alpha _{S}) $
prediction, the data actually agree well with the $ {\mathcal{O}}(\alpha _{S}) $
prediction for the ratio $ \langle \cos \varphi \rangle /\langle \cos 2\varphi \rangle  $,
as shown in Fig.\,\ref{fig:cosphi2cos2phi}. The error bars are the statistical
errors on $ \langle {\cos \varphi }\rangle  $ and $ \langle \cos {2\varphi }\rangle  $
combined in quadrature; this may overestimate the experimental uncertainty if
the two errors are correlated. Since this ratio depends only on the numerators
in Eq.\,(\ref{chgdasym}), which are less singular with respect to $ \wh{z} $
than the denominator, the dependence on the fragmentation functions may be nearly
canceled in the ratio. The good agreement between the $ {\mathcal{O}}(\alpha _{S}) $
prediction and the experimental data for this ratio supports our conjecture
that the fragmentation dynamics has a significant impact on the individual asymmetries
defined in Eq.\,(\ref{chgdasym}). 

The final remark about the azimuthal asymmetries in Eq.\,(\ref{chgdasym}) is that
the $ p_{T} $ (or $ p_{c} $) distributions are not the best observables
to separate the perturbative and nonperturbative effects. The region where multiple
parton radiation effects are important is specified by the condition $ q_{T}^{2}/Q^{2}\ll 1 $.
But the $ p_{T} $ distributions are smeared with respect to the $ q_{T} $
distributions by an additional factor of $ z $, because $ p_{T}=z\, \, q_{T} $.
Thus the whole observable range of $ p_{T} $ is sensitive to the resummation
effects in the region of $ q_{T} $ of the order of several GeV. A better
way to compare the data to the PQCD prediction is to express the azimuthal asymmetries
as a function of $ q_{T} $, not $ p_{T} $. Then the comparison should
be made in the region where the multiple parton radiation is unimportant, \textit{i.e.,}
for $ q_{T}/Q\gtrsim 1 $.

\section{Asymmetry of energy flow}

Next, I describe an alternative test of PQCD, which is less sensitive to the
above theoretical uncertainties: measurement of the azimuthal asymmetries of
the \emph{transverse energy flow}. In the hCM frame, the transverse energy flow
can be written as \cite{zflowdef,Meng1,Meng2,nsy1999,nsy2000}\begin{equation}
\label{angET}
\frac{dE_{T}}{dxdQ^{2}dq_{T}^{2}d\varphi }=\sum _{\rho =1}^{4}\prescr{\rho }V_{E_{T}}(x,Q^{2},q_{T}^{2})A_{\rho }(\psi ,\varphi ).
\end{equation}

Unlike the charged particle multiplicity, the energy flow does not depend on
the final-state fragmentation. According to the results in Chapter \ref{ch:Phenomenology},
a resummation calculation can provide a good description for the experimental
data on the $ \varphi  $-integrated $ E_{T} $-flow. A new class of azimuthal
asymmetries may be defined as \begin{eqnarray}
\langle E_{T}\cos n\varphi \rangle (q_{T})=\frac{\int d\Phi \int _{0}^{2\pi }\cos n\varphi \frac{dE_{T}}{dxdQ^{2}dq_{T}^{2}d\varphi }d\varphi }{\int d\Phi \int _{0}^{2\pi }\frac{dE_{T}}{dxdQ^{2}dq_{T}^{2}d\varphi }d\varphi }.\label{ETasym} 
\end{eqnarray}
 The structure functions $ \prescr{\rho }V_{E_{T}} $ for the $ E_{T} $-flow
can be derived from the structure functions $ \prescr{\rho }V $ for the SIDIS
cross section using Eq.\,(\ref{Vcsz}). Similar to the case of the particle
multiplicities, the asymmetries $ \langle E_{T}\cos \varphi \rangle  $ and
$ \langle E_{T}\cos 2\varphi \rangle  $ receive contributions from $ \prescr{3}V_{E_{T}} $
and $ \prescr{4}V_{E_{T}} $, respectively. But, unlike the previous case,
the denominator in Eq.\,(\ref{ETasym}) is approximated well by the resummed $ E_{T} $-flow.
Thus these asymmetries can be calculated with greater confidence.

\begin{figure}[H]
{\par\centering \resizebox*{!}{0.37\textheight}{\includegraphics{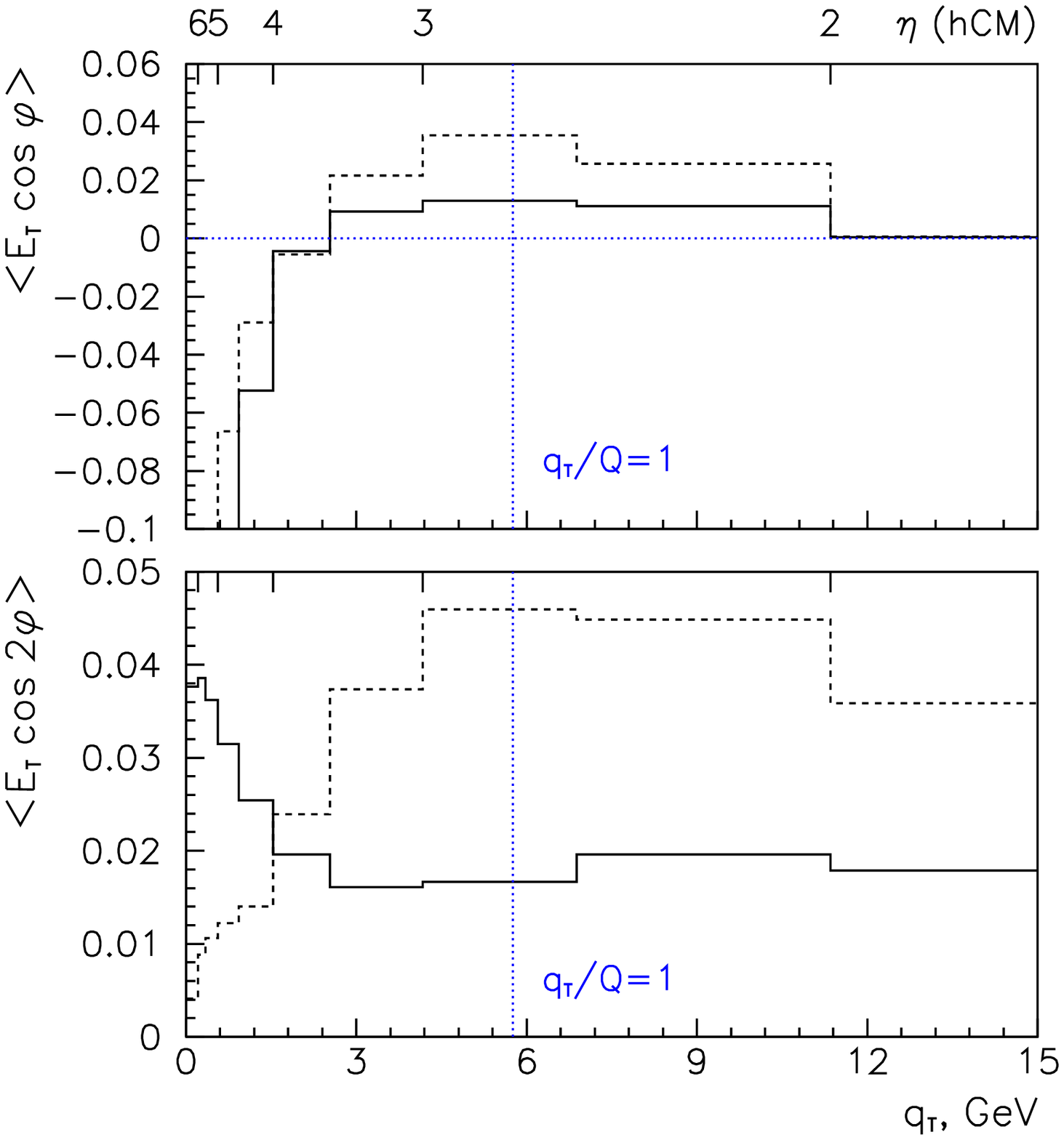}}  \\ a) \\ \par}

{\par\centering \resizebox*{!}{0.37\textheight}{\includegraphics{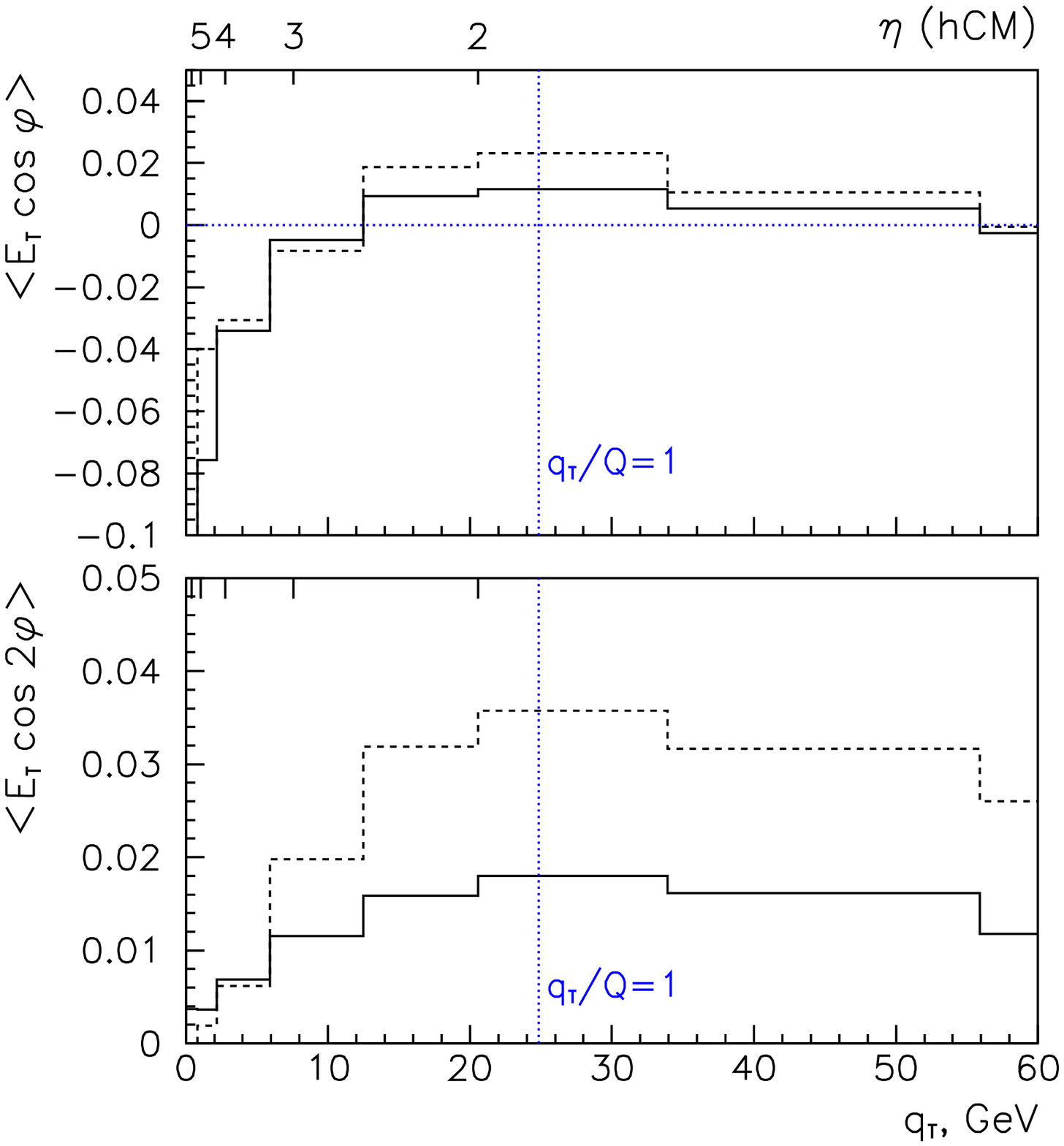}}  \\ b) \\\par}

\caption{\label{fig:ETasymqt} Energy flow asymmetries \protect$ \langle E_{T}\cos \varphi \rangle (q_{T})\protect $
and \protect$ \langle E_{T}\cos {2\varphi }\rangle (q_{T})\protect $ for
(a) \protect$ x=0.0047\protect $, \protect$ Q^{2}=33.2\mbox {\, GeV}^{2}\protect $
and (b) \protect$ x=0.026\protect $, \protect$ Q^{2}=617\mbox {\, GeV}^{2}\protect $.
The Figure shows predictions from the resummed (solid) and the \protect$ {\mathcal{O}}(\alpha _{S})\protect $
(dashed) calculations. }
\end{figure}

Figure \ref{fig:ETasymqt} shows the prediction for the azimuthal asymmetries
$ \langle E_{T}\cos \varphi \rangle  $ and $ \langle E_{T}\cos 2\varphi \rangle  $
as functions of $ q_{T} $ for (a) $ x=0.0047 $, $ Q^{2}=33.2\mbox {\, GeV}^{2} $
in the left plots and (b) $ x=0.026 $, $ Q^{2}=617\mbox {\, GeV}^{2} $
in the right plots. The asymmetries are shown in $ q_{T} $-bins that are
obtained from the experimental pseudorapidity bins for the $ \varphi  $-integrated
$ E_{T} $-flow data from Ref.~\cite{H1z2}. The upper $ x $-axis shows
values of the hCM pseudorapidity $ \eta_{cm} $ that correspond to the values
of $ q_{T} $ on the lower $ x $-axis. For each of the distributions in
Fig.\,\ref{fig:ETasymqt}, the structure functions $ \prescr{3}V_{E_{T}} $
and $ \prescr{4}V_{E_{T}} $ were calculated at leading order in QCD, \textit{i.e.,}
$ {\mathcal{O}}(\alpha _{S}) $. The solid and dashed curves, which correspond
to the resummed and $ {\mathcal{O}}(\alpha _{S}) $ results respectively,
differ because the structure function $ \prescr{1}V_{E_{T}} $ in the denominator
of (\ref{ETasym}) differs for the two calculations. The resummed $ \varphi  $-integrated
$ E_{T} $-flow is closer to the data than the fixed-order result, so that
the predictions made by PQCD for the subleading structure functions $ \prescr{3}V_{E_{T}} $
and $ \prescr{4}V_{E_{T}} $ will be confirmed if the experimental azimuthal
asymmetries agree with the resummed distributions.

The discussion in Section\,\ref{sec:ZFlows} shows that in the region $ q_{T}\sim Q $
the resummed $ \varphi  $-integrated $ E_{T} $-flow is larger than the
$ {\mathcal{O}}(\alpha _{S}) $ prediction. This explains why the asymmetries
for $ q_{T}\sim Q $ are smaller for the resummed denominator than for the
$ {\mathcal{O}}(\alpha _{S}) $ denominator. In the region $ q_{T}/Q\ll 1 $,
the asymmetries are determined by the asymptotic behavior of the fixed-order
and resummed \emph{partonic} structure functions $ \prescr{\rho }\Vh _{E_{T}} $.
As $ q_{T}\rightarrow 0 $, the $ {\mathcal{O}}(\alpha _{S}) $ structure
functions $ (\prescr{1}\Vh _{E_{T}})_{{\mathcal{O}}(\alpha _{S})} $, $ \prescr{3}\Vh _{E_{T}} $,
and $ \prescr{4}\Vh _{E_{T}} $ behave as $ 1/q_{T}^{2} $, $ 1/q_{T} $
and $ 1 $, respectively. Thus, asymptotically, the ratios $ \prescr{3,4}\Vh _{E_{T}}/(\prescr{1}\Vh _{E_{T}})_{{\mathcal{O}}(\alpha _{S})} $
go to zero, although the $ q_{T} $ distribution for the asymmetry $ \langle E_{T}\cos \varphi \rangle  $
is quite large and negative for small, but non-vanishing $ q_{T} $ (\textit{\emph{cf.}}
Fig.\,\ref{fig:ETasymqt}). Resummation of $ \prescr{1}\Vh _{E_{T}} $ changes
the $ q_{T} $-dependence of the denominator, which becomes nonsingular in
the limit $ q_{T}\rightarrow 0 $. Consequently, the asymmetry $ \langle E_{T}\cos \varphi \rangle  $
with the resummed denominator asymptotically grows as $ 1/q_{T} $ (\textit{i.e.,}
in accordance with the asymptotic behavior of $ \prescr{3}\Vh _{E_{T}} $).
Hence neither the fixed-order nor the resummed calculation for $ \langle E_{T}\cos \varphi \rangle  $
is reliable in the low-$ q_{T} $ region, so that higher-order or additional
nonperturbative contributions must be important at $ q_{T}\rightarrow 0 $.
The asymptotic limit for the resummed $ \langle E_{T}\cos 2\varphi \rangle  $
remains finite, with the magnitude shown in Fig.\,\ref{fig:ETasymqt}. Since
the magnitude of $ \langle E_{T}\cos 2\varphi \rangle  $ is predicted not
to exceed a few percent, an experimental observation of a large asymmetry $ \langle E_{T}\cos 2\varphi \rangle  $
at small $ q_{T} $ would signal the presence of some new hadronic dynamics,
\textit{e.g.,} contributions from $ T $-odd structure functions discussed
in \cite{BoerMulders98}.

Figure \ref{fig:ETasymqt} shows that the predicted asymmetry $ \langle E_{T}\cos \varphi \rangle (q_{T}) $
at $ q_{T}/Q=1 $ is about 1--2\% for the resummed denominator, while it is
about 2--4\% for the $ {\mathcal{O}}(\alpha _{S}) $ denominator. The asymmetry
$ \langle E_{T}\cos 2\varphi \rangle (q_{T}) $ at $ q_{T}/Q=1 $ is about
1.5-2\% or 3.5-5\%, respectively. Both asymmetries are positive for $ q_{T}\sim Q $.
According to Fig.\,\ref{fig:ETasymqt}a, the size of the experimental $ q_{T} $
bins (converted from the $ \eta  $ bins in \cite{H1z2}) for low or intermediate
values of $ Q^{2} $ is small enough to reveal the low-$ q_{T} $ behavior
of $ \prescr{3,4}V_{E_{T}} $ with acceptable accuracy. However, for the high-$ Q^{2} $
events in Fig.\,\ref{fig:ETasymqt}b, the experimental resolution in $ q_{T} $
may be insufficient for detailed studies in the low-$ q_{T} $ region. Nonetheless,
it will still be interesting to compare the experimental data to the predictions
of PQCD in the region $ q_{T}/Q\approx 1 $, and to learn about the angular
asymmetries at large values of $ Q^{2} $ and $ x $.

To conclude, the azimuthal asymmetry of the energy flow should be measured as
a function of the scale $ q_{T} $. These measurements would test the predictions
of the PQCD theory more reliably than the measurements of the asymmetries of
the charged particle multiplicity.

\newpage

\newpage

\appendix

\chapter{\label{AppendixPertXSec}The perturbative cross-section,
finite piece and \protect$ z\protect $-flow
distribution}

In this Appendix, I collect the formulas for the NLO parton level cross-sections
$ d\widehat{\sigma
}_{ba}/(d\widehat{x}d\widehat{z}dQ^{2}dq_{T}^{2}d\varphi ) $, which
were originally obtained in \cite{Mendez78}.

According to Eqs.\,(\ref{hadcs},\ref{ang2}), the hadron level cross-section 
$ d\sigma _{BA}/(dxdzdQ^{2}dq_{T}^{2}d\varphi ) $
is related to the parton-level structure functions
$\prescr{\rho}{\Vh}_{ba}$ as  
\begin{eqnarray}
&&\frac{d\sigma _{BA}}{dxdzdQ^{2}dq_{T}^{2}d\varphi }=
\sum_{\rho=1}^4 A_{\rho}(\psi,\varphi)\times \nonumber\\
&\times& \sum _{a,b}\int
_{z}^{1}\frac{d\xi _{b}}{\xi _{b}}D_{B/b}(\xi _{b},\mu)\int
_{x}^{1}\frac{d\xi _{a}}{\xi _{a}}F_{a/A}(\xi
_{a},\mu)\prescr{\rho}{\Vh}_{ba}(\xh,\zh,Q^2,q_T^2,\mu).
\label{NLOXSec}
\end{eqnarray}

At non-zero $ q_{T} $, the parton cross-section receives the contribution
from the real emission diagrams (Fig.~\ref{fig:diags}e-f), so
that  $\prescr{\rho}{\Vh}_{ba}$ can be expressed as
\begin{equation}
\prescr{\rho}{\Vh}_{ba} =  
\frac{\sigma _{0}F_{l}}{4\pi S_{eA}}\frac{\alpha _{S}}{\pi }
\delta \Biggl [\frac{q_{T}^{2}}{Q^{2}}-
\Bigl (\frac{1}{\widehat{x}}-1\Bigr )
\Bigl (\frac{1}{\widehat{z}}-1\Bigr )\Biggr ]
\prescr{\rho }
f_{ba}(\widehat{x},\widehat{z},Q^{2},q_{T}^{2}), 
\label{sighat} 
\end{equation}
with the same notations as in Chapter \ref{ch:Formalism}. In this Equation,
the $q\bar q$ structure functions \nolinebreak  are
\begin{equation}
{}^{\rho }f_{jk}(\widehat{x},\widehat{z},Q^{2},q_{T}^{2})\equiv
2 C_{F}\widehat{x}\widehat{z}e_{j}^{2}\delta _{jk}\,\, {}^{\rho }\bar{f}_{jk},
\end{equation}
where
\begin{eqnarray}
{}^{1}\bar{f}_{jk} & = & \frac{1}{Q^{2}q_{T}^{2}}\biggl (\frac{Q^{4}}{\widehat{x}^{2}\widehat{z}^{2}}+(Q^{2}-q_{T}^{2})^{2}\biggr )+6;\nonumber \\
{}^{2}\bar{f}_{jk} & = & 2 \,\, ({}^{4}\bar{f}_{jk})=4;\nonumber \\
{}^{3}\bar{f}_{jk} & = & \frac{2}{Qq_{T}}\left( Q^{2}+q_{T}^{2}\right).
\end{eqnarray}
The $\stackrel{(-)}{q}g$ structure functions are
\begin{equation}
\prescr{\rho }f_{jg}(\widehat{x},\widehat{z},Q^{2},q_{T}^{2})\equiv
\widehat{x}(1-\widehat{x})e_{j}^{2}\,\, \prescr{\rho }\bar{f}_{jg},
\end{equation}
where
\begin{eqnarray}
\prescr{1}\bar{f}_{jg} & = & \frac{Q^{2}}{q_{T}^{2}}\biggl
(\frac{1}{\widehat{x}^{2}\widehat{z}^{2}}-\frac{2}{\widehat{x}\widehat{z}}+2\biggr
)+
10-\frac{2}{\widehat{x}}-\frac{2}{\widehat{z}};\nonumber \\
\prescr{2}\bar{f}_{jg} & = & 2\,\, (\prescr{4}\bar{f}_{jg})=8;\nonumber \\
\prescr{3}\bar{f}_{jg} & = & \frac{2}{Qq_{T}}\left( 2(Q^{2}+q_{T}^{2})-\frac{Q^{2}}{\widehat{x}\widehat{z}}\right).
\end{eqnarray}
Finally, the $g\stackrel{(-)}{q}$ structure functions are
\begin{equation}
\prescr{\rho }f_{gj}(\widehat{x},\widehat{z},Q^{2},q_{T}^{2})\equiv
2C_{F}\widehat{x}(1-\widehat{z})e_{j}^{2}\,\, \prescr{\rho }\bar{f}_{gj},
\end{equation}
where\begin{eqnarray}
\prescr{1}\bar{f}_{gj} & = & \frac{1}{Q^{2}\tilde{q}_{T}^{2}}\biggl
(\frac{Q^{4}}{\widehat{x}^{2}(1-\widehat{z})^{2}}+(Q^{2}-\tilde{q}_{T}^{2})^{2}\biggr
)+6;\nonumber \\
\prescr{2}\bar{f}_{gj} & = & 2\,\,
(\prescr{4}\bar{f}_{gj})=4;\nonumber \\
\prescr{3}\bar{f}_{gj} & = & \frac{2}{Q\tilde{q}_{T}}\left( Q^{2}+\tilde{q}_{T}^{2}\right) \label{111}.
\end{eqnarray}
 In (\ref{111}), \begin{equation}
\tilde{q}_{T}\equiv \frac{\widehat{z}q_{T}}{1-\widehat{z}}.
\end{equation}
 The indices $ j $ and $ k $ correspond to a quark (antiquark) of a type
$ j $ or $ k $, the index $ g $ corresponds to a gluon.

The finite part $Y_{BA}$ of the ${\cal O}(\alpha_S)$ cross section
(\ref{NLOXSec}) is
\begin{eqnarray}
&&Y_{BA}=
\frac{\sigma _{0}F_{l}}{4\pi S_{eA}}
\frac{\alpha _{S}}{\pi }
\sum_{\rho=1}^4 A_{\rho}(\psi,\varphi) \sum _{a,b}\int
_{z}^{1}\frac{d\xi _{b}}{\xi _{b}}D_{B/b}(\xi _{b},\mu)\int
_{x}^{1}\frac{d\xi _{a}}{\xi _{a}}F_{a/A}(\xi
_{a},\mu)\times\nonumber\\
&\times&
\prescr{\rho}{R}_{ba}(\xh,\zh,Q^2,q_T^2,\mu).
\label{YBA}
\end{eqnarray} 
For $\rho=1$, the functions $\prescr{\rho}{R_{ba}}$ are
\begin{eqnarray}
\prescr{1}R_{jk}(\widehat{x},\widehat{z},Q^{2},q_{T}^{2}) & = & \delta \left[ \frac{q_{T}^{2}}{Q^{2}}-\left( \frac{1}{\widehat{x}}-1\right) \left( \frac{1}{\widehat{z}}-1\right) \right] {}^{1}f_{jk}(\widehat{x},\widehat{z},Q^{2},q_{T}^{2})\nonumber \\
 & - & \frac{1}{q_{T}^{2}}\delta _{jk}e_j^2
\Biggl \{\delta (1-\wh{z})P^{(1)}_{qq}(\wh{x})+P^{(1)}_{qq}(\wh{z})
\delta (1-\wh{x})\nonumber \\
 & + & 2C_{F}\delta (1-\wh{z})\delta (1-\wh{x})\left( \log \frac{Q^{2}}{q_{T}^{2}}-\frac{3}{2}\right) \Biggr \};\\
\prescr{1}R_{jg}(\widehat{x},\widehat{z},Q^{2},q_{T}^{2}) & = & \delta \left[ \frac{q_{T}^{2}}{Q^{2}}-\left( \frac{1}{\widehat{x}}-1\right) \left( \frac{1}{\widehat{z}}-1\right) \right] {}^{1}f_{jg}(\widehat{x},\widehat{z},Q^{2},q_{T}^{2})\nonumber \\
 & - & \frac{1}{q_{T}^{2}}e_j^2\delta (1-\wh{z})P_{qg}^{(1)}(\wh{x});
\end{eqnarray}
\begin{eqnarray}
\prescr{1}R_{gj}(\widehat{x},\widehat{z},Q^{2},q_{T}^{2}) & = & \delta \left[ \frac{q_{T}^{2}}{Q^{2}}-\left( \frac{1}{\widehat{x}}-1\right) \left( \frac{1}{\widehat{z}}-1\right) \right] {}^{1}f_{gj}(\widehat{x},\widehat{z},Q^{2},q_{T}^{2})\nonumber \\
 & - & \frac{1}{q_{T}^{2}}e_j^2 P_{gq}^{(1)}(\wh{z})\delta (1-\wh{x}).
\end{eqnarray}
For $\rho =2,3,4$,
\begin{equation}
\prescr{\rho }R_{ba}(\widehat{x},\widehat{z},Q^{2},q_{T}^{2})=\delta \left[ \frac{q_{T}^{2}}{Q^{2}}-\left( \frac{1}{\widehat{x}}-1\right) \left( \frac{1}{\widehat{z}}-1\right) \right] {}^{\rho }f_{ba}(\widehat{x},\widehat{z},Q^{2},q_{T}^{2}).
\end{equation}

From (\ref{hadcs}), it is possible to derive the perturbative $ z $-flow
distribution, \[
\frac{d\Sigma _{z}}{dxdQ^{2}dq_{T}^{2}d\varphi }\equiv \sum _{B}\int _{z_{min}}^{1}zdz\frac{d\sigma _{BA}}{dxdzdQ^{2}dq_{T}^{2}d\varphi }=\]
\begin{equation}
\label{zx}
\frac{\sigma _{0}F_{l}}{4\pi S_{eA}}\frac{\alpha _{S}}{\pi }\sum _{a,b}\sum _{j}e^{2}_{j}\int _{x}^{1}\frac{d\xi _{a}}{\xi _{a}-x}F_{a/A}(\xi _{a})\widehat{z}^{3}\widehat{x}\sum _{\rho =1}^{4}\prescr{\rho }f_{ba}(\widehat{x},\widehat{z},Q^{2},q_{T}^{2})A_{\rho }(\psi ,\varphi ).
\end{equation}
 It depends on the same functions $ \prescr{\rho }f_{ba}(\widehat{x},\widehat{z},Q^{2},q_{T}^{2}) $,
with the parton variable $ \widehat{z} $ determined by the $ \delta  $-function
in (\ref{sighat}),\begin{equation}
\wh{z}=\frac{1-\wh{x}}{\left( q_{T}^{2}/Q^{2}-1\right) \wh{x}+1}.
\end{equation}

\chapter{\label{ch:resumOas}\protect$ \Oas \protect $ part of the resummed cross
section}

In this Appendix, I demonstrate that the $ {\cal O}(\alpha _{S}) $ part of
the $ \wt{W} $-term in the resummed cross section (\ref{resum}) coincides
with the small-$ q_{T} $ approximation of the factorized $ \Oas  $ fixed-order
cross section. Correspondingly the complete $ \Oas  $ part does not depend
on the scales $ C_{1}/b $ and $ C_{2}Q $ separating collinear-soft and collinear
contributions to the resummed cross section.  

At $ q_{T}\ll Q $, the resummed cross section (\ref{resum}) is dominated
by the $ \wt{W} $-term:
\begin{eqnarray}
\left. \frac{d\sigma _{BA}}{dxdzdQ^{2}dq_{T}^{2}d\varphi }\right| _{resum} & \approx  & \sFs \frac{A_{1}(\psi ,\varphi )}{2}\int \frac{d^{2}b}{(2\pi )^{2}}e^{i\vec{q}_{T}\cdot \vec{b}}\widetilde{W}_{BA}(b)=\nonumber \\
 & = & \sFs \frac{A_{1}(\psi ,\varphi )}{2Q^{2}}\int \frac{d^{2}\beta }{(2\pi )^{2}}e^{i\frac{\vec{q}_{T}\cdot \vec{\beta }}{Q}}\widetilde{W}_{BA}(\frac{\beta }{Q}),\label{resum3} 
\end{eqnarray}
where\begin{equation}
\vec{\beta }\equiv Q\vec{b}.
\end{equation}
 According to Eq.\,(\ref{W}), at $ b\rightarrow 0 $\begin{equation}
\label{W3}
\widetilde{W}_{BA}(b,Q,x,z)=\sum _{j}e_{j}^{2}(D_{B/b}\otimes {\cal C}^{out}_{bj})(z,b)({\cal C}_{ja}^{in}\otimes F_{a/A})(x,b)e^{-S^{P}(b,Q)}.
\end{equation}
The perturbative Sudakov factor $ S^{P} $and the $ {\cal C}- $functions
in $ \wt{W}_{BA}(b,Q,x,z) $ can be expanded up to $ \Oas  $ using Eqs.\,(\ref{SudP},\ref{A1}-\ref{C1out2})\begin{eqnarray}
 &  & S^{P}(b,Q)\equiv \int _{C_{1}^{2}/b^{2}}^{C_{2}^{2}Q^{2}}\frac{d\ov \mu ^{2}}{\ov \mu ^{2}}\left( \ASud (\alpha _{S}(\ov \mu ),C_{1})\ln \frac{C_{2}^{2}Q^{2}}{\ov \mu ^{2}}+\BSud (\alpha _{S}(\ov \mu ),C_{1},C_{2})\right) \approx \nonumber \\
 &  & \approx \frac{\alpha _{S}}{\pi }\left( \frac{\ASud _{1}}{2}\ln ^{2}\frac{C^{2}_{2}Q^{2}b^{2}}{C^{2}_{1}}+\BSud _{1}(\alpha _{S}(\ov \mu ),C_{1},C_{2})\ln \frac{C^{2}_{2}Q^{2}b^{2}}{C^{2}_{1}}\right) +{\cal O}(\alpha _{S}^{2})\nonumber \\
 &  & \approx \frac{\alpha _{S}}{\pi }C_{F}\left( \frac{1}{2}\ln ^{2}\frac{C^{2}_{2}Q^{2}b^{2}}{C^{2}_{1}}-\frac{3}{2}\ln \frac{C^{2}_{2}Q^{2}b^{2}}{C^{2}_{1}}-2\ln \frac{C_{2}b_{0}}{C_{1}}\ln \frac{C^{2}_{2}Q^{2}b^{2}}{C^{2}_{1}}\right) ;\label{SudOas} 
\end{eqnarray}
\begin{eqnarray}
 &  & ({\cal C}_{ja}^{in}\otimes F_{a/A})(x,b,\mu )=F_{a/A}(x,\mu )\left( 1+\frac{\alpha _{S}}{\pi }C_{F}\left( -\ln ^{2}\frac{C_{1}}{C_{2}b_{0}}+\frac{3}{2}\ln \frac{C_{1}}{C_{2}b_{0}}\right) \right) \nonumber \\
 &  & +\frac{\alpha _{S}}{\pi }\left( (c^{(1)in}_{ja}\otimes F_{a/A})(x,\mu )-(\ln \frac{\mu b}{b_{0}}P_{ja}^{(1)}\otimes F_{a/A})(x,\mu )\right) ;\\
 &  & (D_{B/b}\otimes {\cal C}^{out}_{bj})(z,b,\mu )=D_{B/b}(z,\mu )\left( 1+\frac{\alpha _{S}}{\pi }C_{F}\left( -\ln ^{2}\frac{C_{1}}{C_{2}b_{0}}+\frac{3}{2}\ln \frac{C_{1}}{C_{2}b_{0}}\right) \right) \nonumber \\
 &  & +\frac{\alpha _{S}}{\pi }\left( (D_{B/b}\otimes c^{(1)out}_{bj})(z,\mu )-(D_{B/b}\otimes \ln \frac{\mu b}{\wh{z}b_{0}}P_{bj}^{(1)})(z,\mu )\right) ,\label{CoutOas} 
\end{eqnarray}
where the functions $ c_{ba}^{(1)in} $, $ c^{(1)out}_{ba} $ are defined
in Eqs.\,(\ref{Csdis0}-\ref{Csdis}). In these equations the running of
$ \alpha _{S} $, which is an effect of $ {\cal O}(\alpha _{S}^{2}) $,
is neglected. Inserting the $\Oas$ representations (\ref{SudOas}-\ref{CoutOas}) in Eq.\,(\ref{W3}),
we obtain the $ \Oas  $ expression for $ \wt{W}_{BA}(b,Q,x,z): $
\begin{eqnarray}
 &  & \left. \wt{W}_{BA}(\frac{\beta }{Q},Q,x,z)\right| _{\Oas }=
\sum _{j}e_{j}^{2}\times \nonumber \\
 &  & 
\Biggl \{
D_{B/j}(z,\mu )F_{j/A}(x,\mu )\left[ 1-\frac{\alpha _{S}}{\pi }C_{F}\left( \frac{1}{2}\ln ^{2}\frac{\beta ^{2}}{b_{0}^{2}}-\frac{3}{2}\ln \frac{\beta ^{2}}{b_{0}^{2}}\right) \right] \nonumber\\
 &  & +\frac{\alpha _{S}}{\pi }\Biggl [\left( (D_{B/b}\otimes
 c^{(1)out}_{bj})(z,\mu )-(D_{B/b}\otimes \ln \left[ \frac{\mu
 }{\wh{z}Q}\frac{\beta }{b_{0}}\right] P_{bj}^{(1)})(z,\mu )\right)
 F_{j/A}(x,\mu ) \nonumber
\eea
\bea
 &  & +D_{B/j}(z,\mu )\left( (c^{(1)in}_{ja}\otimes F_{a/A})(x,\mu
 )-(\ln \left[ \frac{\mu }{Q}\frac{\beta }{b_{0}}\right]
 P_{ja}^{(1)}\otimes F_{a/A})(x,\mu )\right) \Biggr ]\Biggr
 \}.\nonumber \\
&&\label{Wbeta} 
\end{eqnarray}
This expression does not depend on the constants $ C_{1},C_{2} $, so that
the only factorization scale in Eq.\,(\ref{Wbeta}) is $ \mu  $.
The Fourier-Bessel transform of $ \wt{W}_{BA}(b,Q,x,z) $ to the $ q_{T} $-space
can be realized by using relationships\begin{eqnarray}
\int \frac{d^{2}b}{(2\pi )^{2}}e^{-i\vec{q}_{T}\cdot \vec{b}} & = & \delta (\vec{q}_{T});\\
\int \frac{d^{2}b}{(2\pi )^{2}}e^{-i\vec{q}_{T}\cdot \vec{b}}\ln \frac{b^{2}}{b_{0}^{2}} & = & -\frac{1}{\pi }\left[ \frac{1}{q^{2}_{T}}\right] _{+};\\
\int \frac{d^{2}b}{(2\pi )^{2}}e^{-i\vec{q}_{T}\cdot \vec{b}}\ln ^{2}\frac{b^{2}}{b_{0}^{2}} & = & \frac{2}{\pi }\left[ \frac{\ln q_{T}^{2}}{q^{2}_{T}}\right] _{+},
\end{eqnarray}
where the ``+''-prescription is defined as\begin{equation}
\int d^{2}q_{T}\left[ f(q_{T})\right] _{+}g(\vec{q}_{T})=\int _{0}^{2\pi }d\varphi \int ^{+\infty }_{0}q_{T}dq_{T}f(q_{T})\left( g(\vec{q}_{T})-g(\vec{0})\right) .
\end{equation}
Hence the small-$ q_{T} $ approximation for the $ \Oas  $cross-section
is \begin{eqnarray}
\left. \frac{d\sigma _{BA}}{dxdzdQ^{2}dq_{T}^{2}d\varphi }\right| _{\Oas ,q_{T}\rightarrow 0} & = & \sFs \frac{A_{1}(\psi ,\varphi )}{2}\sum _{j}e_{j}^{2}\times \nonumber \\
 & \times  & \left( \delta (\vec{q}_{T})F_{\delta }(x,z,Q,\mu )+F_{+}(x,z,Q,q_{T},\mu )\right) ,\label{smallqtOas} 
\end{eqnarray}
where
\begin{eqnarray}
 &  & F_{\delta }(x,z,Q,\mu )=D_{B/j}(z,\mu )F_{j/A}(x,\mu )+\nonumber \\
 &  & 
\frac{\alpha _{S}}{\pi }\Biggl \{
\left( (D_{B/b}\otimes c^{(1)out}_{bj})(z,\mu )-(D_{B/b}\otimes \ln \left[ \frac{\mu }{\wh{z}Q}\right] P_{bj}^{(1)})(z,\mu )\right) F_{j/A}(x,\mu )\nonumber \\
 & + & D_{B/j}(z,\mu )\left( (c^{(1)in}_{ja}\otimes F_{a/A})(x,\mu
 )-(\ln \left[ \frac{\mu }{Q}\right] P_{ja}^{(1)}\otimes
 F_{a/A})(x,\mu )\right) \Biggr \};
\eea
\bea
 &  & F_{+}(x,z,Q,q_{T},\mu )=\frac{1}{2\pi Q^{2}}\frac{\alpha _{S}}{\pi }
\times
\nonumber \\
 &  & 
\Biggl \{
2C_{F}D_{B/j}(z,\mu )F_{j/A}(x,\mu )\left( \left[ \frac{Q^{2}}{q_{T}^{2}}\ln \frac{Q^{2}}{q_{T}^{2}}\right] _{+}-\frac{3}{2}\left[ \frac{Q^{2}}{q^{2}_{T}}\right] _{+}\right) +\left[ \frac{Q^{2}}{q_{T}^{2}}\right] _{+}\times \nonumber \\
 & \times  & \left( (D_{B/b}\otimes P_{bj}^{(1)})(z,\mu )F_{j/A}(x,\mu )+D_{B/j}(z.\mu )(P_{ja}^{(1)}\otimes F_{a/A})(x,\mu )\right) \Biggr \}.
\end{eqnarray}
The function $ F_{\delta }(x,z,\mu ) $ contributes at $ q_{T}=0 $; it
receives contributions from the leading order scattering, evolution of the PDFs
between the scales $ \mu  $ and $ Q $, evolution of the FFs between the
scales $ \mu  $ and $ \wh{z}Q $, and $ \Oas  $ coefficient functions
$ c^{(1)in}, $ $ c^{(1)out} $. The function $ F_{+}(x,z,Q,q_{T},\mu ) $
is just a regularized asymptotic part (\ref{NLL2}) of the $ \Oas  $ fixed-order
structure function $\prescr{1}V_{BA}$ (cf. Eq.\,(\ref{NLL2})). 

The $ \Oas  $ cross section (\ref{smallqtOas}) can be integrated over the
lowest bin of $ q_{T} $,\linebreak $ 0\leq q_{T}^{2}\leq \left( q_{T}^{S}\right) ^{2}\ll Q^{2} $.
The resulting integral is\begin{eqnarray}
 &  & \int _{0}^{(q_{T}^{S})^{2}}dq_{T}^{2}\left. \frac{d\sigma _{BA}}{dxdzdQ^{2}dq_{T}^{2}d\varphi }\right| _{\Oas ,q_{T}\rightarrow 0}=\sFs \frac{A_{1}(\psi ,\varphi )}{2\pi }\sum _{j}e_{j}^{2}\times \nonumber \\
 &  & \times \left( F_{\delta }(x,z,Q,\mu )+F^{\prime }_{+}(x,z,\frac{q_{T}^{S}}{Q},\mu )\right) .
\end{eqnarray}
where\begin{eqnarray}
 &  & F^{\prime }_{+}(x,z,\frac{q_{T}^{S}}{Q},\mu )=\nonumber\\
&-&\frac{\alpha _{S}}{2\pi }\Biggl \{D_{B/j}(z,\mu )F_{j/A}(x,\mu )\left[ C_{F}\left( \ln ^{2}\frac{Q^{2}}{(q_{T}^{S})^{2}}-3\ln \frac{Q^{2}}{(q_{T}^{S})^{2}}\right) \right] \nonumber \\
 &+& \ln \frac{Q^{2}}{(q_{T}^{S})^{2}}\Bigl[ (D_{B/b}\otimes
 P_{bj}^{(1)})(z,\mu )F_{j/A}(x,\mu )\nonumber\\
&+&D_{B/j}(z,\mu )(P_{ja}^{(1)}\otimes F_{a/A})(x,\mu )\Bigr] \Biggr \}.
\end{eqnarray}
 This expression agrees with Eqs.\,(\ref{NLOcs}-\ref{V1NLO}). Technically,
this integration can be easily realized by going back to the $ b $-space
and using relationships \begin{eqnarray}
\int _{0}^{q^{S}_{T}}J_{0}(q_{T}b)q_{T}dq_{T} & = & \frac{q_{T}^{S}}{b}J_{1}(q_{T}^{S}b),
\end{eqnarray}
\begin{eqnarray}
\int _{0}^{+\infty }J_{1}(ab)\ln \frac{b^{2}}{b^{2}_{0}}bdb & = & -\ln \left( a^{2}\right) ,
\end{eqnarray}
\begin{eqnarray}
\int _{0}^{+\infty }J_{1}(ab)\ln ^{2}\frac{b^{2}}{b^{2}_{0}}bdb & = & \ln ^{2}\left( a^{2}\right) .
\end{eqnarray}

\addtocontents{toc}{\vspace{1\baselineskip} \bf Bibliography \hfill 140}

\end{document}